\documentclass[twocolumn]{aastex631}
\usepackage{color, colortbl}
\usepackage{array}

\newcommand\helios{\texttt{HELIOS}}
\newcommand\pandexo{\texttt{Pandexo}}
\newcommand\platon{\texttt{PLATON}}

\newcommand\OtwoCOtwo{O$_2\vert$CO$_{2(100{\rm ppm})}$}
\newcommand\NtwoCHfour{N$_2\vert$CH$_{4(1\%)}$}
\newcommand\OtwoSOtwo{O$_2\vert$SO$_{2(100{\rm ppm})}$}
\newcommand\OtwoHtwoO{O$_2\vert$H$_2$O$_{(1{\rm ppm})}$}

\definecolor{Gray}{gray}{0.9}
\definecolor{LightCyan}{rgb}{0.88,1,1}
\definecolor{Black}{gray}{0}
\definecolor{green}{rgb}{0.88, 1, 0.88}
\definecolor{purple}{rgb}{0.96, 0.88, 1}

\shorttitle{The Detectability of Rocky Planet Surface and Atmosphere}
\shortauthors{Whittaker et al.}

\begin{document}

\title{The Detectability of Rocky Planet Surface and Atmosphere Composition with JWST: The Case of LHS 3844b}

\correspondingauthor{Emily A. Whittaker}
\email{emilywhittaker@g.ucla.edu}

\author[0000-0002-1518-7475]{Emily A. Whittaker}
\affiliation{University of California, Los Angeles, Department of Earth, Planetary, and Space Sciences, Los Angeles, CA 90095-1567, USA}
\affiliation{University of Maryland, Department of Astronomy, College Park, MD 20742, USA}

\author[0000-0002-2110-6694]{Matej Malik}
\affiliation{University of Maryland, Department of Astronomy, College Park, MD 20742, USA}

\author[0000-0003-2775-653X]{Jegug Ih}
\affiliation{University of Maryland, Department of Astronomy, College Park, MD 20742, USA}

\author[0000-0002-1337-9051]{Eliza M.-R. Kempton}
\affiliation{University of Maryland, Department of Astronomy, College Park, MD 20742, USA}

\author[0000-0003-4241-7413]{Megan Mansfield}
\affiliation{Steward Observatory, University of Arizona, Tucson, AZ 85721, USA}

\author[0000-0003-4733-6532]{Jacob L.\ Bean}
\affiliation{Department of Astronomy \& Astrophysics, University of Chicago, Chicago, IL 60637, USA}

\author{Edwin S. Kite}
\affiliation{Department of the Geophysical Sciences, University of Chicago, Chicago, IL 60637, USA}

\author{Daniel D. B. Koll}
\affiliation{Department of Earth, Atmospheric, and Planetary Sciences, MIT, Cambridge, MA 02139, USA}

\author{Timothy W. Cronin}
\affiliation{Department of Earth, Atmospheric, and Planetary Sciences, MIT, Cambridge, MA 02139, USA}

\author[0000-0003-2215-8485]{Renyu Hu}
\affiliation{Jet Propulsion Laboratory, California Institute of Technology, Pasadena, CA 91109, USA}
\affiliation{Division of Geological and Planetary Sciences, California Institute of Technology, Pasadena, CA 91125, USA}

\begin{abstract}

The spectroscopic characterization of terrestrial exoplanets will be made possible for the first time with JWST.  One challenge to characterizing such planets is that it is not known \textit{a priori} whether they possess optically thick atmospheres or even any atmospheres altogether.  But this challenge also presents an opportunity — the potential to detect the surface of an extrasolar world. This study explores the feasibility of characterizing the atmosphere and surface of a terrestrial exoplanet with JWST, taking LHS 3844b as a test case because it is the highest signal-to-noise rocky thermal emission target among planets that are cool enough to have non-molten surfaces. We model the planetary emission, including the spectral signal of both atmosphere and surface, and we explore all scenarios that are consistent with the existing Spitzer 4.5 $\mu$m measurement of LHS 3844b from \citet{Kreidberg_2019}. In summary, we find a range of plausible surfaces and atmospheres that are within 3 $\sigma$ of the observation — less reflective metal-rich, iron oxidized and basaltic compositions are allowed, and atmospheres are restricted to a maximum thickness of 1 bar, if near-infrared absorbers at $\gtrsim$ 100 ppm are included. We further make predictions on the observability of surfaces and atmospheres, perform a Bayesian retrieval analysis on simulated JWST data and find that a small number, $\sim$ 3, of eclipse observations should suffice to differentiate between surface and atmospheric features. However, the surface signal may make it harder to place precise constraints on the abundance of atmospheric species and may even falsely induce a weak H$_2$O detection.

\end{abstract}


\section{Introduction} \label{sec:intro}

\subsection{Characterizing Terrestrial Exoplanets in the Era of JWST}

During the next few years the James Webb Space Telescope (JWST) will serve as the prime observatory enabling us to characterize rocky exoplanets using spectroscopic information about the planetary thermal emission. The challenges to characterizing such planets include their smaller sizes and increased potential for atmospheric diversity, compared to more well-studied hot Jupiters.  Another less-appreciated challenge is that we lack \textit{a priori} knowledge of whether individual terrestrial exoplanets possess atmospheres at all.  This leads to a potential source of confusion in characterizing rocky worlds --- it is not initially clear whether a planet's emission originates from its surface, or from its optically thick atmosphere.  (A third possibility is a semi-optically thin atmosphere, for which the surface is seen directly at certain wavelengths corresponding to transparent windows through the atmosphere.)  To optimize the scientific yield of observations of rocky exoplanets with JWST, it is therefore crucial to understand the spectroscopic fingerprints of both their atmospheres and surfaces and how these translate to an emission signal recorded by the telescope. 

Terrestrial exoplanets that are found to not possess atmospheres could present the first opportunities to characterize the surface composition of planets beyond our solar system. The analogs of airless or nearly airless rocky bodies in the solar system are Mercury, Mars, the Moon, and various asteroids. Before they could be spatially well resolved, many of these objects were studied using the spectra of reflected solar radiation or of thermal emission. The lunar mare are generally dark and absorb strongly at 1 $\micron$ and 2 $\micron$, indicating that they are basaltic in composition \citep{Pieters_1978}, while the lunar highlands are bright and relatively featureless, pointing to a plagioclase feldspar composition \citep{Pieters_1986}.  The spectral features of the Martian surface, together with its visually red appearance and its polarization properties, show that it is rich in ferric oxides \citep{McCord_1971}. Spectra of Mercury's surface initially indicated that it had a similar composition to that of the lunar highlands due to its relatively flat spectrum \citep{Blewett_1997}. However, recent observations from spacecraft of Mercury have pointed to a surface which is closer to ultramafic \citep{Nittler_2011} --- the more recent revision demonstrates the challenges in using remote sensing data to uniquely constrain planetary surface composition.  Present-day Earth possesses a granitoid surface, a tertiary crust that is a record of plate tectonics and the incorporation of water in subducted crustal materials. All of these solar system examples present plausible surface compositions for terrestrial exoplanets. Additional possibilities include a metal-rich surface, which would be indicative of a world with the mantle stripped off \citep{Hu_2012}, among others, as discussed in Section~4.2 of \citealt{Mansfield_2019}).

At the limit of airless planets, \citep{Hu_2012} proposed to use reflected and emitted light spectra of terrestrial exoplanets to constrain their surface compositions. However, the surface characterization may be complicated by the fact that such planets could possess substantial atmospheres, which could obscure the surface from view. We therefore need to explore whether an exoplanet's surface composition can be constrained in the absence of an atmosphere, and whether degeneracies exist between determining atmospheric and surface properties for cases in which an overlying atmosphere is present. This is the purpose of our current work.

\subsection{The Rocky Super-Earth LHS 3844b}

In this study, we focus on the specific test case of the hot, rocky exoplanet LHS 3844b.  We assess our ability to characterize the planet with JWST with a focus on plausible atmospheric and surface compositions, given our current knowledge of the planet's properties. LHS~3844b, discovered by the Transiting Exoplanet Survey Satellite (TESS) in 2018, is a presumed tidally-locked super-Earth with a radius of 1.303 $\pm$ 0.022 $R_{Earth}$ and orbits its red dwarf M5-type host star every 11 hours \citep{Vanderspek_2019}. With an emission spectroscopy metric (ESM) of 28 \citep{Kempton_2018}, LHS 3844b provides one of the strongest emission signals for any exoplanet that is believed to have a dayside made of solid rock (in contrast to even hotter magma worlds), making it a prime target for future characterization. 

Due to the very high irradiation level from the host star which promotes atmospheric escape any hydrogen-dominated primordial atmosphere of LHS 3844b is thought to have been lost over the course of its lifetime \citep{owen13, Owen_2017}. However, the presence of a secondary atmosphere is difficult to predict by theory as such can be produced through many pathways during the planet's evolution, e.g., degassing during accretion \citep{elkins-tanton08}, vaporization of a molten mantle \citep{schaefer12}, reactions between nebula gas and magma \citep{kite21}, high-energy impacts \citep{lupu14} and volcanism \citep{gaillard14}. However, there are also a number of potential conditions that would prevent the formation and retention of a thick secondary atmosphere on LHS 3844b or similar planets, including (i) a volatile-poor formation and associated inhibition of late-stage degassing \citep{kite20}, (ii) stripping of the atmosphere due to impacts \citep{schlichting15}, (iii) primordial atmosphere loss, as heavier molecules will be dragged away alongside H$_2$ when the flux is high \citep{hunten87}, and (iv) a high stellar UV flux preventing the revival of an atmosphere \citep{kite20}. 

Empirical data appear to strengthen the picture that LHS 3844b does not possess a thick atmosphere. Namely, Spitzer phase curve observations with IRAC Channel 2 at 4.5 $\micron$ by \citet{Kreidberg_2019} measured a high dayside brightness temperature of $1040\pm40$~K together, an undetectably small emission from the nightside and a phase curve offset consistent with zero. Since the main driver of heat redistribution is large-scale atmospheric circulation, the finding of a negligible longitudinal heat transport points toward the absence of a thick atmosphere \citep{showman13, koll22}. Furthermore, comparing atmospheric model predictions to the measured eclipse depth of $380\pm40$ ppm \citet{Kreidberg_2019} narrowed the possible surface pressure to 10 bars or less. Simulating atmospheric mass loss over the planet's history, \citet{Kreidberg_2019} further concluded that only with an initial water inventory of at least $\sim 100$ Earth oceans could atmospheric erosion be prevented. Additionally, more recent ground-based observations using 13 transits in the optical to near-infrared by \citet{diamond-lowe21} reveal a flat transmission spectrum, which is also consistent with a bare rock scenario. Specifically, the transmission spectrum observations ruled out a hydrogen-dominated atmosphere with high confidence (unless there are high-altitude clouds that would also flatten the spectrum) and disfavored a water-steam atmosphere with a surface pressure $\geq 0.1$ bar.

The most promising strategy to constrain the surface composition of rocky exoplanets is through secondary eclipse observations which probe the planetary reflection and thermal emission spectrum. In contrast to the transmission spectrum, the planetary surface signal is imprinted on the emission spectrum in wavelengths where the atmosphere is optically thin. In addition, as the thermal emission signal scales with the planetary temperature, secondary eclipse spectroscopy appears to be especially well suited for atmospheric detection and characterization of hot planets, such as LHS 3844b \citep{morley17, koll19a}. Due to these reasons, we model the planetary emission spectrum and make predictions on the detectability of LHS 3844b's atmosphere and surface using secondary eclipse observations in the current work.

\subsection{Outline of This Study}

First, we constrain the parameter space of surface and atmospheric properties of LHS 3844b that is consistent with the measured Spitzer eclipse depth (Sect.~\ref{sec:spitzer_constraints}). It turns out that the consistent surfaces are metal-rich, basaltic and iron oxidized crusts (see Fig.~\ref{fig:spitzer_surface_comp}) and atmospheres up to $\sim 1$ bar surface pressure, if a near-infrared absorber at $ \geq 100$ ppm is included (see Fig.~\ref{fig:kreidberg_plot} and Table~\ref{table:max_pressures}). Among all surface and atmosphere combinations that are consistent with observations we focus on two limiting cases. On one end, we assume that there is no atmosphere and the planetary signal stems solely from the surface; this is the {\it no atmosphere limit} (Sect.~\ref{sec:no_atmo}). On the other end, we include an atmosphere overlying a rocky surface, exploring a large number of compositions and thicknesses. We assess the maximum surface pressure that is consistent with Spitzer, and for each tested atmospheric background and gas absorber pair we pick the model that provides the largest gas absorber features in the planetary spectrum; this the {\it thick atmosphere limit} (Sect.~\ref{sec:thick_atmo}). As shown in Fig.~\ref{fig:spitzer_eclipse_spectra} the atmosphere models in this limit consist of the following compositions and surface pressures: O$_2$ with 100 ppm CO$_2$ and 0.1 bar, O$_2$ with 100 ppm SO$_2$ and 1 bar, O$_2$ with 1 ppm H$_2$O and 10 bar and N$_2$ with CH$_4$ and 1 bar. For the no and thick atmosphere limits we assess the observability with JWST, determining to what extent JWST will help characterize this planet under optimal conditions in terms of surface and atmosphere visibility within the constraint of the Spitzer measurement (Sect.~\ref{sec:observability}). These results are shown in Figures~\ref{fig:surface_bb}, \ref{fig:surface_bb_implausible}, \ref{fig:atmo_vs_noatmo_isolated} and Tables~\ref{table:distinguish_surf_R3}, \ref{table:distinguish_surf_isolated}, \ref{table:atmo_v_noatmo}, illustrating that the more reflective surfaces and all of the atmospheric setups will be detectable with less than a handful of eclipse observations. Finally, using a Bayesian framework we explore how well atmospheric abundances may be constrained from observations and what impact the surface signal has on the atmospheric retrieval (Sect.~\ref{sec:retrieval}). We find that, while the presence of atmospheric species can be recovered with 5 eclipses, it will not be straightforward to pinpoint their precise abundance. On top of that, the surface signal makes it harder for the retrieval model to disfavor the presence of atmospheric species and may even mimic atmospheric H$_2$O (see Figures~\ref{fig:retrieval_no_surf_no_atmo}, \ref{fig:retrieval_o2_h2o} and Tables~\ref{table:platon_atmo}, \ref{table:platon_noatmo}). We conclude and summarize in Sect.~\ref{sec:discussion}, discussing new constraints for the surface and atmosphere of this planet based on our results and how these constraints are narrower than the previous findings of \citet{Kreidberg_2019}. Lastly, we discuss implications for characterizing terrestrial exoplanet surfaces and atmospheres in the coming years with JWST.

\section{Methods} 

\subsection{Radiative Transfer Modeling}

We generate model spectra of LHS 3844b with different assumed atmospheric species and surfaces with the open-source, 1D radiative transfer (RT) code \helios{}\footnote{https://github.com/exoclime/helios} \citep{Malik_2017, Malik_2019a, Malik_2019b}, which simulates the atmosphere in radiative-convective equilibrium and the corresponding planetary emission. In contrast to most radiative transfer models used in the exoplanet field, \helios{} takes into account the radiative effects of the surface on both the atmosphere and the resulting planetary spectrum. Previously, only a grey surface albedo, constant over all wavelengths, could be modeled with \helios{}, as described in detail in \citet{Malik_2019b}. We have since then expanded the code's functionality to include a surface albedo as a function of wavelength, enabling the consideration of realistic surface albedos in the model. We describe updates to the \helios{} code and the new method of modeling the surface in Appendix~\ref{app:surface}. 

In \helios{} the radiative transfer calculation is performed twice.  First, the atmospheric temperature profile and the surface temperature in radiative-convective equilibrium are obtained. For this step, we use the k-distribution method with 420 wavelength bins  between 0.245 $\mu$m and 10$^5$ $\mu$m (distributed evenly in wavenumber space) and 20 Gaussian points within each bin. Since in this work each atmospheric set-up consists of only one major absorbing species, it is not necessary to assume any correlation between opacities from different sources (as is done in the correlated-k approximation). Once the equilibrium state is found, \helios{} is used in post-processing mode to generate the emission spectrum using opacity sampling with a resolution of $R$ = 4000. 

Convection is treated in \helios{} via convective adjustment with the adiabatic coefficient $\nabla_{\rm ad} = (\partial \log T / \partial \log P)_S$ -- where $T$ is the temperature, $P$ is the pressure and $S$ is the entropy -- set to 2/7 (1/4) if a diatomic (tri-atomic) molecule acts as the atmospheric bulk gas, according to the ideal gas approximation. We model dayside-averaged conditions and use the scaling theory for heat redistribution of \citet{koll22}, their Eq.~(10), making the day-to-night heat transport dependent on the thickness of the atmosphere. Note that the scaling of the heat redistribution subtracts the total heat that is assumed to be transported to the nightside from the incoming stellar flux in order to estimate the dayside heat content, but it does not include any vertical dependency of the day-to-night heat flow. In the no-atmosphere case, we assume no global heat transport and set the redistribution parameter accordingly to $f = 2/3$ \citep{burrows08, hansen08}. Lastly, a numerical limitation of the \helios{} code is that it is not possible to model a surface without an overlaying atmosphere, and so we approximate this situation by including an atmosphere with $P_{\rm surf} = 2 \times 10^{-9}$ bar which has a $\sim 10^{-10}$ Planck mean optical depth.\footnote{According to our tests, an atmosphere with $P_{\rm surf} < 10^{-6}$ bar does not noticeably affect the surface temperature or planetary emission found in the model.}

We use a PHOENIX stellar model \citep{Husser_2013} to simulate the spectrum of the host star, downloaded from the online spectral library\footnote{ftp://phoenix.astro.physik.uni-goettingen.de/HiResFITS/} and interpolated for the stellar parameters of LHS 3844. Further input parameters include the planetary surface gravity, semi-major axis, radius of the planet, radius of the star, temperature of the star, stellar surface gravity and stellar metallicity. The adopted numerical values for these parameters are displayed in Table~\ref{table:parameters}.

\begin{table*}[ht!]
\small
\begin{center}
\begin{tabular}{ |p{2cm}|p{2cm}|p{2cm}|p{2cm}|p{2cm}|p{2.3cm}|p{1.7cm}|}
 \multicolumn{7}{c}{{\bf Planetary and Stellar Parameters}} \\
 \hline
 $R_{\mathrm{pl}}$ ($R_{\mathrm{Earth}}$) & $g_{\mathrm{pl}}$ (cm s$^{-2}$) & $a$ (AU) & $R_{\mathrm{star}}$ ($R_\odot$) & $T_{\mathrm{star}}$ (K) & $\log g_\star$ (cm s$^{-2}$) & [M/{\rm H}]$_\star$ \\
 \hline
 1.246\tablenotemark{\scriptsize{\rm a}} & 1600\tablenotemark{\scriptsize{\rm b}} & 0.00622\tablenotemark{\scriptsize{\rm a}} & 0.178\tablenotemark{\scriptsize{\rm b}} & 3036\tablenotemark{\scriptsize{\rm a}} & 5.136 & 0\tablenotemark{\scriptsize{\rm a}} \\
 \hline
\end{tabular}
\end{center}
\textsuperscript{a}\citet{Vanderspek_2019}, \textsuperscript{b}\citet{Kreidberg_2019}
\caption{Numerical values for the planetary and stellar parameters used in this study. As the planet's mass is unknown, the surface gravity is approximated assuming an Earth-like bulk density. Note that, the surface gravity affects the emission spectrum more weakly than the planetary transmission spectrum. The stellar surface gravity is calculated via Newton's law of gravity using the stellar mass $M_\star$ = 0.158 $M_\odot$ \citep{Kreidberg_2019}.}
\label{table:parameters}
\end{table*}

\subsection{Atmospheric and Surface Compositions Considered}

For the atmospheric models of LHS 3844b we choose N$_2$, O$_2$ and CO$_2$ as potential dominant atmospheric species. From an empirical perspective, there are two Solar system rocky bodies with a thick atmosphere that have N$_2$ as the main constituent: Earth, which is of similar size to LHS 3844b, and Titan. There are a number of hypothetical pathways that may lead to substantial amounts of atmospheric N$_2$ (see discussion in \citealt{lammer19}). For instance, N$_2$ outgassing can potentially counteract nitrogen fixation by lightning, meteoritic impacts, high-energy UV or flares, leading to atmospheric N$_2$ buildup \citep{chameides81, mikhail14, airapetian16}. Another theoretical prediction is an O$_2$ dominated atmosphere which is a potential product of an initially water-rich atmosphere that lost most of its hydrogen due to high irradiation from the host star \citep{Wordsworth_2014, luger15, Wordsworth_2018}. The high stellar irradiation places LHS 3844b close to the ``cosmic shoreline" \citep{zahnle17}, assuming an Earth-like bulk density. The idea here is that if this planet started with a thick envelope and a sufficient water abundance, it is expected to have undergone a Runaway Greenhouse effect at one point during its history, possibly ending up as a Venus-analogue \citep{kane14}.

Additional absorbers in our models include CO$_2$, CO, SO$_2$, CH$_4$ and H$_2$O. They are all found in the atmospheres of Earth, Mars and Venus and they are predicted to be ubiquitous and major constituents in rocky exoplanet atmospheres as well, be it from mantle degassing \citep{schaefer11, schaefer12, liggins21}, meteoritic accretion \citep{lupu14, zahnle20}, photochemistry \citep{gao15} or late-stage volcanism \citep{gaillard14}. Additionally, all of these species exhibit strong absorption bands in the infrared, providing an opportunity to be detected with JWST.

Our atmospheric grid consists of the following gas combinations. On one side we explore oxygen-rich and oxidizing atmospheres: bulk O$_2$ with H$_2$O, bulk O$_2$ with CO$_2$, bulk O$_2$ with SO$_2$, and a pure CO$_2$ scenario; and on the other side we explore nitrogen-rich and reducing atmospheres: bulk N$_2$ with CO, bulk N$_2$ with CH$_4$, bulk N$_2$ with CO$_2$, and a pure CO scenario.

\begin{figure}
\centering\includegraphics[width=0.5\textwidth]{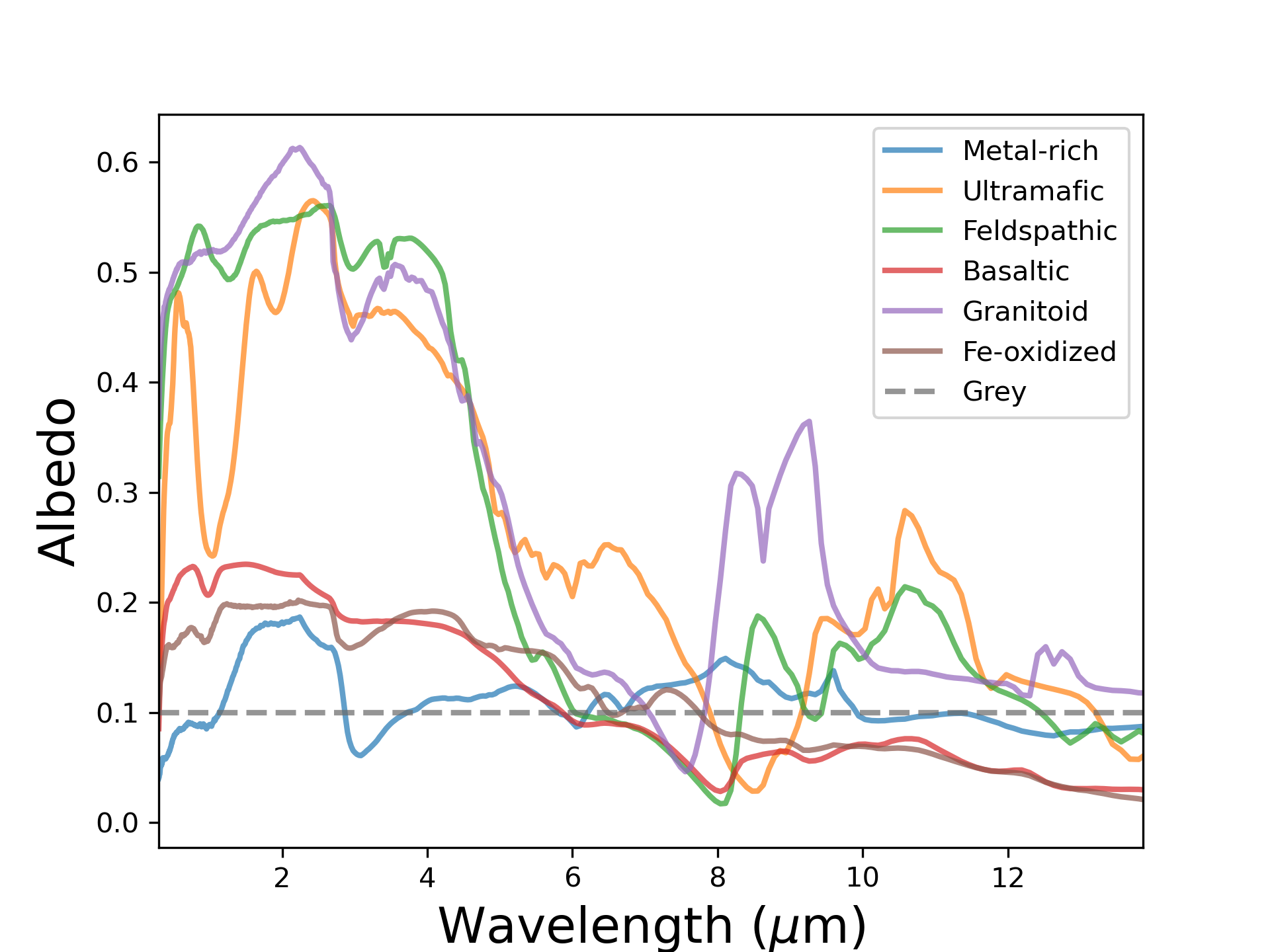}
\caption{Non-grey albedo vs. wavelength for each of the 6 tested surfaces as well as the grey control case with a constant 0.1 albedo (dashed). The non-grey albedo extends over the wavelength range from 0.3 to 25 $\mu$m and thus captures the physically correct amount of reflected stellar light as well as thermal emissivity.} The albedo data is from \citet{Hu_2012}.
\label{fig:albedo_spectra}
\end{figure}

For the surface contribution, six realistic surfaces compositions are tested: metal-rich (pyrite), feldspathic (97\% Fe-plagioclase, 3\% augite), ultramafic (60\% olivine, 40\% enstatite), basaltic (76\% plagioclase, 8\% augite, 6\% enstatite, 5\% glass, 1\% olivine), granitoid (40\% K-feldspar, 35\% quartz, 20\% plagioclase, 5\% biotite), and iron oxidized (50\% nanophase hematite, 50\% basalt), which are ubiquitous in the Solar System and common products of geological processes (see Sect.~\ref{sec:intro} and further discussion in \citealt{Hu_2012}). We use the geometric albedo spectra from \citet{Hu_2012} for these surfaces, pre-tabulated between 0.3 and 25 \micron{}, which were obtained from experimental data taken on rock powders and extended by radiative transfer modeling. Although these albedo spectra are mineralogically realistic, spectra of the surface of Mars and the Moon differ from that of mineral mixtures in the laboratory due to various effects \citep[e.g.,][]{carli15, bishop19}. We extrapolate the tabulated albedo data to the whole wavelength range of the RT calculation by using the albedo value at 0.3 \micron{} for all smaller and the value at 25 \micron{} for all larger wavelengths. Lastly, we include a uniform grey surface with a geometric albedo of 0.1 as control to test the impact of the non-grey albedo variation of the realistic surfaces. The albedo spectra used in this work are shown in Figure~\ref{fig:albedo_spectra}. 

\subsection{Opacities \& Scattering Cross-sections}

Molecular opacities are calculated with \helios-K \citep{Grimm_2015, grimm21}. Included are O$_2$ \citep{gordon17} H$_2$O \citep{polyansky18}, CO \citep{li15}, CO$_2$ \citep{rothman10}, CH$_4$ \citep{yurchenko17} and SO$_2$ \citep{underwood16}. The opacities are calculated on a wavenumber grid with a resolution of 0.01 cm$^{-1}$, assuming a Voigt profile with a wing cut-off at 100 cm$^{-1}$. For H$_2$O, CH$_4$ and SO$_2$ we include the default pressure broadening coefficients provided by the Exomol database \footnote{https://exomol.com/data/molecules/}. For O$_2$, CO and CO$_2$ we use the HITRAN broadening formalism with self-broadening only. Further included is collision induced absorption (CIA) by O$_2$-O$_2$, O$_2$-CO$_2$, CO$_2$-CO$_2$, N$_2$-N$_2$, and N$_2$-CH$_4$ \citep{richard12} and Rayleigh scattering of H$_2$O, O$_2$, N$_2$, CO$_2$ and CO \citep{cox00, sneep05, wagner08, thalman14}.

\subsection{Simulating JWST observations}

To simulate observations with JWST the python package \pandexo{}\footnote{https://exoctk.stsci.edu/pandexo/} \citep{Batalha_2017} is used. We focus on two JWST instruments that will be suitable and are planned to be utilized for rocky exoplanet characterization, MIRI LRS (Low Resolution Spectrometer) and NIRSpec G395H. Being on the longwave end of the JWST capabilities, the  bandpass of MIRI LRS (5.02 - 13.86 $\mu$m) maximizes the recorded secondary eclipse depth (planet-to-star contrast), since the emission of the hotter star falls off faster with wavelength than the planetary emission. Although the relative eclipse depth is smaller in the NIRSpec G395H range (2.87 – 5.18 $\mu$m), a larger stellar flux at these wavelengths leads to a smaller overall photon noise. Additionally, many atmospheric species possess strong spectral signatures in the near-infrared range, making NIRSpec G395H one of the preferred choices for exoplanet observations. 

When simulating observations with \pandexo{}, we insert the model values from the \helios{} spectrum (consistent with noise-free data), set a given instrument, resolution, and number of secondary eclipse measurements, and add the simulated statistical noise as error bars. For all tests done over the entire instrument wavelength range, points are rebinned using \pandexo{}'s integrated functionality which defines new wavelength bins of a constant resolution and uses the included ``uniform\_tophat\_mean" function to calculate the new bin-averaged function values together with the their uncertainty. In some cases, in addition to using the whole instrument range, we also focus on the detectability of individual spectral features by sampling over the width of these features, manually applying \pandexo{}'s rebinning procedure. As for the systematic noise floor, we assume a value of 30 ppm based on tentative estimates \citep{matsuo19, schlawin20, schlawin21}. To test this assumption, the same analysis as shown in \ref{fig:atmo_v_noatmo} has been conducted with the noise floor set to 0 ppm, which has led to indistinguishable results as our nominal setup. Therefore, we assume that the choice of noise floor has negligible impact on our results.  

\subsection{Bayesian Retrieval Modeling}
\label{sec:method_retrieval}

We retrieve on the simulated JWST spectrum using a modified version of the Bayesian retrieval code \platon{} \citep{zhang18, zhang20}. Our version, first utilized in \citet{ih21}, differs from the main branch of \platon{} in that it allows for measuring abundances of multiple species during retrieval (in a so-called ``free'' retrieval). \platon{} natively supports custom abundance profiles during forward models, but is configured to only perform equilibrium chemistry retrievals, in which the mixing ratios of all species are prescribed by the metallicity and C-to-O ratio defined globally, and temperature and pressure per layer. We relax this restriction by allowing the abundances of each species to be included as a retrieved parameters. All other details regarding radiative transfer remain the same as the original implementation.

In our free retrieval, the atmosphere is parameterized as follows. For the composition of the atmosphere, we assume two possible background gases that do not have a spectral signature (O$_2$ and N$_2$). We then assign a parameter corresponding to the vertically fixed log-abundance of each species other than the background gases and a (linear) abundance for one of the background gases. The abundance of the other background gas is then derived as the remainder from unity. The thermal structure of the atmosphere is described by using the parametric T-P profile as given by \citet{madhusudhan09}. Having been developed with gas planets in focus, the retrieval code does not offer the possibility to include a solid surface. Still, we mimic the location of the surface in the retrieval modeling by the pressure level above which an isotherm is assumed, set by the parameter $P_3$ in the formalism of \citet{madhusudhan09}. Since a surface radiating as a blackbody is equivalent to an optically thick atmosphere of a single temperature, $P_3$ should in theory correspond to a limit on the surface pressure. Approximating the surface with a blackbody does not allow us to make any statements about a non-zero surface albedo and thus we do not retrieve on the surface but merely on the atmosphere in this work.

We perform our retrievals on the spectra generated using \pandexo{} by binning (resampling) the data down to a resolution of $R=100$ and using uncertainties based on five simulated secondary eclipse observations. 

\section{Results}

\subsection{LHS 3844b Spitzer eclipse depth constraint}
\label{sec:spitzer_constraints}

In the following we first explore which of the tested surface types agree with the Spitzer eclipse depth measurement without considering an atmosphere ({\it no atmosphere limit}). Then, analogously, we find the atmospheric models (varying composition and thickness) consistent with the Spitzer measurement. Among those we determine the atmospheres with the largest spectral features that may allow for characterization with JWST ({\it thick atmosphere limit}). 

Whether or not a model is in agreement with the Spitzer observation is determined by comparing the \helios{}-generated spectrum of the planet for each surface with the observed Spitzer 4.5 $\mu$m eclipse depth. The model is considered consistent with the data point if the simulated emission over the Spitzer bandpass, obtained by convolving the model spectrum with the Spitzer IRAC channel 2 bandpass function\footnote{We use the Spitzer IRAC.I2 filter transmission function from http://svo2.cab.inta-csic.es/theory/fps/.}, is within a 3 $\sigma$ confidence interval from the observed value.

\subsubsection{No atmosphere limit}
\label{sec:no_atmo}

\begin{figure}[ht]
\includegraphics[width=0.5\textwidth]{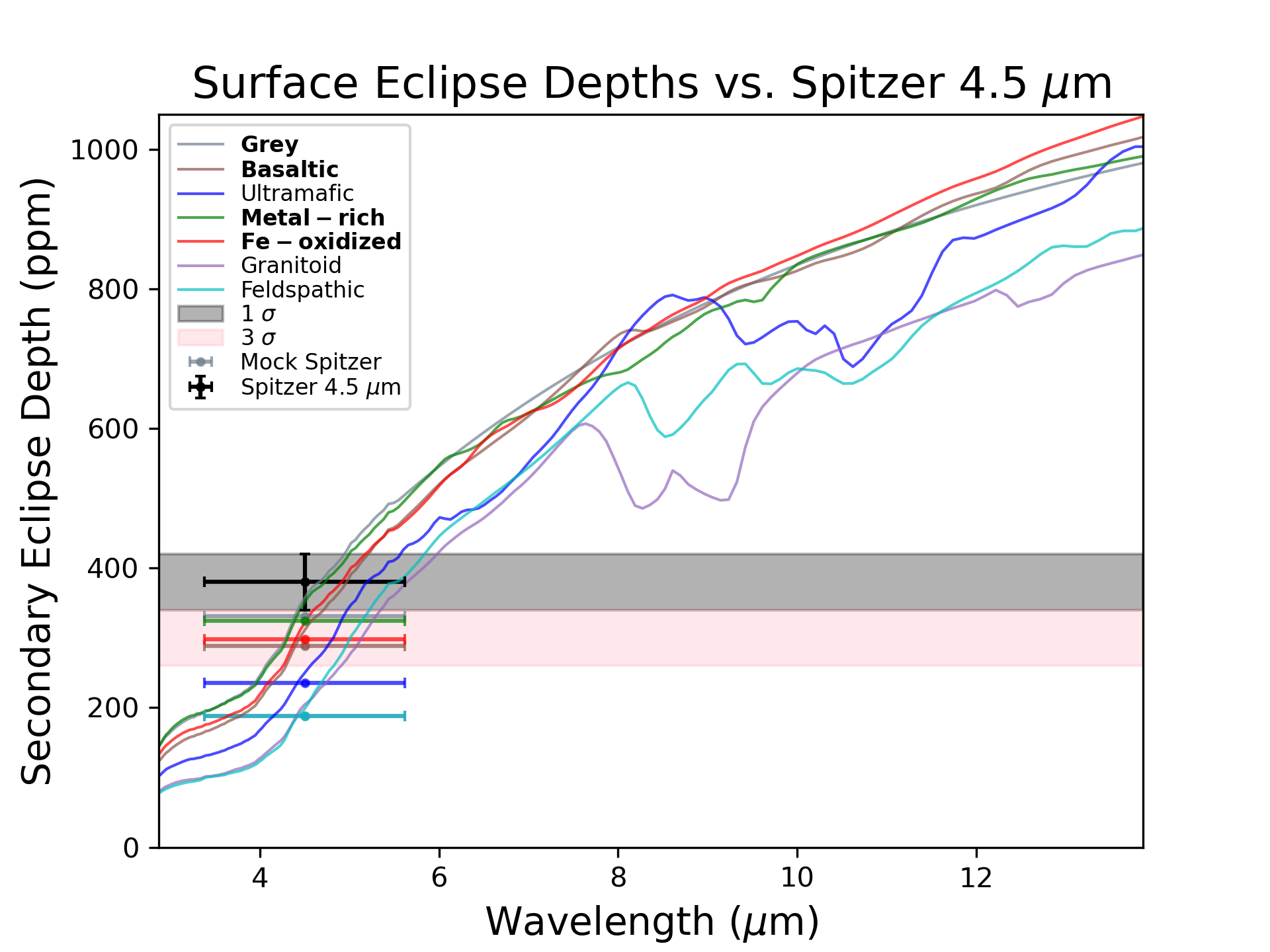}
\caption{Secondary eclipse spectra for each of the tested surfaces and the grey albedo control case, shown at $R$ = 100.  The actual Spitzer 4.5 $\micron$ measurement is shown in black, where the y-axis error bar represents the uncertainty and the x-axis error bar represents the width of the bandpass. Modeled Spitzer points for each surface are shown in their respective color. We find the grey, metal-rich, iron oxidized and basaltic surfaces to be consistent (bold), i.e., $< 3$ $\sigma$ from the observation, and the ultramafic, granitoid and feldspathic surfaces to be inconsistent with the Spitzer measurement. The granitoid and feldspathic mock Spitzer points are almost directly overlapping.}
\label{fig:spitzer_surface_comp}
\end{figure}

In the no atmosphere limit we generate planetary emission spectra including the surface albedo signal only. Each model assumes that the whole planetary dayside is covered by one of the considered surface compositions. The secondary eclipse spectra of the surface-only models are shown in Figure~\ref{fig:spitzer_surface_comp}. Based on this analysis, the Spitzer data point is most consistent with a metal-rich surface (1.39 $\sigma$ from Spitzer measurement), followed by an iron oxidized (2.05 $\sigma$) and a basaltic (2.28 $\sigma$) surface. The ultramafic (3.61 $\sigma$), feldspathic (4.79 $\sigma$) and granitoid (4.80 $\sigma$) surfaces are excluded by the data based on the assumed confidence limit. Curiously, all explored models predict a significantly smaller eclipse depth than Spitzer with no single model prediction within 1 $\sigma$ of the observation. First, this could be due to stochastic noise (although unlikely), but it could also indicate that the error of the Spitzer observation is underestimated due to an unknown systematic bias. As the reanalysis of the raw Spitzer phase curve of \citet{Kreidberg_2019} is beyond the scope of this work, we limit ourselves to taking the Spitzer measurement at its face value. Second, it should be noted that not only is the albedo affected by grain size but even more the shortwave albedo can be lowered by small amounts of minor surface constituents, e.g., carbon, like on Mercury's surface \citep{izenberg14}, and metallic iron, produced by space weathering. Thus, by increasing the absorption of light in the shortwave and consequently boosting the thermal emission in the infrared, minor modeling assumptions can in theory affect the ordering of surfaces as well as the answer to whether they are consistent with the Spitzer data point. Testing these assumptions is again beyond the scope of this work. 

Compared to the previous analysis of \citet{Kreidberg_2019} we obtain consistently shallower eclipse depths for the same surface compositions and also find the ultramafic crust inconsistent with the observation, which is in contrast to the modeling result in \citet{Kreidberg_2019}. This discrepancy is discussed in Sect.~\ref{sec:kreidberg_comp}.

\subsubsection{Including an atmosphere}
\label{sec:with_atmo}

\begin{figure*}
\centering\includegraphics[width=17cm]{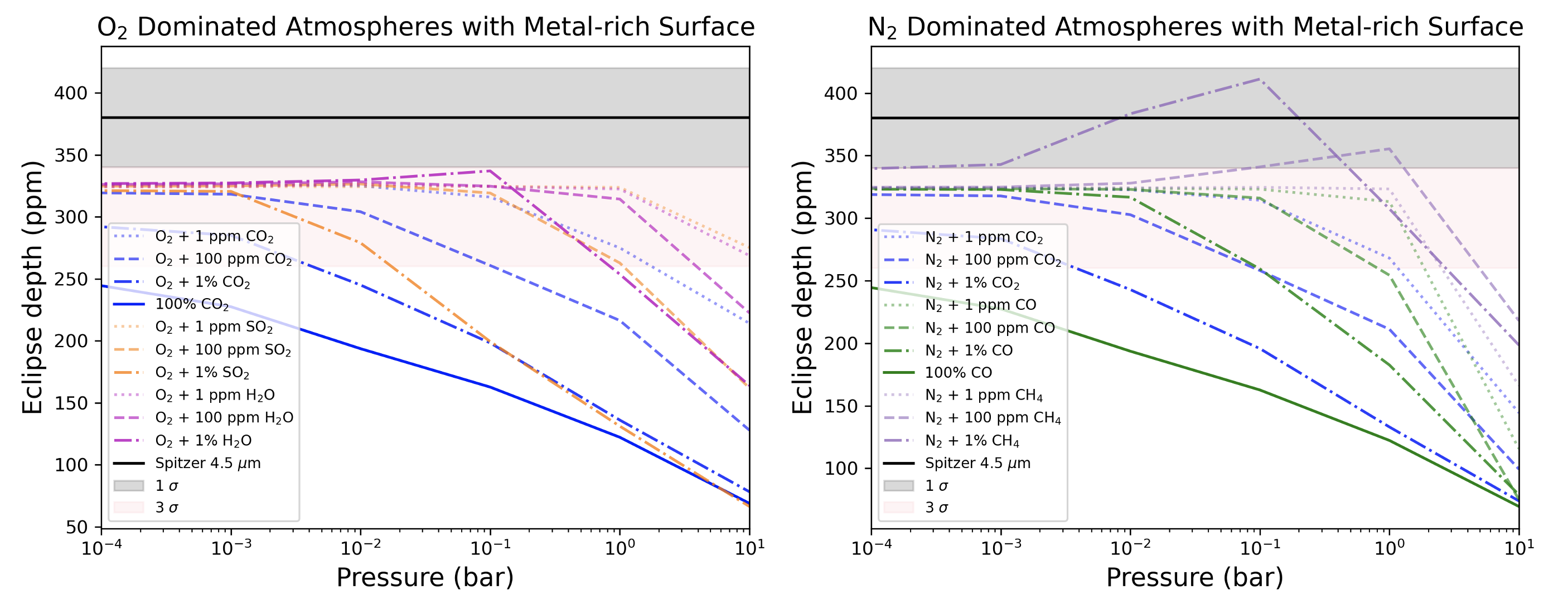}
\caption{Predicted Spitzer 4.5 $\mu$m eclipse depth for the explored atmospheric models in combination with a metal-rich surface as a function of surface pressure compared to the Spitzer measurement (black horizontal line). Oxidizing/O$_2$-dominated atmospheres are on the left and reducing/N$_2$-dominated atmospheres on the right. The grey shaded area corresponds to 1 $\sigma$ uncertainty and the pink shaded area to 3 $\sigma$ uncertainty in the negative direction from the observed value.}
\label{fig:kreidberg_plot}
\end{figure*}

As bottom boundary our nominal atmosphere models use the metal-rich surface. Being most consistent with Spitzer, this choice maximizes the allowed atmospheric parameter space. Following the style of Figure 3 from \citet{Kreidberg_2019}, our Figure~\ref{fig:kreidberg_plot} shows the modeled eclipse depths for each tested atmospheric composition plotted across the range of probed surface pressures. The atmospheric temperature-pressure (T-P) profiles for the 10 bar and 1 mbar atmospheres are shown in Fig.~\ref{fig:tp}. The associated maximum surface pressures consistent with Spitzer for the 20 combinations of gases and their abundances are listed in Table~\ref{table:max_pressures}. We find that the modeled eclipse depths result from a combination of three distinct effects: day-night heat transport, greenhouse warming and strength of gaseous absorption at 4.5 $\micron$. Optically thick atmospheres transport more heat from the day to the night side than thinner atmospheres and consequently lead to a smaller dayside emission \citep{koll22}. This mechanism translates to a decreasing eclipse depth with increasing surface pressure and also a decreasing eclipse depth with increasing absorber abundance. The second effect is the greenhouse effect. As all of the explored absorbers are greenhouse gases, the T-P profiles increase with pressure in optically thick atmospheric regions. (Note that the surface is smoothly connected to the first atmospheric layer via convective stability, i.e., temperature jumps at surface are not permitted.) If the planetary emission originates at or near the surface, increasing surface pressure consequently leads to an increased thermal emission and larger eclipse depth. Similarly, increasing the  greenhouse gas abundance increases the strength of the greenhouse effect which warms the surface leading to a larger eclipse depth in spectral window regions (the total energy budget remains constant in radiative equilibrium). Finally, some species (CO$_2$, CO) exhibit a strong absorption band around 4.5 $\mu$m whereas others do not (H$_2$O, SO$_2$) or even exhibit an absorption window (CH$_4$) which affects the eclipse depth at that wavelength location. These combined effects lead to non-monotonic trends in the eclipse depth versus surface pressure or absorber abundance (as visible in Fig.~\ref{fig:kreidberg_plot}) and also non-monotonic trends in the maximum surface pressure consistent with Spitzer versus the absorber abundance (as visible in Table~\ref{table:max_pressures}).

Overall, see Table~\ref{table:max_pressures}, we find that some atmospheres with a surface pressure of 10 bar are consistent with the Spitzer measurement, but those possess a mixing ratio of only 1 ppm of a near-infrared absorber, which in the vast majority of cases has a marginal effect on the atmospheric extinction (1 ppm of H$_2$O being the exception among our models). Once absorbing species with a mixing ratio of at least 100 ppm are included, the maximum surface pressure for any tested atmosphere is 1 bar, obtained for compositions with CH$_4$, SO$_2$ or H$_2$O. With CO$_2$ or CO as absorber, the maximum surface pressure allowed decreases to 0.1 bar. 

The eclipse depths of the atmospheric models with O$_2$ and CO$_2$ deviate from the model predictions in \citet{Kreidberg_2019} and also the overall maximum surface pressure is lower than the previously obtained maximum limit of 10 bar previously found. These differences are discussed in Sect.~\ref{sec:kreidberg_comp}.

Lastly, exploring the impact of the surface on the allowed atmospheric parameter space, Figure~\ref{fig:kreidberg_plot_basaltic} shows the analogous suite of atmospheric models as Fig.~\ref{fig:kreidberg_plot} but with a basaltic surface. As expected, the modeled eclipse depths for each atmospheric composition are generally more consistent with Spitzer for the metal-rich surface than the basaltic surface since the latter is overall more reflective and thus less consistent with the Spitzer observation.

\subsubsection{Determining the thick atmosphere limit}
\label{sec:thick_atmo}

\begin{figure*}
\hskip-1cm\centering\includegraphics[width=13cm]{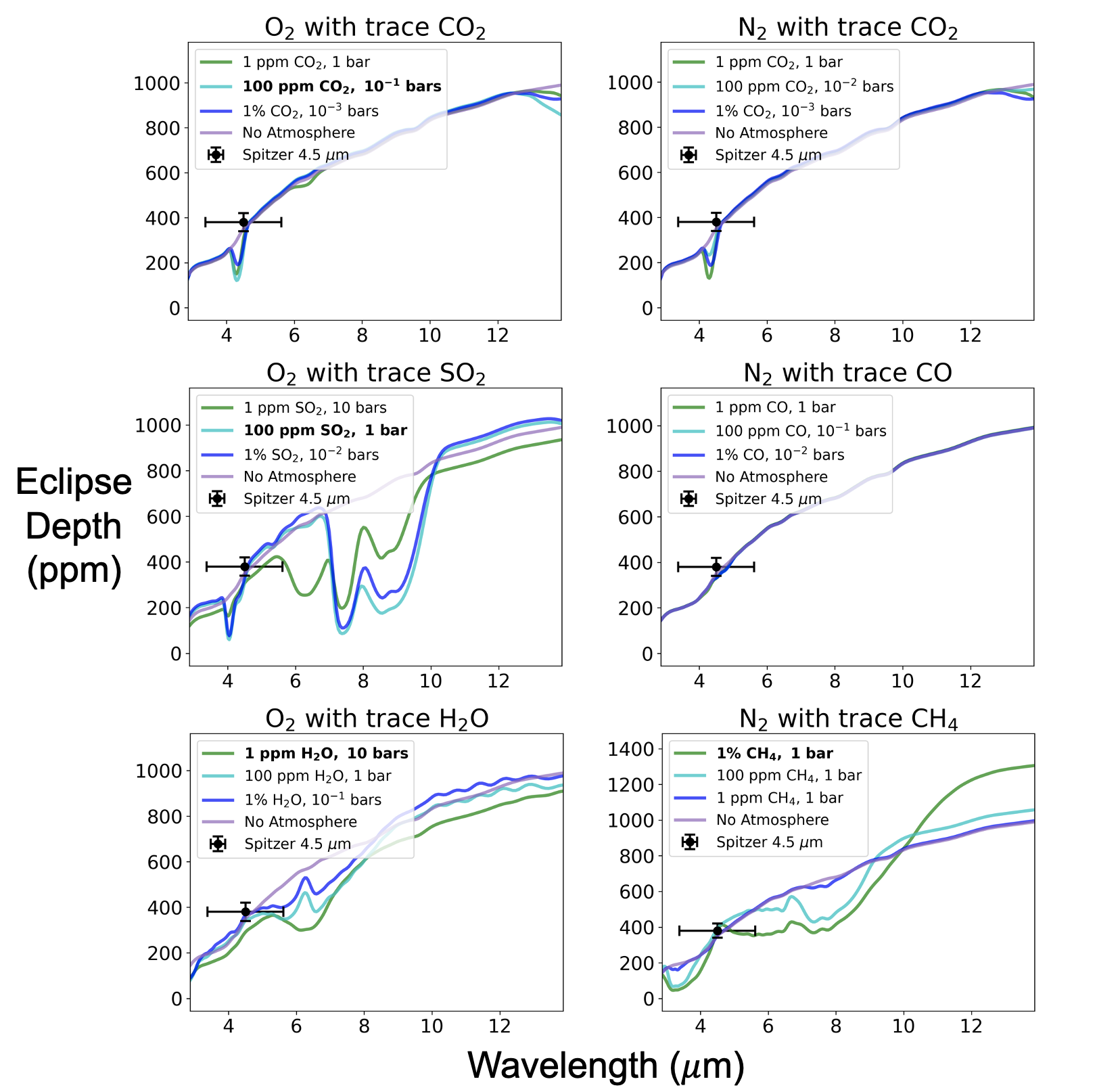}
\caption{Secondary eclipse spectra for each atmosphere \& metal-rich surface model at the maximum surface pressure consistent with the Spitzer observation. On each plot, the spectrum labelled "No Atmosphere" represents the surface-only spectrum for the metal-rich surface for comparison. The Spitzer 4.5 $\micron$ point is shown in black, where the y-axis error bar represents the uncertainty and the x-axis error bar represents the width of the bandpass. The atmospheric models with the largest features (bold) are used in the JWST observability analysis. No models are picked for the N$_2$ with CO$_2$ and the N$_2$ with CO cases as the former is very similar to O$_2$ with CO$_2$ and the latter does not exhibit any noticeable features.}
\label{fig:spitzer_eclipse_spectra}
\end{figure*}

In the following we pick for each gas pair (background and infrared absorber) the atmospheric model that provides the largest spectral features. Those models are then used for the JWST observability analysis presented in Sect.~\ref{sec:jwst_thick_atmo} and the retrieval analysis in Sect.~\ref{sec:retrieval}. The eclipse spectra of the atmospheric models that are consistent with the Spitzer measurement are shown in Figure~\ref{fig:spitzer_eclipse_spectra}. Since the atmospheric optical thickness depends on both the amount of absorbers in the atmosphere and the overall atmospheric density, the size of absorption features is degenerate in those parameters. As CO$_2$ and CO are strongly absorbing in the Spitzer bandpass, the only atmospheric models not excluded by Spitzer are those that either have no significant amount of these absorbers or low surface pressures. Hence, model spectra including those two absorbers have only weak features. The models including CO$_2$ appear very similarly, independent of the background gas, as the only noticeable absorption features are due to CO$_2$. Thus, we use only one of these compositions for further analysis, choosing the O$_2$ with 100 ppm CO$_2$ model. No model with CO exhibits visible features and thus no CO models are used for further analysis. In contrast, H$_2$O, SO$_2$ and CH$_4$ have no absorption feature directly at 4.5$\micron$ which allows for larger atmospheric abundances of these species, still satisfying the Spitzer constraint. Among the corresponding models, we find that O$_2$ with 100 ppm SO$_2$ and N$_2$ with 1\% CH$_4$ provide the largest spectral features. Specifically, SO$_2$ possesses a large double-peaked feature at 7 - 9 $\micron$ and a smaller one at 4 $\micron$ and CH$_4$ has strong absorption bands at 2 - 3 $\micron$ and from 4 to 8 $\micron$. Lastly, the O$_2$ with 1 ppm H$_2$O model provides the largest spectral feature among all models that include H$_2$O. Interestingly, the striking feature at 5.5 - 7.5 $\micron$ is not due to H$_2$O alone, but also has a contribution from O$_2$-O$_2$ CIA, which covers the small dip at 6.3 $\micron$ in the H$_2$O opacity.

The four atmospheric models picked for the JWST observability analysis in the thick atmosphere limit are O$_2$ with 100 ppm SO$_2$ and $P_{\rm surface} = 1$ bar (hereafter \OtwoSOtwo{}), N$_2$ with 1$\%$ CH$_4$ and $P_{\rm surface} = 1$ bar (\NtwoCHfour{}), O$_2$ with 100 ppm CO$_2$ and $P_{\rm surface} = 10^{-1}$ bar (\OtwoCOtwo{}) and O$_2$ with 1 ppm H$_2$O and $P_{\rm surface} = 10$ bar (\OtwoHtwoO{}).

\begin{table*}
\small
\hskip-1.5cm\begin{tabular}{|c|c|c|c|c|c|c|}
 \multicolumn{7}{c}{{\bf Maximum Pressure (bar) Consistent with Spitzer Eclipse Depth}} \\
 \hline
  & N$_2$ with CO$_2$ & N$_2$ with CO & N$_2$ with CH$_4$ & O$_2$ with CO$_2$ & O$_2$ with SO$_2$ & O$_2$ with H$_2$O\\
 \hline
 1 ppm & 1 & 1 & 1 & 1 & 10 & \textbf{10}\\
     100 ppm & $10^{-2}$ & $10^{-1}$ & 1 & $\bf{10^{-1}}$ & \textbf{1} & 1\\
 1$\%$ & $10^{-3}$ & $10^{-2}$ & \textbf{1} & $10^{-3}$ & $10^{-2}$ & $10^{-1}$\\
 100$\%$ & No Soln. & No Soln. & Not Modeled & No Soln. & Not Modeled & Not Modeled\\
 \hline
\end{tabular}
\caption{Maximum surface pressure (bar) consistent with the Spitzer 4.5 $\mu$m eclipse depth at 3 $\sigma$, for each atmospheric composition modeled. Abundances listed in the left hand column correspond to abundance of the secondly-listed gas. The overall maximum surface pressure consistent with Spitzer is 10 bar among all set-ups and 1 bar once an infrared absorber at $\geq$ 100 ppm is included. The atmospheric models in bold possess the largest features and are used in the JWST observability analysis. If a value is listed as ``No Soln.", then there was no solution which was consistent with Spitzer.}
\label{table:max_pressures}
\end{table*}

\subsection{Observability with JWST}
\label{sec:observability}   

In the following, we first explore the observability of surfaces without an overlying atmosphere (no atmosphere limit). For that we use all six surface types shown in Fig.~\ref{fig:albedo_spectra} including the three surfaces excluded by the Spitzer measurement for general applicability of our results (see Sect.~\ref{sec:no_atmo}). In contrast, the observability of atmospheres is assessed only for the atmospheric models that are consistent with the recorded Spitzer eclipse depth and that provide the largest signal for each background composition and gas absorber (thick atmosphere limit, see Sect.~\ref{sec:thick_atmo}).

\subsubsection{No Atmosphere limit}

First, we test how many secondary eclipse observations would be necessary with JWST in order to distinguish each surface emission from a blackbody spectrum using the MIRI LRS or the NIRSpec G395H instrument. For the two models to be considered ``distinguishable", a chi-square test is conducted and a p-value is obtained for a given number of eclipses, where the p-value represents the probability that the simulated data from the model to be tested are consistent with a reference model. We begin with one eclipse and increase in integer steps until the two models are distinguishable by 3 $\sigma$, given by a p-value of less than 0.0027.  If the number of eclipses exceeds 30 we declare the models indistinguishable under realistic observation times. The blackbody spectrum is obtained by fitting the mock JWST spectrum using SciPy's ``curve\_fit" function, with uncertainties on each data point at the given resolution obtained by \pandexo{}.  The mock data used as input for the blackbody fitting are down-sampled to a resolution of 3, which we have found provides the most precise fit (not shown).

\begin{figure*}
\centering\includegraphics[width=7cm]{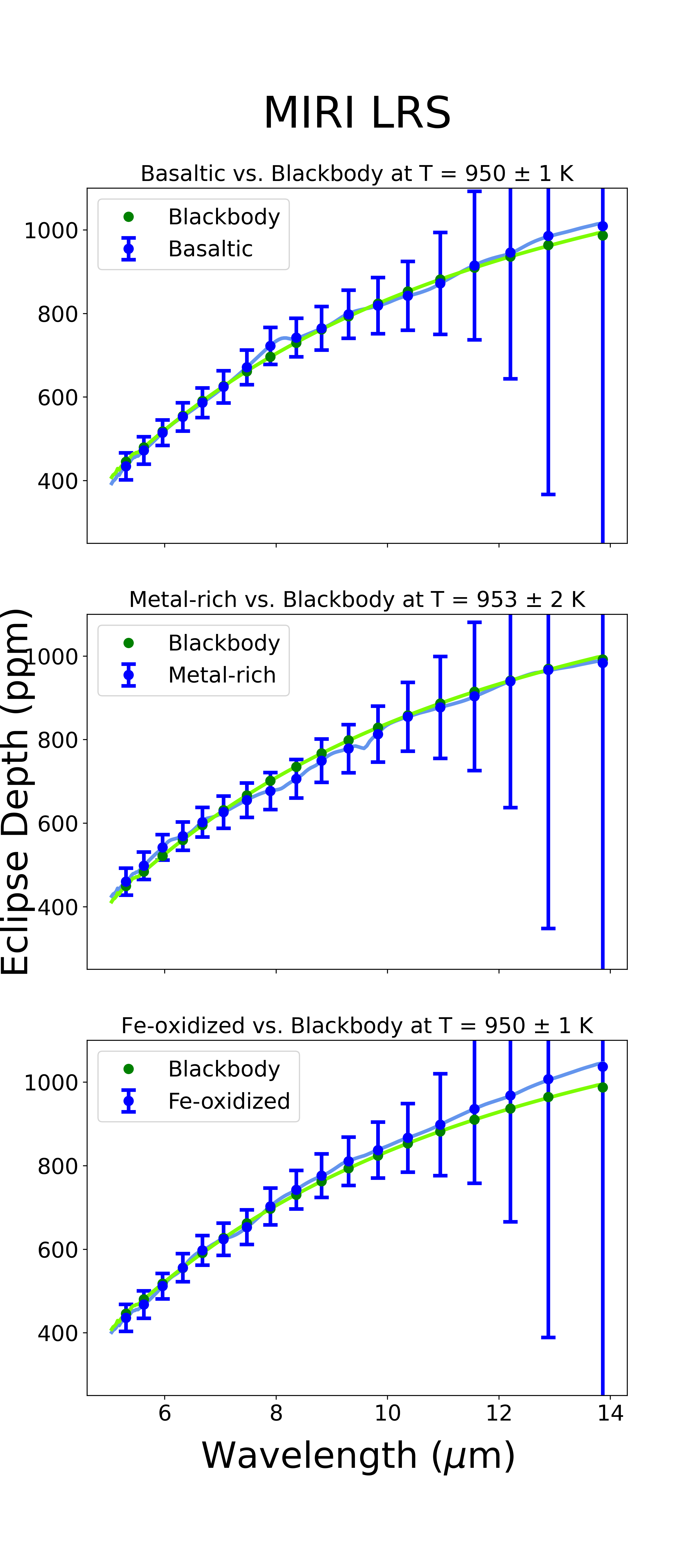}
\centering\includegraphics[width=7cm]{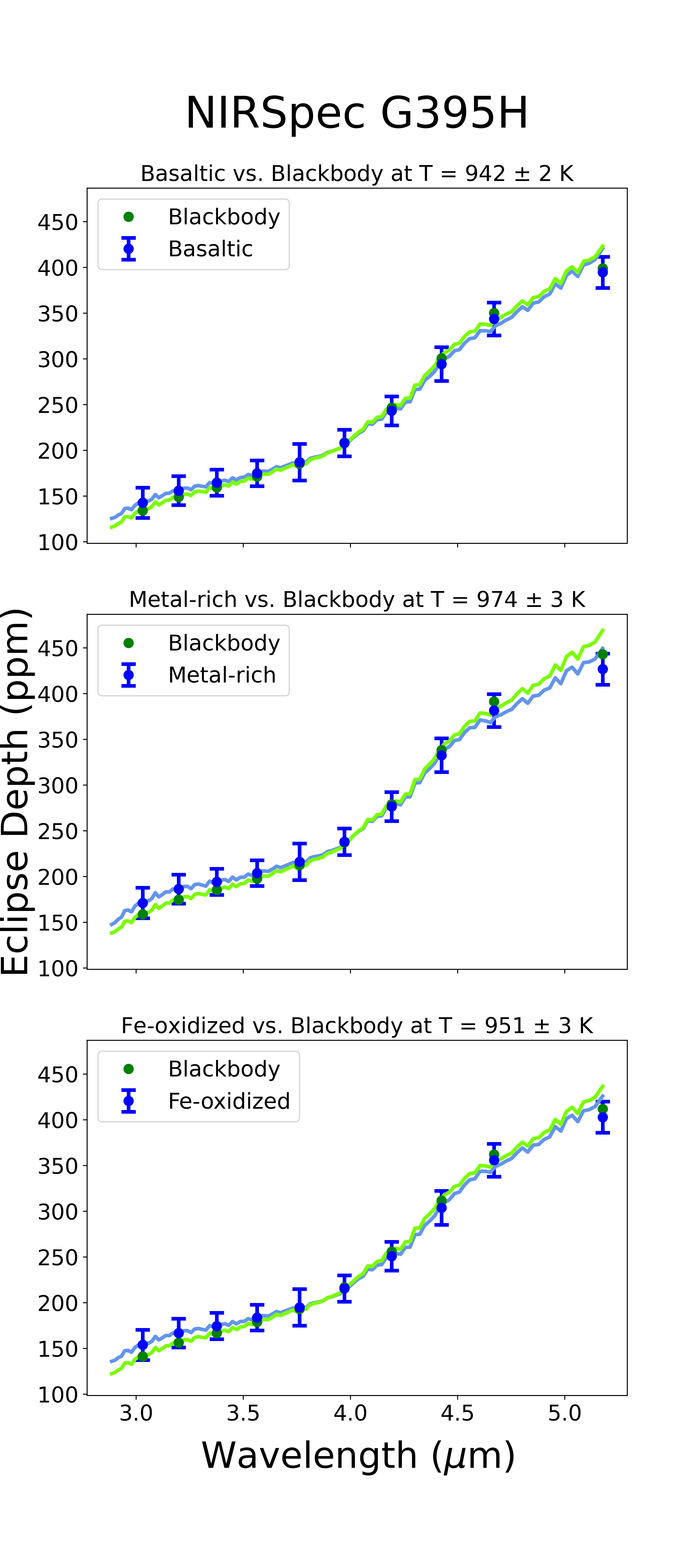}
\caption{Secondary eclipse spectra of the surface-only models, mock data points for observations with JWST MIRI LRS (left) and NIRSpec G395H (right) and the corresponding blackbody fits. Note that the blackbody models don't appear as smooth curves because the emitted flux is divided by the stellar spectrum model to obtain the eclipse depth. Only the surfaces consistent with the Spitzer 4.5 $\mu$m eclipse depth are shown here. The mock data points are rebinned to $R$ = 10 with error bars corresponding to 5 eclipse observations. The original model spectra are downsampled from their native resolution to $R$ = 100 for clarity.}
\label{fig:surface_bb}
\end{figure*}

\begin{figure*}
\centering\includegraphics[width=7cm]{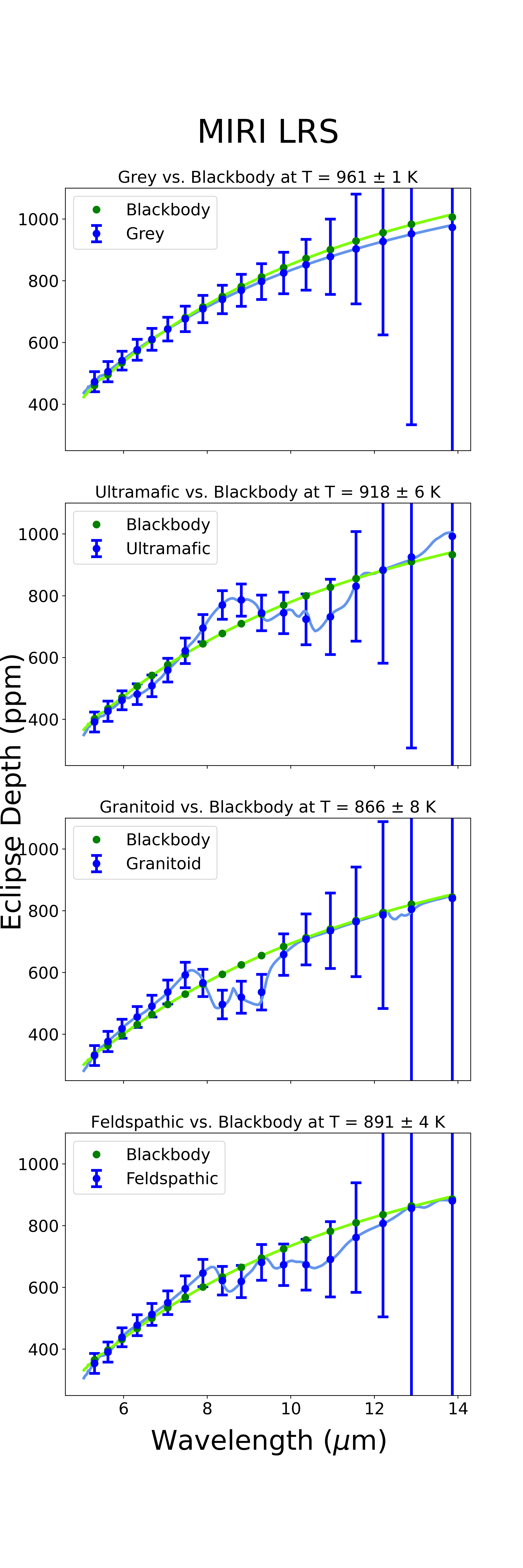}
\centering\includegraphics[width=7cm]{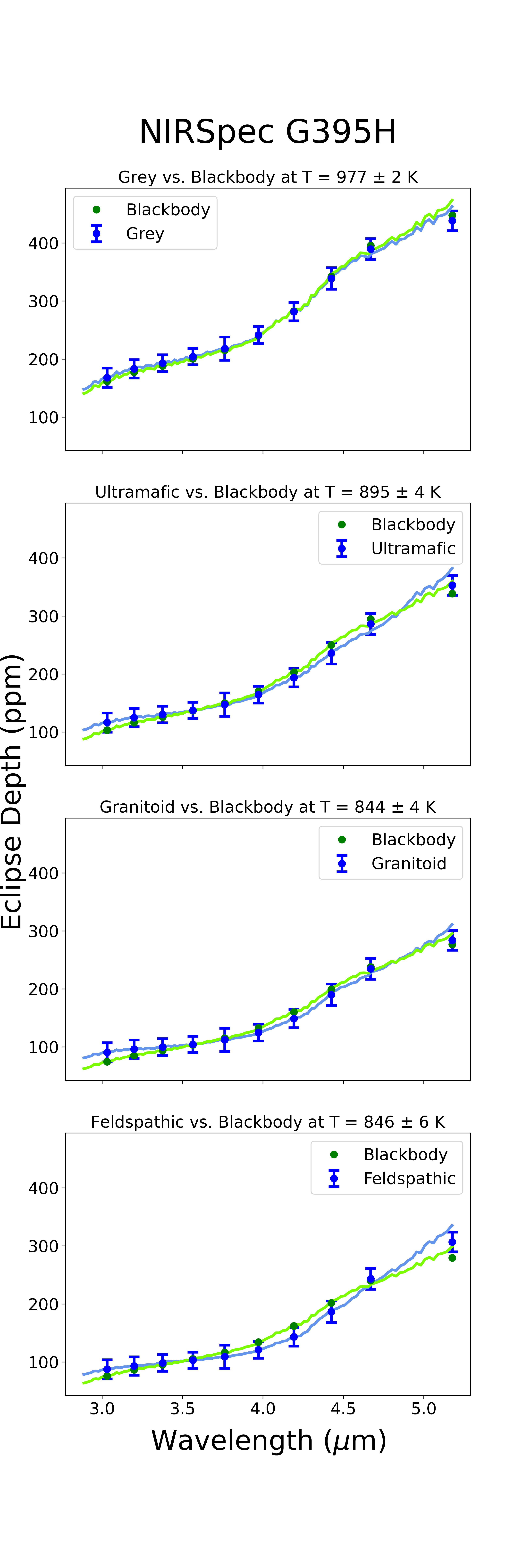}
\caption{Secondary eclipse spectra of the surface-only models, mock data points for observations with JWST MIRI LRS (left) and NIRSpec G395H (right) and the corresponding blackbody fits. Shown here are the surfaces inconsistent with the Spitzer 4.5 $\mu$m eclipse depth. The mock data points are rebinned to $R$ = 10, with error bars corresponding to 5 eclipse observations. The model spectra are downsampled from their native resolution to $R$ = 100 for clarity.}
\label{fig:surface_bb_implausible}
\end{figure*}

The number of eclipses needed to distinguish each surface from a blackbody is listed in the first column and the first row of Table~\ref{table:distinguish_surf_R3}, depending on the JWST instrument used. In addition, the surface spectra for all plausible (i.e., consistent with Spitzer) surfaces together with the blackbody fit are displayed in Figure~\ref{fig:surface_bb} and the inconsistent-with-Spitzer surface compositions are displayed in Figure~\ref{fig:surface_bb_implausible}. We find that overall NIRSpec is more viable than MIRI for the purpose of distinguishing surfaces from a blackbody. Still, only the surfaces inconsistent with the Spitzer measurement are predicted to be distinguishable with less than 10 eclipses. That is because these surfaces exhibit more pronounced albedo features and higher albedos in general, making them more distinct from a blackbody spectrum. Specifically, we find that the ultramafic, granitoid and feldspathic surfaces should be distinguishable with 4, 4, and 6 eclipses, respectively. Perhaps surprisingly, the grey surface, although having no albedo features, differs from pure blackbody emission as well. This is a consequence of assuming a zero albedo for a blackbody compared to the 0.1 albedo in the grey model, creating a ``tilt'' between the two corresponding spectra. With MIRI, no single surface will be distinguishable from a blackbody with a feasible number of eclipses.

Additionally, we assess the number of eclipses needed to distinguish each surface from one another, again using MIRI or NIRSpec. After testing a variety of binning resolutions, we have found that the models are maximally distinguishable for a resolution of $R = 3$. It appears that a small number of data points are sufficient, since the surface features tend to be broad and the comparison appears to be driven by the overall eclipse depth rather than the shape of individual features.

Table~\ref{table:distinguish_surf_R3} shows the number of eclipses needed to distinguish surface pairs using an $R = 3$ binning. Secondary eclipse spectra for each pair of surface spectra are shown in Fig.~\ref{fig:distinguish_surf}. It can be seen that many of the surfaces are more easily distinguishable from each other in the NIRSpec G395H band than in the MIRI LRS band. For example, the metal-rich and basaltic surfaces are indistinguishable in the MIRI LRS band, but are distinguishable with 6 secondary eclipses in the NIRSpec G395H band. Additionally, while the metal-rich and iron oxidized surfaces are indistinguishable with MIRI LRS, they are distinguishable with 10 secondary eclipses using NIRSpec G395H. Overall, we find that less reflective materials are harder to disentangle due to their small features and eclipse depth variations than the more reflective ultramafic, granitoid and feldspathic surfaces.

In addition to utilizing the entire instrument wavelength range and constant binning for the pairwise distinguishability, we also conduct the same exploration focusing on the most prominent surface feature on either of the two surfaces and using optimized binning over the range of that feature. This is only done using the MIRI LRS instrument, since there are no prominent surface features within the NIRSpec G395H wavelength range (see Fig.~\ref{fig:distinguish_surf} on the right). Using qualitative ranking, we choose three prominent features to be included in this analysis: The double-feature around 10 $\mu$m for the ultramafic, the double-feature at 8.5 $\mu$m for the granitoid, and the double-feature around 10 $\mu$m for the feldspathic surface. The grey, basaltic, metal-rich, and iron oxidized surfaces do not exhibit any sufficiently significant features for this analysis. Table~\ref{table:distinguish_surf_isolated} shows the number of eclipses needed to distinguish the ultramafic, granitoid and feldspathic from all other surfaces using the described method, while the corresponding pairs of surface spectra are shown in Figure~\ref{fig:distinguish_surf_isolated}.

Using this optimized binning over the most prominent surface feature, the number of eclipses needed to distinguish between three of the surface combinations was reduced, meaning that this method represented an improvement upon constant binning over the entire instrument wavelength range. More specifically, the number of eclipses was reduced from 6 to 3 for the granitoid and ultramafic surfaces, from 27 to 11 for the granitoid and feldspathic surfaces, and from 16 to 10 for the feldspathic and ultramafic surfaces. However, all other surface combinations tested using this optimized binning either remained unchanged or needed a greater number of eclipses than in the runs using the whole instrument range. As pointed out before, this result indicates that the constraining of surface compositions is largely driven by the Bond albedo of the surface (overall eclipse depth offset over the width of an instrument bandpass) rather than the spectral variation of features.

\begin{table*}
\begin{center}
\hskip-2.5cm\begin{tabular}{|c|cccccccc|}
\multicolumn{9}{c}{{\bf Surface Detectability using Whole Instrument Range}} \\
\hline
 & BB fit & grey & Basaltic & Ultramafic & Metal-rich & Fe-oxidized & Granitoid & Feldspathic \\
\hline
BB fit & & \cellcolor{Gray}9 & \cellcolor{Gray}$\geq$ 30 & \cellcolor{Gray}6 & \cellcolor{Gray}$\geq$ 30 & \cellcolor{Gray}$\geq$ 30 & \cellcolor{Gray}6 & \cellcolor{Gray}4 \\
\hline
grey & \cellcolor{LightCyan}$\geq$ 30 & & \cellcolor{Gray}5 & \cellcolor{Gray}1 & \cellcolor{Gray}$\geq$ 30 & \cellcolor{Gray}8 & \cellcolor{Gray}1 & \cellcolor{Gray}1 \\
\hline
Basaltic & \cellcolor{LightCyan}$\geq$ 30 & \cellcolor{LightCyan}$\geq$ 30 & & \cellcolor{Gray}3 & \cellcolor{Gray}6 & \cellcolor{Gray}$\geq$ 30 & \cellcolor{Gray}1 & \cellcolor{Gray}1\\
\hline
Ultramafic & \cellcolor{LightCyan}25 & \cellcolor{LightCyan}7 & \cellcolor{LightCyan}13 & & \cellcolor{Gray} 1 & \cellcolor{Gray} 3 & \cellcolor{Gray} 4 & \cellcolor{Gray} 4\\
\hline
Metal-rich & \cellcolor{LightCyan}$\geq$ 30 & \cellcolor{LightCyan}$\geq$ 30 & \cellcolor{LightCyan}$\geq$ 30 & \cellcolor{LightCyan}9 & & \cellcolor{Gray}10 & \cellcolor{Gray}1 & \cellcolor{Gray}1\\
\hline
Fe-oxidized & \cellcolor{LightCyan}$\geq$ 30 & \cellcolor{LightCyan}$\geq$ 30 & \cellcolor{LightCyan}$\geq$ 30 & \cellcolor{LightCyan}11 & \cellcolor{LightCyan}$\geq$ 30 & & \cellcolor{Gray}1 & \cellcolor{Gray}1 \\
\hline
Granitoid & \cellcolor{LightCyan}$\geq$ 30 & \cellcolor{LightCyan}2 & \cellcolor{LightCyan}3 & \cellcolor{LightCyan}6 & \cellcolor{LightCyan}3 & \cellcolor{LightCyan}3 & & \cellcolor{Gray}$\geq$ 30 \\
\hline
Feldspathic & \cellcolor{LightCyan}$\geq$ 30 & \cellcolor{LightCyan}4 & \cellcolor{LightCyan}5 & \cellcolor{LightCyan}16 & \cellcolor{LightCyan}5 & \cellcolor{LightCyan}5 & \cellcolor{LightCyan}27 & \\
\hline
\end{tabular}
\end{center}
\caption{Number of secondary eclipse observations needed with JWST to distinguish each surface from a blackbody fit and from another surface at 3 $\sigma$. The number of eclipses using MIRI LRS is shown in cyan while the number of eclipses using NIRSpec G395H is shown in grey. A constant binning with $R$ = 3 and the whole wavelength range of each instrument is used for this test. The detectability is not analyzed beyond 30 eclipses.}
\label{table:distinguish_surf_R3}
\end{table*}

\begin{table*}
\begin{center}
\hskip-2.5cm\begin{tabular}{|c|cccccc|}
\multicolumn{7}{c}{{\bf Distinguishing Surfaces with Optimal Binning}} \\
\hline
 & grey & Basaltic & Ultramafic & Metal-rich & Fe-oxidized & Granitoid \\
\hline
Ultramafic & \cellcolor{LightCyan}$\geq$ 30 & \cellcolor{LightCyan}$\geq$ 30 & & \cellcolor{LightCyan}$\geq$ 30 & \cellcolor{LightCyan} 25 & \\
\hline
Granitoid & \cellcolor{LightCyan}3 & \cellcolor{LightCyan}3 & \cellcolor{LightCyan}3 & \cellcolor{LightCyan}3 & \cellcolor{LightCyan}3 & \\
\hline
Feldspathic & \cellcolor{LightCyan}7 & \cellcolor{LightCyan}7 & \cellcolor{LightCyan}10 & \cellcolor{LightCyan}10 & \cellcolor{LightCyan}6 & \cellcolor{LightCyan}11\\
\hline
\end{tabular}
\end{center}
\caption{Number of secondary eclipse observations needed with JWST to distinguish surfaces at 3 $\sigma$, focusing on isolated features with optimal binning. Only MIRI LRS is used for this test since there are no sufficiently pronounced features in the NIRSpec G395H bandpass for any of the tested surface compositions.}
\label{table:distinguish_surf_isolated}
\end{table*}

\subsubsection{Inferring planetary albedo and temperature}

In the previous section we have fit the model spectra with a blackbody to determine whether surface features are detectable. However, the blackbody fit can also be used to determine the brightness temperature over a certain wavelength range. The brightness temperature in turn allows to infer the planetary Bond albedo, which is a useful quantity that can be used to place constraints on the presence and properties of an atmosphere, as elucidated in detail in \citet{Mansfield_2019}. However, as pointed out in that work, inferring the Bond albedo from MIRI observations is prone to be biased due to the non-grey shape of the surface albedo. In the following we re-enact their assessment of this issue for LHS 3844b using the non-grey radiative transfer code \helios{} and extend it to NIRSpec observations as well. First we assume that the brightness temperature is equivalent to the temperature of the blackbody fit for each surface. The inferred albedo is then calculated according to

\begin{equation}
\label{eq:inferred_albedo}
\alpha_{in} = 1 - \frac{\epsilon}{f}\left(\frac{T_{bb}}{T_{star}}\right)^4\left(\frac{a}{R_{star}}\right)^2 ,
\end{equation}

where $\alpha_{in}$ is the inferred albedo, $T_{bb}$ is the temperature from the blackbody fit, $T_{star}$ is the temperature of the star, $a$ is the semi-major axis of the planetary orbit, and $R_{star}$ is the radius of the star. We set the heat redistribution factor $f$ to $2/3$ as predicted for a ``vanishingly thin" atmosphere. The wavelength-integrated emissivity $\epsilon$ is a priori unknown and thus set to 1 in our reference models, which is equivalent to assuming that the surface radiates as a blackbody. We also discuss the effects of assuming $\epsilon < 1$ in Appendix~\ref{app:emissivity}. We calculate the true Bond albedo for each surface by integrating the flux reflected from the planet over all wavelengths, and dividing by the total stellar flux received by the planet.

The upper two rows of Table~\ref{table:inferred_albedo} list the temperatures of best fit associated with each of these blackbody models based on the MIRI or NIRSpec spectra, while the third row of Table~\ref{table:inferred_albedo} lists the true surface temperatures obtained in the \helios{} models.

The Bond albedos and inferred albedos for each surface are listed in the bottom part of Table~\ref{table:inferred_albedo}. Each surface has a separate inferred albedo for each JWST instrument, in contrast to the the Bond albedo which is a single quantity resulting from the interplay between the planetary surface and the stellar irradiation. 

\begin{figure*}
\centering\includegraphics[width=17cm]{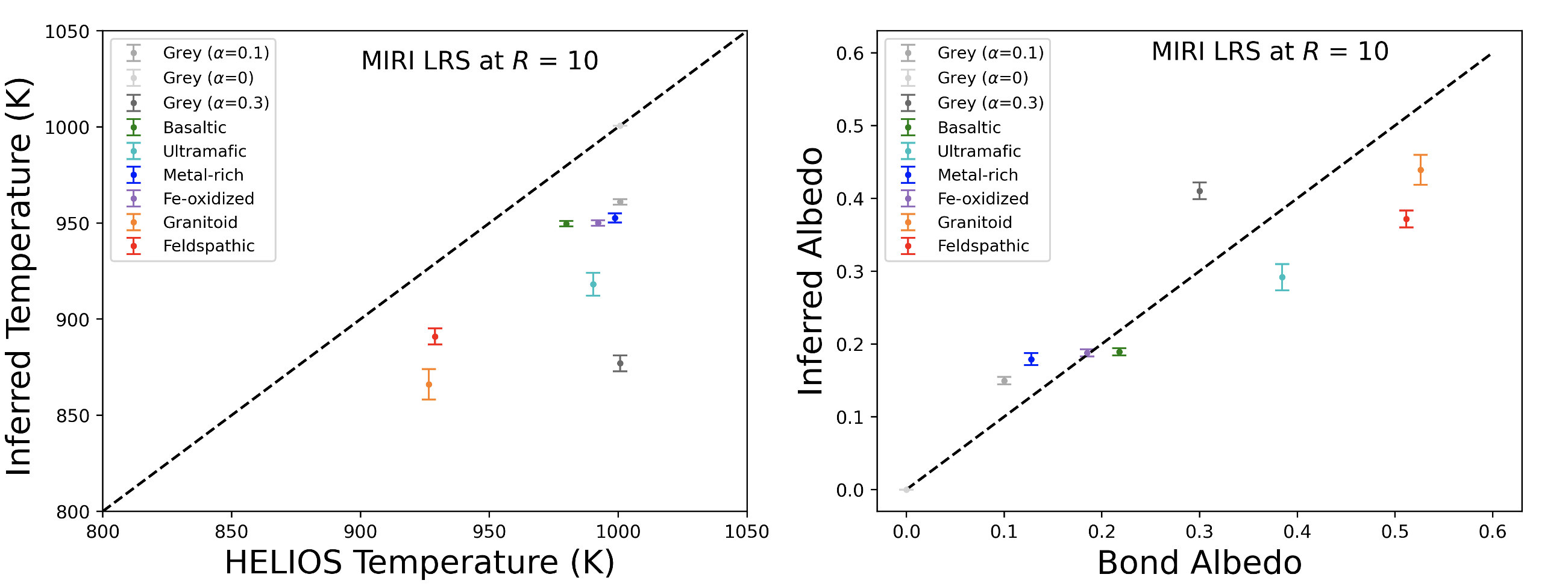}
\caption{{\bf Left:} Surface temperature obtained with \helios{} vs. inferred temperature of the blackbody fit for simulated MIRI LRS data at $R$ = 10. The black dotted line represents the location where the inferred and \helios{} temperatures are equal. The error bars correspond to 5 eclipse observations. {\bf Right:} Analogous to the left panel but showing the Bond albedo vs. Inferred albedo. The inferred albedo is calculated with Eq.~\ref{eq:inferred_albedo} using the inferred temperature of the blackbody fit.}
\label{fig:inferred_albedo}
\end{figure*}

Figure~\ref{fig:inferred_albedo} displays the inferred brightness temperatures against the true model temperatures (left panel) and the inferred albedos against the Bond albedo (right panel) for each surface based on simulated MIRI observations. We find that the surface temperature of the planet is consistently underestimated by the blackbody fitting. This is because the fitting routine assumes an emissivity of 1 (equivalent to an albedo of 0), which assumes that the planet radiates heat more efficiently than it does in reality, finding a lower temperature for a given amount of thermal heat. The right panel shows that the majority of surfaces have an inferred albedo in the MIRI LRS band which is lower than the true Bond albedo. This can be explained by the non-grey albedo variation, discussed in detail in \citet{Mansfield_2019}. As a realistic, non-grey surface albedo exhibits strong variations in the near to mid-infrared, the planetary emission is muted in the albedo peaks, and amplified in the albedo troughs, in accordance with Kirchhoff's law of radiation. Since the majority of the surfaces tested here have a lower albedo within the MIRI bandpass than outside of it, emission within the bandpass is amplified, leading to an underestimation of the inferred albedo. In contrast, the albedo of the Fe-oxidized surface is determined accurately and moreover the metal-rich albedo is overestimated. Here, the wavelength variation effect is countered by another bias induced by the assumption of an emissivity of unity which leads to an overestimation of the albedo, best visible for the grey surfaces added as control cases (for more details see Appendix~\ref{app:emissivity}). Lastly, another bias on the inferred albedo stems from the reflected stellar light being superimposed on the planetary thermal emission, mimicking a higher planetary emission and lower albedo. However, as the fraction of reflected light in the MIRI bandpass is minor we find that this effect is relatively small, see Fig.~\ref{fig:inf_albedo_emitted_only}, left panel.

Interestingly, compared to the previous MIRI predictions in \cite{Mansfield_2019} our new calculation predicts a consistently smaller observational bias for the albedo. After comparing the current and previous methodologies, we have found that the choice of stellar spectrum plays a significant role for the calculation of the inferred albedo. In the present study we use a PHOENIX stellar spectrum in all steps of the inferred albedo calculation. In \cite{Mansfield_2019}, the planetary temperature was determined using a PHOENIX stellar spectrum, but when inferring the brightness temperature from the secondary eclipse depth a blackbody was assumed for the host star. This seemingly minor inconsistency appears to be responsible for the stark difference between the current and previously obtained albedo values.

\begin{figure*}
\centering\includegraphics[width=17cm]{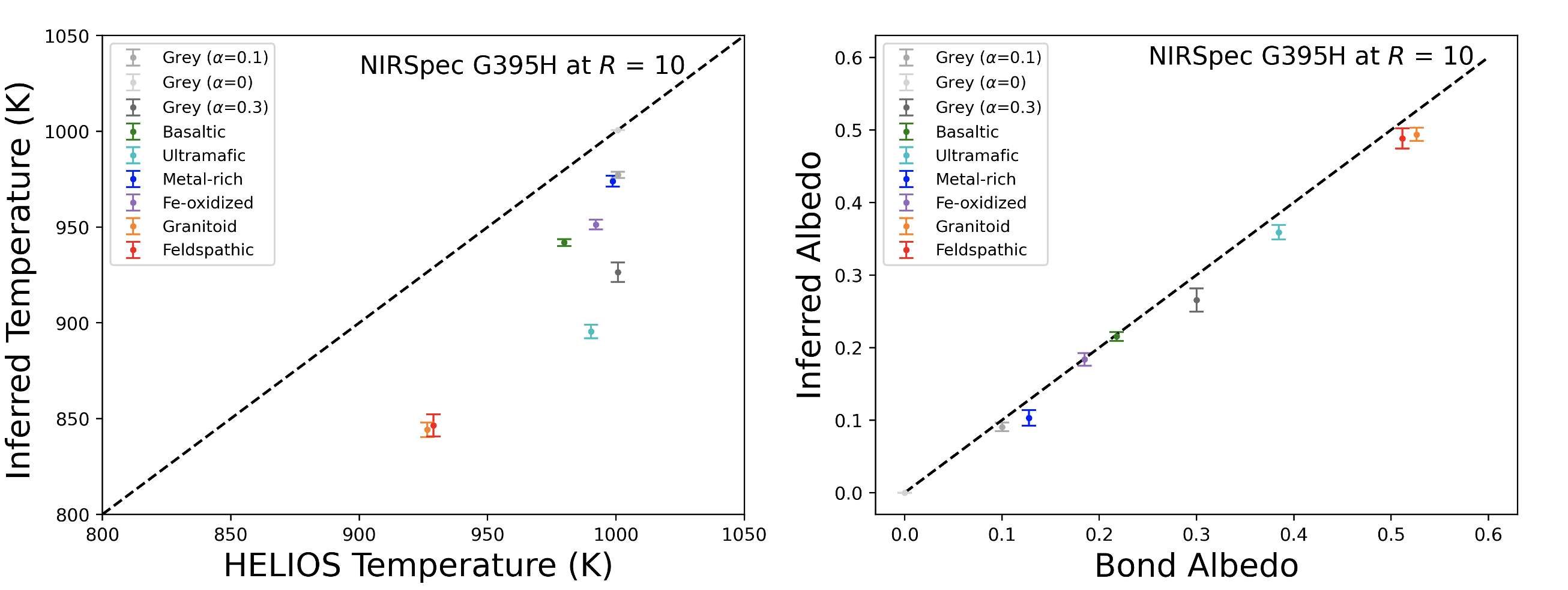}
\caption{Same as Fig.~\ref{fig:inferred_albedo} but using simulated observations with the NIRSpec G395H instrument at $R$ = 10.}
\label{fig:bond_NIRSpec}
\end{figure*}

With the NIRSpec instrument the albedo can be inferred more accurately than with MIRI, even though the temperature of the blackbody fit is similarly underestimated, see Fig.~\ref{fig:bond_NIRSpec}. The albedo is in general, just as with MIRI, somewhat underestimated. However, here the main reason for the bias is the scattered light, the amount of which in the NIRSpec bandpass is significant for the more reflective surfaces. If the scattered light is removed from the analysis, the inferred albedo values become higher and predominantly overestimated, see Fig.~\ref{fig:inf_albedo_emitted_only}, right panel. The emissivity effect, a major source of inference bias when using MIRI, is only significant for the higher-albedo surface in the case of NIRSpec, as is further explained in Appendix~\ref{app:emissivity}.

\begin{table*}
\begin{center}
\begin{tabular}{ |c|c|c|c|c|c|c|c|}
\multicolumn{8}{c}{\textbf{Inferred Temperature and Albedo for Each Surface}} \\
\hline
\multicolumn{8}{|c|}{Inferred Temperature / Blackbody Fit Temperature (K)} \\
\hline
& Grey (0.1) & Basaltic & Ultramafic & Metal-rich & Fe-oxidized & Granitoid & Feldspathic \\
\hline
MIRI & 961 $\pm$ 1 & 950 $\pm$ 1 & 918 $\pm$ 6 & 953 $\pm$ 2 & 950 $\pm$ 1 & 866 $\pm$ 8 & 891 $\pm$ 4 \\
\hline
NIRSpec & 977 $\pm$ 2 & 942 $\pm$ 2 & 895 $\pm$ 4 & 974 $\pm$ 3 & 951 $\pm$ 3 & 844 $\pm$ 4 & 846 $\pm$ 6\\
\hline
\multicolumn{8}{|c|}{\helios{} Surface Temperature (K)} \\
\hline
 & 1000.78 & 979.831 & 990.324 & 998.682 & 992.145 & 926.571 & 928.914 \\
\hline
\multicolumn{8}{|c|}{Inferred Albedo} \\
\hline
& Grey (0.1) & Basaltic & Ultramafic & Metal-rich & Fe-oxidized & Granitoid & Feldspathic \\
\hline
MIRI & 0.150 $\pm$ 0.005 & 0.189 $\pm$ 0.005 & 0.292 $\pm$ 0.018 & 0.179 $\pm$ 0.008 & 0.188 $\pm$ 0.005 & 0.439 $\pm$ 0.021 & 0.372 $\pm$ 0.012 \\
\hline
NIRSpec & 0.091 $\pm$ 0.006 & 0.215 $\pm$ 0.006 & 0.359 $\pm$ 0.010 & 0.103 $\pm$ 0.011 & 0.184 $\pm$ 0.009 & 0.494 $\pm$ 0.009 & 0.488 $\pm$ 0.014 \\
\hline
\multicolumn{8}{|c|}{Bond Albedo} \\
\hline
 & 0.100 & 0.218 & 0.384 & 0.128 & 0.185 & 0.526 & 0.511 \\
\hline
\end{tabular}
\end{center}
\caption{{\bf Top:} Inferred brightness temperature corresponding to the blackbody fit temperature for each surface based on simulated MIRI LRS or NIRSpec G395H observations at $R$ = 10, compared to the actual surface temperature found by the \helios{} radiative transfer code Uncertainties correspond to 5 eclipse observations. {\bf Bottom:} Inferred albedo derived from the best-fitting blackbody temperatures, listed on the top, and the true Bond albedo for each surface.}
\label{table:inferred_albedo}
\end{table*}

\subsubsection{Thick Atmosphere limit}
\label{sec:jwst_thick_atmo}

In the following we explore whether the atmospheres in the thick limit (recall Sect.~\ref{sec:thick_atmo}) are detectable with JWST, i.e., if they can be distinguished from a pure surface spectrum. First, we utilize the whole instrument bandpasses of MIRI LRS and NIRSpec G395H and choose the resolution $R$ = 10 as this represents a good compromise between statistical noise and sampling of the spectral features. The number of secondary eclipse observations needed to detect the atmosphere in each tested scenario is listed in Table~\ref{table:atmo_v_noatmo}, top rows. We find that all of the atmospheres are detectable with $\leq$ 9 eclipse observations if using the preferred instrument, with NIRSpec being preferable at finding CO$_2$ and CH$_4$, and MIRI preferable at finding SO$_2$ and H$_2$O. The detections with the lowest time requirement are \OtwoSOtwo{} with MIRI and \NtwoCHfour{} with NIRSpec, both requiring 3 eclipse observations. The \OtwoHtwoO{} atmosphere is detectable with MIRI and 7 eclipses. As a worst case observing scenario, with MIRI the \OtwoCOtwo{} atmosphere is indistinguishable from a surface. The spectra of the atmospheric models and the associated surface-only spectra can be seen in Figure~\ref{fig:atmo_v_noatmo}.

\begin{table*}
\begin{center}
\hskip-1.5cm\begin{tabular}{|c|c|c|c|c|}
\multicolumn{5}{c}{{\bf Distinguishing Atmospheres from the Surface-only Case}} \\
\hline
 & O$_2$ with 100 ppm CO$_2$ & O$_2$ with 100 ppm SO$_2$ & O$_2$ with 1 ppm H$_2$O & N$_2$ with 1\% CH$_4$ \\
\hline
\multicolumn{5}{|c|}{Utilizing Whole Instrument Bandpass} \\
\hline
MIRI LRS & $\geq$ 30 & 3 & 7 & 6 \\
\hline
NIRSpec G395H & 9 & 7 & 10 & 3 \\
\hline
\multicolumn{5}{|c|}{Focusing on Individual Gas Features} \\
\hline
MIRI LRS & No Discernible Features & 1 & 2 & 2 \\
\hline
NIRSpec G395H & 3  & 3 & No Discernible Features & 1 \\
\hline
\end{tabular}
\end{center}
\caption{{\bf Top:} Number of secondary eclipses needed to distinguish an atmospheric model in the ``thick limit" (see Sect.~\ref{sec:thick_atmo}) from a surface-only spectrum at 3 $\sigma$ using MIRI and NIRSpec mock observations at $R$ = 10. The metal-rich surface is used for this test. The distinguishability of models is not analyzed beyond 30 secondary eclipses. {\bf Bottom:} Same as above but focusing on the strongest gas features and absorption bands in isolation. No discernible features are found within the MIRI LRS range for the O$_2$ with 100 ppm CO$_2$ case and within the NIRSpec G395H range for the O$_2$ with 1 ppm H$_2$O case.}
\label{table:atmo_v_noatmo}
\end{table*}

\begin{figure*}
\centering\includegraphics[width=7cm]{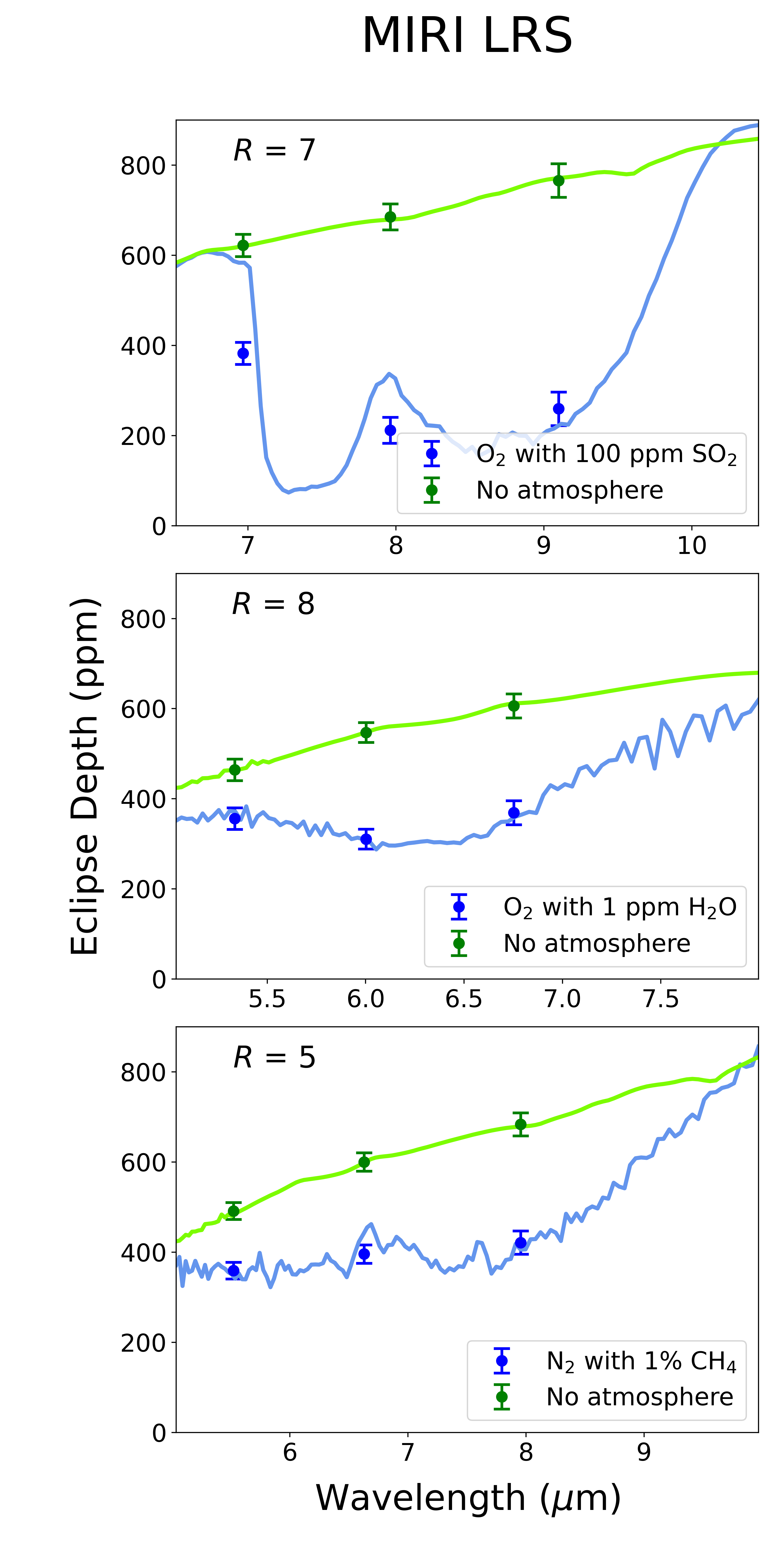}
\centering\includegraphics[width=7cm]{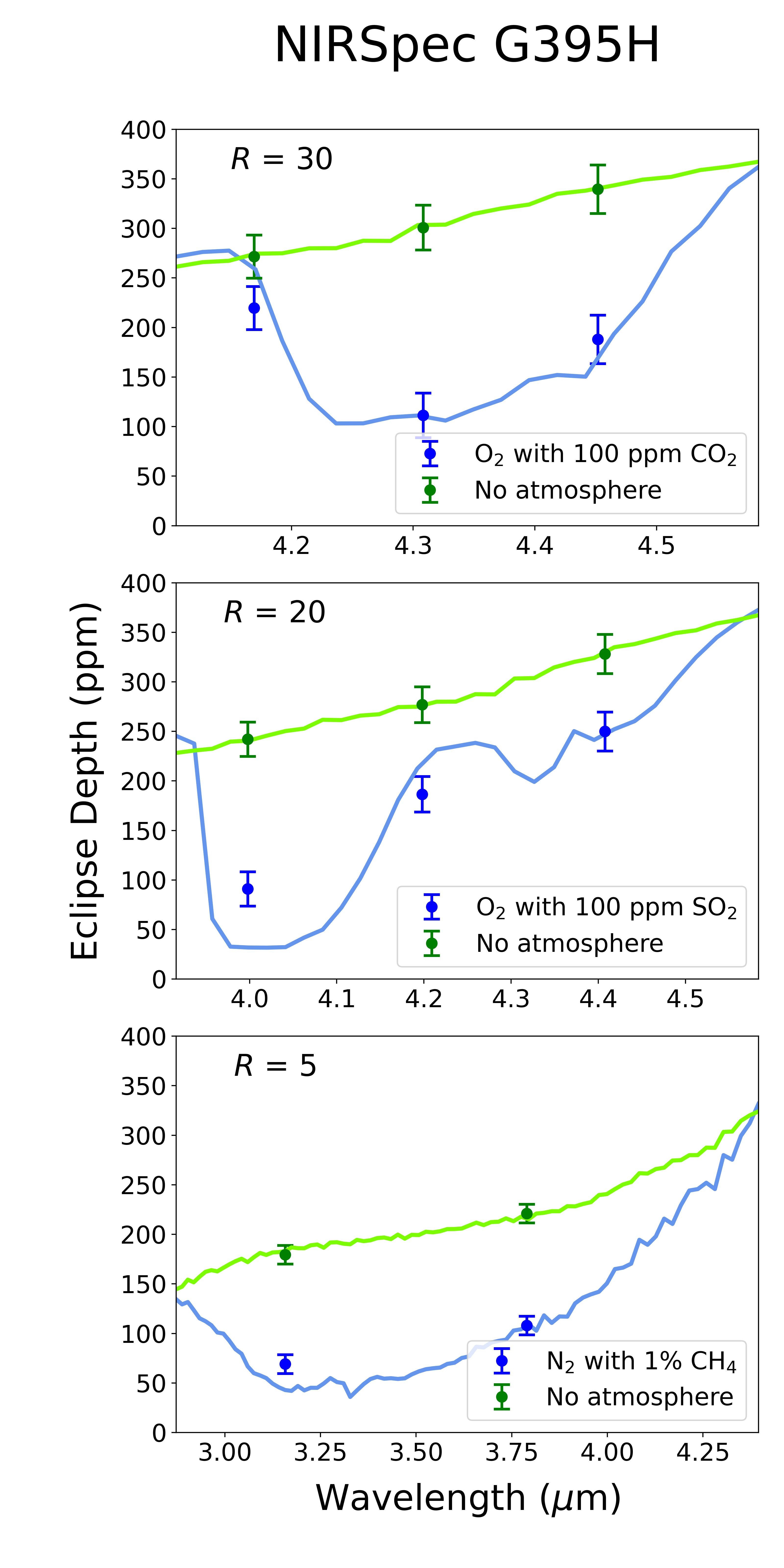}
\caption{Secondary eclipse spectra of atmospheric models vs. surface-only models over the range of the strongest absorption bands in the respective models, overlaid with MIRI LRS (left) and NIRSpec G395H (right) simulated data points. The metal-rich surface is used for this test. Error bars show uncertainties for 5 eclipse observations. Model spectra are downsampled from the native resolution to $R$ = 100 for clarity.}
\label{fig:atmo_vs_noatmo_isolated}
\end{figure*}

Similar as in the surface observability analysis, we conduct another examination focusing on individual atmospheric absorption features or bands in isolation instead of taking the whole instrument range. Specifically, for SO$_2$ we test the feature at 4 $\mu$m and the double-peaked band around 8 $\mu$m, for CH$_4$ we test the feature at 3.5 $\mu$m and the band around 7 $\mu$m and for CO$_2$ we test the feature at 4.3 $\mu$m. There are no discernible features within the MIRI instrument range for the atmosphere with CO$_2$ and, analogously, no discernible features withing the NIRSpec range for the atmosphere with H$_2$O. The resolution and binning of each spectrum is chosen to ensure that there are two to three data points sampling the feature being probed. The number of secondary eclipses needed to distinguish each atmospheric feature from the surface-only spectrum is listed in Table~\ref{table:atmo_v_noatmo}, bottom rows. The resulting atmospheric model spectra and their no-atmosphere counterparts are shown in Figure~\ref{fig:atmo_vs_noatmo_isolated}, with error bars representing the uncertainty after 5 secondary eclipse measurements. We find that isolating the gaseous features increases the detectability significantly to the extent that all explored atmospheres can be distinguished from a surface with at most 3 eclipses if using the preferred instrument for each case. Most promisingly, the CH$_4$ and SO$_2$ cases can be detected with only a single eclipse using NIRSpec or MIRI, respectively. The \OtwoHtwoO{} atmosphere is detectable with MIRI and 2 eclipses and the \OtwoCOtwo{} atmosphere with NIRSpec and 3 eclipses.

\subsection{Atmospheric Retrieval}
\label{sec:retrieval}

In addition to the chi-square tests presented in the last section, we further run atmospheric retrieval models in order to determine how JWST observations may help constrain the atmospheric composition of LHS 3844b using a Bayesian framework. To this end, we use the simulated JWST data from both NIRSpec G395H and MIRI LRS instruments with errors equivalent to 5 eclipses as input for the retrieval modeling. Note that, as described in Sect.~\ref{sec:method_retrieval}, our code generating the forward models, \helios{}, adds the non-grey surface to the spectrum, but the retrieval code \platon{} does not include the surface and only assumes an isotherm extending to the infinite, which should mimic retrieving the pressure where the surface is expected. Hence this analysis further allows us to explore the impact of the realistic surface albedo on the results of the retrieval modeling.

Instead of immediately using the forward models that include a non-grey surface, we first test the retrieval performance on pure atmosphere models, i.e., models with the atmosphere and a zero albedo surface. This test is indicative of whether the atmospheric signal alone is sufficient to make any statements about the gas inventory of the planetary envelope. Secondly, we then use the same atmospheric models and include the albedo signal of various surfaces. We include the crusts that are consistent with the Spitzer measurement, basaltic, metal-rich and iron oxidized with the addition of ultramafic in order to have an example of a more reflective surface. Finally, the last set of retrievals uses the surface spectrum only in order to explore whether the surface albedo variation itself can be interpreted as a false atmospheric signal by the retrieval algorithm in the case of the planet having no atmosphere at all.

\begin{figure*}
\centering
\includegraphics[width=12cm]{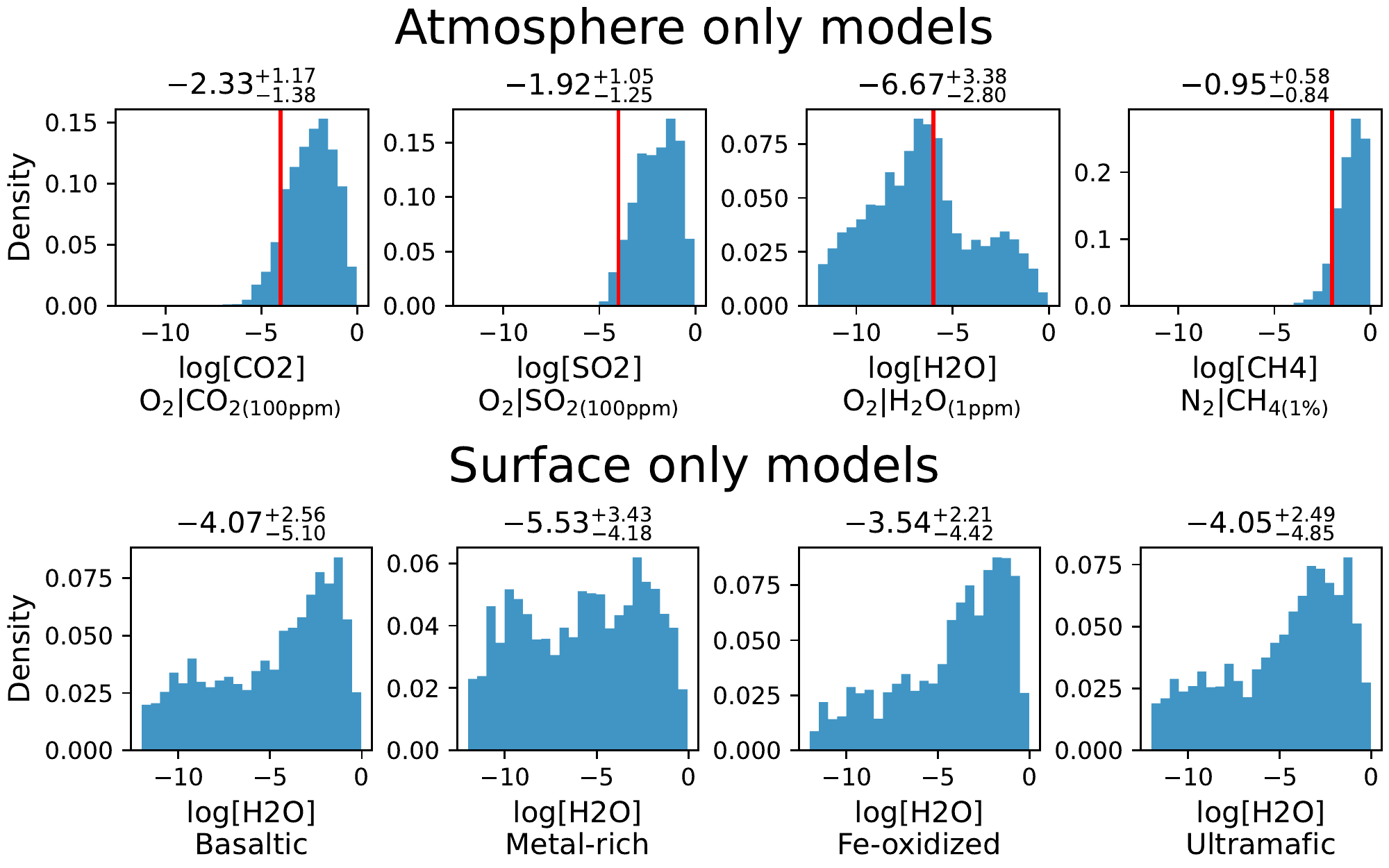}
\caption{{\bf Top:} Posterior distributions of the retrieved volume mixing ratios of CO$_2$, SO$_2$, H$_2$O and CH$_4$ using atmospheric forward models without a non-grey surface. The retrieved mean and 1-sigma width is given on top of each panel and the input mixing ratio (the ``true'' value) is shown as red, vertical line. The atmospheric composition used in the forward model is listed below the horizontal axis label. All atmospheric species are recovered in the retrieval, although the retrieved mixing ratio tends to be somewhat overpredicted for CO$_2$, SO$_2$ and CH$_4$. {\bf Bottom:} Posterior distributions of the retrieved H$_2$O volume mixing ratios from the surface-only (no atmosphere) forward models. The surface crust used in the forward model is listed below the horizontal axis label. The basaltic, fe-oxidized and ultramafic surfaces lead to a relatively pronounced posterior distribution that may erroneously be interpreted as a weak detection of atmospheric H$_2$O.}
\label{fig:retrieval_no_surf_no_atmo}
\end{figure*}

\begin{figure*}
\centering
\plotone{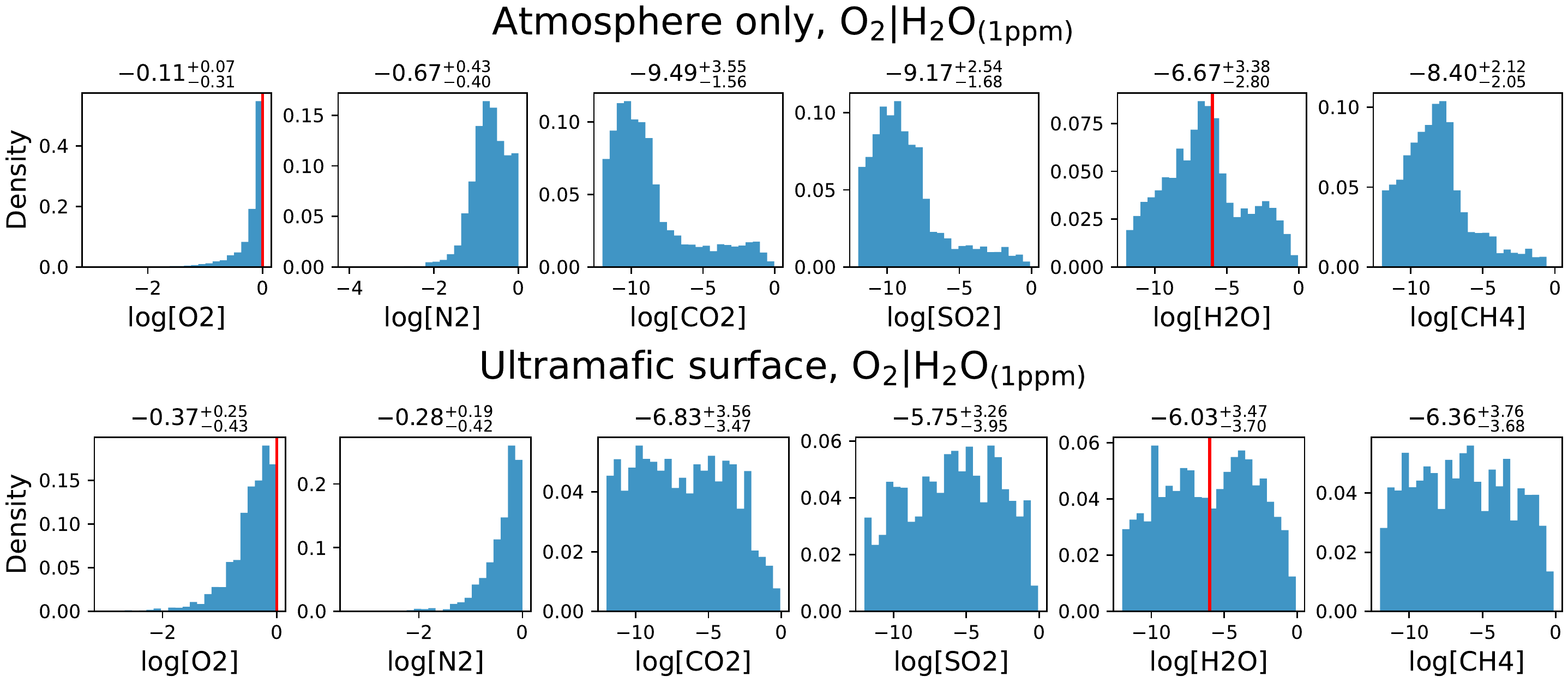}
\caption{Posterior distributions of volume mixing ratios using the O$_2$ with 1 ppm H$_2$O atmosphere as input for the retrieval modeling, once without a non-grey surface (top) and once with an ultramafic surface (bottom). The retrieved mean and 1-sigma width is given on top of each panel and the input mixing ratio (the ``true'' value) is shown as red, vertical line. Compared to the atmosphere-only case, adding the ultramafic signal leads to ``washed-out'' posteriors making it harder to place limits on the mixing ratios. Note that O$_2$ and N$_2$ are treated as background gases in the retrieval modeling and are not directly retrieved.}
\label{fig:retrieval_o2_h2o}
\end{figure*}

The results of the atmospheric retrievals with and without surface contribution are listed in Table~\ref{table:platon_atmo} and the surface-only retrievals in Table~\ref{table:platon_noatmo}. Striking results that are discussed separately and presented in Figures~\ref{fig:retrieval_no_surf_no_atmo} and \ref{fig:retrieval_o2_h2o} are highlighted in magenta. Note that the species O$_2$ and N$_2$, not explicitly listed in the tables, are treated as background gases in the retrieval and their mixing ratio always adds up to unity with the other absorbers.

In the atmosphere-only set-up the retrieval model manages to recover the absorbing species in each case it is present (see Fig.~\ref{fig:retrieval_no_surf_no_atmo}, top row). Curiously, only the true H$_2$O mixing ratio of 1 ppm lies within the 1 $\sigma$ region of the retrieved posterior, $\log f_{\rm H_2O} = -6.67^{+3.38}_{-2.80}$. The mixing ratios of the other absorbers, CO$_2$, SO$_2$ and CH$_4$ are somewhat overestimated by the retrieval algorithm compared to their true values. This is likely a consequence of a degeneracy between the mixing ratio of an absorber and the parameterized temperature profile used in the retrieval. Specifically, what may be happening is that the retrieval code sets the photosphere predominantly at too low pressures (and gas densities) so that a higher mixing ratio of an absorber is needed to provide the optical depth that is necessary for a given spectroscopic signal. This photosphere mismatch may in turn be a consequence of \helios{} and \platon{} having radiative transfer solvers and ingredients that are not identical, i.e., they use different opacity line lists and scattering prescriptions, which could result in the calculation of slightly different optical depths for the same input conditions. Lastly, just as the retrieval algorithm succeeds at recovering present species, it also correctly disfavors absent species, as the retrieved mean mixing ratios of the latter do not exceed $\sim 0.1$ ppm in any of the tested cases (see Fig.~\ref{fig:retrieval_grid_atmo_only} for a full grid of atmosphere-only model posteriors). 

We find that the addition of a realistic surface crust generally appears not to impede our ability to detect absorbing species that are present in the atmosphere and sufficiently visible in the planetary spectrum. In all setups the posterior distributions of absorbers that are present are consistent between the cases with and without a non-gray surface signal. However, we find a tendency that a non-grey surface makes it harder for the retrieval code to reject species that are absent in the forward model. This phenomenon is best visible in the case of the \OtwoHtwoO{} atmosphere (see Fig.~\ref{fig:retrieval_o2_h2o}). Whereas the posteriors of the atmosphere-only setup appear well-behaved, with a clear upper limit on the mixing ratio (see CO$_2$, SO$_2$ and CH$_4$ panels of Fig.~\ref{fig:retrieval_o2_h2o}), the posterior distributions in the case with the ultramafic surface remain broad and undetermined. Technically, it appears that the retrieval algorithm tries to match the surface features in the planetary spectrum with combinations of small amounts of atmospheric absorbers leading to many degenerate solutions. For instance, for SO$_2$ higher mixing ratios of $\gtrsim 1$ ppm are strongly disfavored in the no-surface case, ($\log f_{\rm SO_2} = -9.17^{+2.54}_{-1.68}$), but once the ultramafic signal is added ($\log f_{\rm SO_2} = -5.75^{+3.26}_{-3.95}$) even mixing ratios of $\sim$ 1 \% cannot be ruled out.

This last finding is further supported by our results from retrieving on the surface-only models. In general we find that the surface signatures imprinted on the planetary spectrum mislead the retrieval algorithm to falsely allow for the presence of atmospheric gases, with possible mixing ratios given by broad posterior distributions with a mean around $\sim 1$ ppm. The extreme case is H$_2$O for which the retrieved mixing ratios are higher with a mean around $\sim 100$ ppm in the cases of the basaltic, fe-oxidized and ultramafic crusts (see Fig.~\ref{fig:retrieval_no_surf_no_atmo}, bottom row). Also, in these cases the posterior distributions are more clearly defined and could misleadingly point to a weak H$_2$O detection. Figure~\ref{fig:retrieval_grid_surf_only} shows the full grid of surface-only model posteriors.

\begin{table*}
\footnotesize
\renewcommand{\arraystretch}{1.3}
\begin{center}
\hskip-3cm\begin{tabular}{ |p{2cm}|c|c|c|c|c|c|c|c|}
\multicolumn{9}{c}{{\bf Retrieval Results for Atmosphere Models with and without Surface}} \\
\hline
 & \multicolumn{2}{|c|} {CO$_2$} & \multicolumn{2}{|c|} {SO$_2$} & \multicolumn{2}{|c|} {H$_2$O} & \multicolumn{2}{|c|} {CH$_4$} \\
\hline
No~Surface,$\;\;$ \OtwoCOtwo{} & -4 & \cellcolor{purple}$-2.33^{+1.17}_{-1.38}$ & N Inc & $-8.24^{+2.24}_{-2.35}$ & N Inc & $-7.46^{+2.91}_{-2.81}$ & N Inc & $-8.45^{+2.70}_{-2.33}$ \\ 
\hline
No~Surface,$\;\;$ \OtwoSOtwo{} & N Inc & $-8.19^{+2.25}_{-2.32}$ & -4 & \cellcolor{purple}$-1.92^{+1.05}_{-1.25}$ & N Inc & $-8.30^{+2.64}_{-2.35}$ & N Inc & $-8.13^{+2.44}_{-2.36}$ \\ 
\hline
No~Surface,$\;\;$ \OtwoHtwoO{} & N Inc & \cellcolor{purple}$-9.49^{+3.55}_{-1.56}$ & N Inc & \cellcolor{purple}$-9.17^{+2.54}_{-1.68}$ & -6 & \cellcolor{purple}$-6.67^{+3.38}_{-2.80}$ & N Inc & \cellcolor{purple}$-8.40^{+2.12}_{-2.05}$ \\ 
\hline
No~Surface,$\;\;$ \NtwoCHfour{} & N Inc & $-7.16^{+1.93}_{-3.15}$ & N Inc & $-6.79^{+2.66}_{-2.90}$ & N Inc & $-7.02^{+3.23}_{-3.28}$ & -2 & \cellcolor{purple}$-0.95^{+0.58}_{-0.84}$ \\ 
\hline
Basaltic, \OtwoCOtwo{} & -4 & $-2.31^{+1.21}_{-1.63}$ & N Inc & $-8.17^{+2.02}_{-2.25}$ & N Inc & $-7.28^{+3.23}_{-2.80}$ & N Inc & $-7.65^{+2.72}_{-2.65}$ \\ 
\hline
Basaltic, \OtwoSOtwo{} & N Inc & $-8.27^{+2.58}_{-2.40}$ & -4 & $-1.66^{+0.97}_{-1.20}$ & N Inc & $-7.77^{+2.76}_{-2.58}$ & N Inc & $-7.23^{+2.49}_{-2.85}$ \\ 
\hline
Basaltic, \OtwoHtwoO{} & N Inc & $-6.62^{+4.00}_{-3.60}$ & N Inc & $-6.31^{+3.52}_{-3.55}$ & -6 & $-5.93^{+3.56}_{-3.75}$ & N Inc & $-6.12^{+3.37}_{-3.51}$ \\ 
\hline
Basaltic, \NtwoCHfour{} & N Inc & $-8.12^{+2.26}_{-2.03}$ & N Inc & $-7.75^{+3.10}_{-2.78}$ & N Inc & $-7.54^{+2.97}_{-2.82}$ & -2 & $-1.21^{+0.65}_{-0.92}$ \\ 
\hline
Metal-rich, \OtwoCOtwo{} & -4 & $-2.33^{+1.19}_{-1.56}$ & N Inc & $-8.17^{+2.28}_{-2.40}$ & N Inc & $-7.73^{+3.05}_{-2.66}$ & N Inc & $-7.81^{+2.71}_{-2.55}$ \\ 
\hline
Metal-rich, \OtwoSOtwo{} & N Inc & $-8.41^{+2.51}_{-2.25}$ & -4 & $-2.10^{+1.24}_{-1.27}$ & N Inc & $-8.05^{+2.70}_{-2.22}$ & N Inc & $-8.15^{+2.37}_{-2.30}$ \\ 
\hline
Metal-rich, \OtwoHtwoO{} & N Inc & $-7.54^{+4.99}_{-2.99}$ & N Inc & $-7.53^{+4.52}_{-2.87}$ & -6 & $-6.86^{+3.70}_{-3.24}$ & N Inc & $-7.36^{+4.13}_{-2.87}$ \\ 
\hline
Metal-rich, \NtwoCHfour{} & N Inc & $-7.80^{+2.19}_{-2.33}$ & N Inc & $-7.47^{+2.53}_{-2.34}$ & N Inc & $-6.68^{+2.79}_{-3.35}$ & -2 & $-1.16^{+0.61}_{-0.75}$ \\ 
\hline
Fe-oxidized, \OtwoCOtwo{} & -4 & $-2.06^{+1.10}_{-1.68}$ & N Inc & $-7.86^{+2.42}_{-2.55}$ & N Inc & $-7.35^{+3.07}_{-2.92}$ & N Inc & $-7.75^{+2.73}_{-2.50}$ \\ 
\hline
Fe-oxidized, \OtwoSOtwo{} & N Inc & $-7.92^{+2.19}_{-2.47}$ & -4 & $-1.71^{+0.97}_{-1.34}$ & N Inc & $-8.32^{+2.94}_{-2.40}$ & N Inc & $-7.67^{+2.57}_{-2.59}$ \\ 
\hline
Fe-oxidized, \OtwoHtwoO{} & N Inc & $-6.31^{+3.58}_{-3.62}$ & N Inc & $-5.99^{+3.47}_{-3.55}$ & -6 & $-5.77^{+3.40}_{-3.91}$ & N Inc & $-6.09^{+3.64}_{-3.56}$ \\ 
\hline
Fe-oxidized, \NtwoCHfour{} & N Inc & $-8.23^{+1.88}_{-2.08}$ & N Inc & $-7.77^{+2.59}_{-2.10}$ & N Inc & $-8.13^{+3.83}_{-2.69}$ & -2 & $-1.39^{+0.71}_{-0.98}$ \\ 
\hline
Ultramafic, \OtwoCOtwo{} & -4 & $-2.62^{+1.45}_{-2.01}$ & N Inc & $-8.30^{+2.45}_{-2.12}$ & N Inc & $-6.09^{+3.24}_{-3.71}$ & N Inc & $-7.14^{+2.42}_{-2.96}$ \\ 
\hline
Ultramafic, \OtwoSOtwo{} & N Inc & $-8.04^{+2.61}_{-2.31}$ & -4 & $-1.64^{+0.90}_{-1.30}$ & N Inc & $-7.61^{+2.84}_{-2.68}$ & N Inc & $-7.67^{+3.03}_{-2.66}$ \\ 
\hline
Ultramafic, \OtwoHtwoO{} & N Inc & \cellcolor{purple}$-6.83^{+3.56}_{-3.47}$ & N Inc & \cellcolor{purple}$-5.75^{+3.26}_{-3.95}$ & -6 & \cellcolor{purple}$-6.03^{+3.47}_{-3.70}$ & N Inc & \cellcolor{purple}$-6.36^{+3.76}_{-3.68}$ \\ 
\hline
Ultramafic, \NtwoCHfour{} & N Inc & $-8.40^{+2.39}_{-2.35}$ & N Inc & $-7.44^{+2.57}_{-2.40}$ & N Inc & $-6.31^{+2.82}_{-2.97}$ & -2 & $-1.18^{+0.65}_{-0.84}$ \\ 
\hline
\end{tabular}
\end{center}
\caption{Logarithm of input and retrieved gas volume mixing ratios for the atmospheres in the ``thick'' limit (see Sect.~\ref{sec:thick_atmo}) without and with realistic surface crust.``N Inc" means that the species is not included in the forward model. Individual results discussed in Sect.~\ref{sec:retrieval} and presented in Figures~\ref{fig:retrieval_no_surf_no_atmo} and \ref{fig:retrieval_o2_h2o} are highlighted in magenta. Note that O$_2$ and N$_2$, treated as background gases in the retrieval modeling, are not directly retrieved and thus their mixing ratios are not listed here.}
\label{table:platon_atmo}
\end{table*}

\begin{table*}
\footnotesize
\renewcommand{\arraystretch}{1.5}
\begin{center}
\hskip-2.5cm\begin{tabular}{ |p{3cm}|c|c|c|c|}
\multicolumn{5}{c}{{\bf Retrieval Results for Surface-only Models}} \\
\hline
 & CO$_2$ & SO$_2$ & H$_2$O & CH$_4$ \\
\hline
Basaltic, No Atmo. & $-6.79^{+3.58}_{-3.11}$ & $-6.11^{+3.20}_{-3.53}$ & \cellcolor{purple}$-4.07^{+2.56}_{-5.10}$ & $-6.57^{+3.41}_{-3.18}$ \\ 
\hline
Metal-rich, No Atmo. & $-6.67^{+3.83}_{-3.31}$ & $-6.86^{+3.55}_{-3.15}$ & \cellcolor{purple}$-5.53^{+3.43}_{-4.18}$ & $-6.32^{+3.55}_{-3.55}$ \\ 
\hline
Fe-oxidized, No Atmo. & $-6.57^{+3.77}_{-3.40}$ & $-6.95^{+3.62}_{-3.06}$ & \cellcolor{purple}$-3.54^{+2.21}_{-4.42}$ & $-6.99^{+3.85}_{-2.98}$ \\ 
\hline
Ultramafic, No Atmo. & $-7.03^{+3.62}_{-3.12}$ & $-6.04^{+3.70}_{-3.91}$ & \cellcolor{purple}$-4.05^{+2.49}_{-4.85}$ & $-6.10^{+3.82}_{-3.64}$ \\ 
\hline
\end{tabular}
\end{center}
\caption{Logarithm of retrieved gas volume mixing ratios for surface models without including an atmosphere. Individual results discussed in Sect.~\ref{sec:retrieval} and presented in Figure~\ref{fig:retrieval_no_surf_no_atmo} are highlighted in magenta. Note that O$_2$ and N$_2$, treated as background gases in the retrieval modeling, are not directly retrieved and thus their mixing ratios are not listed here.}
\label{table:platon_noatmo}
\end{table*}

\section{Discussion \& Conclusions \label{sec:discussion}}

\subsection{New Constraints on the Surface and Atmosphere of LHS 3844b and their Observability}

In this study we have explored the feasibility of characterizing the atmosphere and surface of the rocky super-Earth LHS 3844b with JWST. To find the parameter space of atmospheres and surface types that are plausible for LHS 3844b, we have modeled the planetary emission of LHS 3844b, including the spectral signal of both atmosphere and surface, and exhaustively explored all scenarios that are consistent with the existing Spitzer 4.5 $\mu$m measurement of \citet{Kreidberg_2019}. For the surface we have assumed six crust compositions that are common in the solar system and found that the surfaces that are consistent with Spitzer are metal-rich, iron oxidized and basaltic crusts. In contrast, inconsistent with the data are ultramafic, granitoid and feldspathic surfaces, whose high albedos lead to a planetary thermal emission too far below the recorded one. Our atmospheric models consist of O$_2$, N$_2$, CO$_2$ and CO dominated atmospheres with H$_2$O, CO$_2$, CO, CH$_4$ and SO$_2$ as additionally included near-infrared absorbers of various mixing ratios. We have found that in order to be consistent with Spitzer the maximum surface pressure is $10^{-1}$ bars for the models including CO$_2$ or CO and 1 bar for the models including H$_2$O, CH$_4$ or SO$_2$ if the mixing ratio of the included infrared absorber $\geq$ 100 ppm.

Next, we have conducted a JWST observability analysis exploring two limits; on one end we have assumed there is no atmosphere and on the other end we have investigated the atmospheric scenarios that provide the largest gas features while being consistent with the Spitzer data. In the no-atmosphere limit our analysis predicts that with the MIRI LRS instrument it will be very difficult to disentangle specific surface types from a blackbody null-hypothesis. However, with NIRSpec G395H the more reflective surfaces (albeit implausible given the Spitzer constraint) could be differentiated from a blackbody with $\leq$ 6 eclipse observations. Making use of not only the feature strength but also the total offset of the surface emission (or reflection) over an instrument bandpass, we have found that among the plausible surfaces the basaltic and metal-rich surfaces are distinguishable with 6 eclipse measurements and the metal-rich and iron oxidized surfaces are distinguishable with 10 eclipse measurements using NIRSpec. The more reflective surfaces disfavored by Spitzer, the granitoid, feldspathic and ultramafic crusts, would be distinguishable from most of the other surfaces with 1 - 4 eclipses using NIRSpec.

Exploring the limit with an atmosphere, we have found that each atmospheric model is distinguishable from a surface-only spectrum in under 3 eclipse observations if the better-suited JWST instrument, MIRI or NIRSpec, is used in each case. The most amenable cases are the atmospheres with SO$_2$ and CH$_4$, each requiring only a single eclipse to be constrained using MIRI or NIRSpec, respectively.

Lastly, we have run a suite of atmospheric retrieval models to determine how JWST may help constrain the atmospheric composition of LHS 3844b with a Bayesian framework. We have also explored whether surface albedo variations could bias the retrieval algorithm or even be mistaken for atmospheric signatures. Using uncertainties based on 5 eclipse observations, we have been able to detect all four of the included gas absorbers, albeit three of the retrieved mixing ratios are somewhat overpredicted. Furthermore, we have found that the non-grey surface has negligible effect on an atmospheric species' retrieved mixing ratio if the species' spectral signature is sufficiently large. However, the surface contamination of the atmospheric retrieval may lead to a false, weak detection of atmospheric H$_2$O and may also make it harder to disfavor the presence of gas species.

LHS 3844b will be observed during 3 secondary eclipses with JWST MIRI/LRS as part of Cycle 1 of the General Observer program. First, these observations will extend our knowledge about the presence and thickness of an atmosphere and also verify the existing Spitzer data point, testing the measured eclipse depth at 4.5 \micron{} and its derived error. Furthermore, according to our predictions these MIRI observations will indicate whether the planet's surface is granitoid in nature, but the allocated time will not suffice to place any constrains on the other potential surface crusts. Also, if present, any significant amounts of H$_2$O, SO$_2$ or CH$_4$ should be detected, provided the planetary atmosphere is thick enough.

\subsection{Comparison to \citet{Kreidberg_2019} and Implications for Surface and Atmosphere}
\label{sec:kreidberg_comp}

Our current study expands the LHS 3844b phase curve analysis of \citet{Kreidberg_2019} (herein after K19) on multiple fronts. First, we have extended the number of tested surface types from ultramafic, feldspathic, basaltic and granitoid to further include a metal-rich and an iron oxidized surface. Additionally, while K19 assumed atmospheres made up of O$_2$, N$_2$ and CO$_2$, we have tested a wider range of atmospheres by including CO, CH$_4$, SO$_2$ and H$_2$O as absorbers.  Finally, our modeling of LHS 3844b is done using a radiative-convective two-stream radiative transfer code that additionally accounts for energy balance at the planet's solid surface. The analysis of LHS 3844b in K19 was performed against atmosphere models using the techniques described in \citet{morley17}, which utilize a simplified temperature prescription and do not natively account for a surface.  

While our results are in broad agreement with the LHS 3844b phase curve analysis of K19, we also find differences between our and their model predictions that are worth discussion. First, the wavelength-varying secondary eclipse depths and consequently the modeled Spitzer 4.5 $\mu$m eclipse depths of the bare-rock models in our work is significantly lower than predicted by K19 for the same conditions. Second, analogously to the bare-rock models, the eclipse depths of the models with (optically) thin atmospheres are also significantly lower than found by K19 for the same conditions. In contrast, the optically thick atmospheres provide eclipse depths that are larger than in K19. For example, K19's calculated eclipse depths of the O$_2$ + 1 ppm CO$_2$ atmospheres are larger than ours for all surface pressures $\leq$ 10 bar. However, K19's models with 100\% CO$_2$ provide a shallower eclipse depth throughout almost all surface pressures (apart from 1 bar) compared to ours.

We believe the discrepancy in the atmosphere models can at least partially be attributed to the sophistication of the atmospheric radiative transfer modeling as we utilize self-consistent radiative-convective equilibrium models whereas they relied on parameterized temperature profiles which do not take the non-grey radiative feedback of gas absorbers into account. Yet, the modeling treatment of the atmosphere can only explain the cases with optically thick atmospheres, as otherwise the atmosphere has a marginal impact on the planetary spectrum. The discrepancy in the optically thin atmosphere and no-atmosphere models we believe is due to the treatment of the host star radiation. In K19, they scaled the stellar spectrum model to match the measured stellar flux density over the Spitzer 4.5 $\mu$m bandpass. However, this absolute flux measurement is not consistent with the parameters of LHS 3844 derived from SED fitting \citep{Vanderspek_2019, Kreidberg_2019}. In the present work, we have chosen to follow the literature and use a PHOENIX stellar spectrum for the parameters given in K19 without any additional spectrum scaling, as it remains unclear whether the eclipse modeling should be guided by the single-band photometric measurement (that could be prone to an unknown systematic error) or the parameters derived from the overall stellar fit. Ultimately, if multiple flux measurements of the star cannot be reconciled, the correct modeling procedure remains ambiguous. This highlights the need for accurate stellar flux measurements as any uncertainty in the treatment of the star directly affects the secondary eclipse depth.

A shallower eclipse depth prediction for the bare-rock models compared to K19 directly translates to tighter constraints on the type of surfaces that are possible for this planet when taking the Spitzer data point at face value. We recover their result that a basaltic surface is consistent with Spitzer at 3 $\sigma$, but our ultramafic model is outside of this confidence interval in contrast to their modeling. 

In terms of atmospheric thickness, our new modeling sets the top limit on the surface pressure at $\sim 1$ bar down from the previous limit of 10 bar, if an infrared absorbing gas is included at $\geq 100$ ppm. Their best-fit models ($< 1$ $\sigma$ deviation from observation) that include a non-marginal amount of near-infrared absorber at $\geq 10$ ppm require a thin atmosphere with a surface pressure $< 0.1$ bar. Interestingly, our models that fall within 1 $\sigma$ allow atmospheres with up to 1 bar surface pressure if CH$_4$ is included. That is because CH$_4$ is not only a strong greenhouse gas, thus warming the deep atmosphere and surface, but also CH$_4$ is weakly absorbing in a spectral window around 4.5 $\mu$m, enabling a higher thermal emission from the deep atmosphere in the Spitzer bandpass.

\subsection{General Implications for the Characterization of Rocky Exoplanets}

Expanding the study of exoplanet atmospheres into the terrestrial planet regime is a major goal in the JWST era.  Yet such planets will still be challenging atmospheric characterization targets, even with the improved observing capabilities of JWST.  It is therefore necessary to establish robust strategies for extracting meaningful constraints on terrestrial planet atmospheres.  Among such targets, hot rocky planets orbiting bright stars, such as LHS 3844b present the best opportunities for detailed characterization.

A zeroth order question to answer for rocky exoplanet targets is whether they possess atmospheres at all, and if so, how thick said atmospheres are \citep{koll19a, Kreidberg_2019}.  The next natural follow-on question is to establish the composition of the planet's atmosphere and surface.  In this work, we have demonstrated how to go about addressing questions of surface and atmosphere composition for the thick (or thin) and no atmosphere case, as applied to the most amenable rocky target for thermal emission characterization, LHS 3844b.  We have quantified the amount of observing time required and the limitations to which types of surfaces and atmospheres can be uniquely distinguished.  We have also identified shortcomings in our current retrieval capabilities for measuring the properties of terrestrial planet atmospheres.  The approach that we have outlined here can be applied, in principle, to any rocky planet target that presents sufficiently high signal-to-noise thermal emission.    

One particularly subtle challenge that we have discussed at length in this work is that of measuring a planet's albedo and its surface temperature.  When interpreting the measured planetary emission one should be aware that due to the less-than-unity emissivity of the realistic surfaces, the inferred brightness temperature is substantially lower than the true surface temperature for both NIRSpec and MIRI observations. However, while inferring the planetary albedo \citep[using the techniques described in this paper and in][]{Mansfield_2019} with NIRSpec is relatively accurate, using MIRI the inferred albedo is somewhat underestimated for the higher-albedo surfaces and overestimated for the lower-albedo surfaces.

The \citet{Kreidberg_2019} Spitzer phase curve observations and our interpretation in this paper have already provided a phenomenal degree of insight into the properties of LHS 3844b.  If the planet possesses an atmosphere at all, it is thinner than the Earth's.  Yet this conclusion is based only on photometric measurements in a single bandpass.  JWST will, for the first time, open the door to \textit{spectroscopic} characterization of LHS 3844b as well as a wider range of rocky planets, ultimately providing much deeper insight into the fundamental properties of terrestrial planet surfaces and atmospheres. We look forward to the era of rocky planet characterization that is about to begin. 

\begin{acknowledgements}
This study has been conducted as part of the NASA Exoplanets Research Program, grant No.~80NSSC20K0269 (PI: M.Malik).  E.M.-R.K. and J.I. acknowledge support from the National Science Foundation under CAREER Grant No.1931736 and from the AEThER program, funded by the Alfred P. Sloan Foundation under grant G202114194. M. Mansfield acknowledges support from the NASA Hubble Fellowship grant HST-HF2-51485.001-A awarded by STScI. Part of the research was carried out at the Jet Propulsion Laboratory, California Institute of Technology, under a contract with the National Aeronautics and Space Administration. This research has made use of the SVO Filter Profile Service (http://svo2.cab.inta-csic.es/theory/fps/) supported from the Spanish MINECO through grant AYA2017-84089.
\end{acknowledgements}

\software{
\texttt{astropy} \citep{astropy},
\texttt{CUDA} \citep{cuda},
\texttt{HELIOS} \citep{Malik_2017, Malik_2019a, Malik_2019b},
\texttt{\helios{}-K} \citep{Grimm_2015, grimm21},
\texttt{Matplotlib} \citep{matplotlib},
\texttt{NumPy} \citep{numpy},
\texttt{\pandexo{}} \citep{batalha17a},
\texttt{\platon{}} \citep{zhang19,zhang20}
\texttt{PyCUDA} \citep{pycuda}
\texttt{SciPy} \citep{scipy},
}

\bibliography{sample631, bib_matej}{}

\begin{thebibliography}{}
\expandafter\ifx\csname natexlab\endcsname\relax\def\natexlab#1{#1}\fi
\providecommand{\url}[1]{\href{#1}{#1}}
\providecommand{\dodoi}[1]{doi:~\href{http://doi.org/#1}{\nolinkurl{#1}}}
\providecommand{\doeprint}[1]{\href{http://ascl.net/#1}{\nolinkurl{http://ascl.net/#1}}}
\providecommand{\doarXiv}[1]{\href{https://arxiv.org/abs/#1}{\nolinkurl{https://arxiv.org/abs/#1}}}

\bibitem[{{Airapetian} {et~al.}(2016){Airapetian}, {Glocer}, {Gronoff},
  {H{\'e}brard}, \& {Danchi}}]{airapetian16}
{Airapetian}, V.~S., {Glocer}, A., {Gronoff}, G., {H{\'e}brard}, E., \&
  {Danchi}, W. 2016, Nature Geoscience, 9, 452, \dodoi{10.1038/ngeo2719}

\bibitem[{Batalha {et~al.}(2017)Batalha, Mandell, Pontoppidan, Stevenson,
  Lewis, Kalirai, Earl, Greene, Albert, \& Nielsen}]{Batalha_2017}
Batalha, N.~E., Mandell, A., Pontoppidan, K., {et~al.} 2017, Publications of
  the Astronomical Society of the Pacific, 129, 064501,
  \dodoi{10.1088/1538-3873/aa65b0}

\bibitem[{{Batalha} {et~al.}(2017){Batalha}, {Mandell}, {Pontoppidan},
  {Stevenson}, {Lewis}, {Kalirai}, {Earl}, {Greene}, {Albert}, \&
  {Nielsen}}]{batalha17a}
{Batalha}, N.~E., {Mandell}, A., {Pontoppidan}, K., {et~al.} 2017, \pasp, 129,
  064501, \dodoi{10.1088/1538-3873/aa65b0}

\bibitem[{{Bishop} {et~al.}(2019){Bishop}, {Bell}, \& {Moersch}}]{bishop19}
{Bishop}, J.~L., {Bell}, James~F., I., \& {Moersch}, J.~E. 2019, {Remote
  Compositional Analysis: Techniques for Understanding Spectroscopy,
  Mineralogy, and Geochemistry of Planetary Surfaces.},
  \dodoi{10.1017/9781316888872}

\bibitem[{Blewett {et~al.}(1997)Blewett, Lucey, Hawke, Ling, \&
  Robinson}]{Blewett_1997}
Blewett, D.~T., Lucey, P.~G., Hawke, B., Ling, G., \& Robinson, M.~S. 1997,
  Icarus, 129, 217, \dodoi{10.1006/icar.1997.5785}

\bibitem[{{Burrows} {et~al.}(2008){Burrows}, {Budaj}, \& {Hubeny}}]{burrows08}
{Burrows}, A., {Budaj}, J., \& {Hubeny}, I. 2008, \apj, 678, 1436,
  \dodoi{10.1086/533518}

\bibitem[{{Carli} {et~al.}(2015){Carli}, {Serventi}, \& {Sgavetti}}]{carli15}
{Carli}, C., {Serventi}, G., \& {Sgavetti}, M. 2015, Geological Society of
  London Special Publications, 401, 139, \dodoi{10.1144/SP401.19}

\bibitem[{Chameides \& Walker(1981)}]{chameides81}
Chameides, W., \& Walker, J. 1981, Origins of Life, 11, 291,
  \dodoi{10.1007/BF00931483}

\bibitem[{{Cox}(2000)}]{cox00}
{Cox}, A.~N. 2000, {Allen's astrophysical quantities} ((New York: AIP Press))

\bibitem[{{Diamond-Lowe} {et~al.}(2021){Diamond-Lowe}, {Charbonneau}, {Malik},
  {Kempton}, \& {Youngblood}}]{diamond-lowe21}
{Diamond-Lowe}, H., {Charbonneau}, D., {Malik}, M., {Kempton}, E., \&
  {Youngblood}, A. 2021, in American Astronomical Society Meeting Abstracts,
  Vol.~53, American Astronomical Society Meeting Abstracts, 302.05

\bibitem[{{Elkins-Tanton} \& {Seager}(2008)}]{elkins-tanton08}
{Elkins-Tanton}, L.~T., \& {Seager}, S. 2008, \apj, 685, 1237,
  \dodoi{10.1086/591433}

\bibitem[{{Gaillard} \& {Scaillet}(2014)}]{gaillard14}
{Gaillard}, F., \& {Scaillet}, B. 2014, Earth and Planetary Science Letters,
  403, 307, \dodoi{10.1016/j.epsl.2014.07.009}

\bibitem[{{Gao} {et~al.}(2015){Gao}, {Hu}, {Robinson}, {Li}, \& {Yung}}]{gao15}
{Gao}, P., {Hu}, R., {Robinson}, T.~D., {Li}, C., \& {Yung}, Y.~L. 2015, \apj,
  806, 249, \dodoi{10.1088/0004-637X/806/2/249}

\bibitem[{Gordon {et~al.}(2017)Gordon, Rothman, Hill, Kochanov, Tan, Bernath,
  Birk, Boudon, Campargue, Chance, Drouin, Flaud, Gamache, Hodges, Jacquemart,
  Perevalov, Perrin, Shine, Smith, Tennyson, Toon, Tran, Tyuterev, Barbe,
  Császár, Devi, Furtenbacher, Harrison, Hartmann, Jolly, Johnson, Karman,
  Kleiner, Kyuberis, Loos, Lyulin, Massie, Mikhailenko, Moazzen-Ahmadi,
  Müller, Naumenko, Nikitin, Polyansky, Rey, Rotger, Sharpe, Sung, Starikova,
  Tashkun, Auwera, Wagner, Wilzewski, Wcisło, Yu, \& Zak}]{gordon17}
Gordon, I., Rothman, L., Hill, C., {et~al.} 2017, Journal of Quantitative
  Spectroscopy and Radiative Transfer, 203, 3 ,
  \dodoi{https://doi.org/10.1016/j.jqsrt.2017.06.038}

\bibitem[{Grimm \& Heng(2015)}]{Grimm_2015}
Grimm, S.~L., \& Heng, K. 2015, The Astrophysical Journal, 808, 182,
  \dodoi{10.1088/0004-637x/808/2/182}

\bibitem[{{Grimm} {et~al.}(2021){Grimm}, {Malik}, {Kitzmann},
  {Guzm{\'a}n-Mesa}, {Hoeijmakers}, {Fisher}, {Mendon{\c{c}}a}, {Yurchenko},
  {Tennyson}, {Alesina}, {Buchschacher}, {Burnier}, {Segransan}, {Kurucz}, \&
  {Heng}}]{grimm21}
{Grimm}, S.~L., {Malik}, M., {Kitzmann}, D., {et~al.} 2021, \apjs, 253, 30,
  \dodoi{10.3847/1538-4365/abd773}

\bibitem[{{Hansen}(2008)}]{hansen08}
{Hansen}, B. M.~S. 2008, \apjs, 179, 484, \dodoi{10.1086/591964}

\bibitem[{Hu {et~al.}(2012)Hu, Ehlmann, \& Seager}]{Hu_2012}
Hu, R., Ehlmann, B.~L., \& Seager, S. 2012, The Astrophysical Journal, 752, 7,
  \dodoi{10.1088/0004-637x/752/1/7}

\bibitem[{{Hunten} {et~al.}(1987){Hunten}, {Pepin}, \& {Walker}}]{hunten87}
{Hunten}, D.~M., {Pepin}, R.~O., \& {Walker}, J.~C.~G. 1987, \icarus, 69, 532,
  \dodoi{10.1016/0019-1035(87)90022-4}

\bibitem[{Hunter(2007)}]{matplotlib}
Hunter, J.~D. 2007, Computing In Science \& Engineering, 9, 90,
  \dodoi{10.1109/MCSE.2007.55}

\bibitem[{Husser {et~al.}(2013)Husser, von Berg, Dreizler, Homeier, Reiners,
  Barman, \& Hauschildt}]{Husser_2013}
Husser, T.-O., von Berg, S.~W., Dreizler, S., {et~al.} 2013, Astronomy {\&}
  Astrophysics, 553, A6, \dodoi{10.1051/0004-6361/201219058}

\bibitem[{{Ih} \& {Kempton}(2021)}]{ih21}
{Ih}, J., \& {Kempton}, E. M.~R. 2021, \aj, 162, 237,
  \dodoi{10.3847/1538-3881/ac173b}

\bibitem[{Izenberg {et~al.}(2014)Izenberg, Klima, Murchie, Blewett, Holsclaw,
  McClintock, Malaret, Mauceri, Vilas, Sprague, Helbert, Domingue, Head,
  Goudge, Solomon, Hibbitts, \& Dyar}]{izenberg14}
Izenberg, N.~R., Klima, R.~L., Murchie, S.~L., {et~al.} 2014, Icarus, 228, 364,
  \dodoi{https://doi.org/10.1016/j.icarus.2013.10.023}

\bibitem[{{Kane} {et~al.}(2014){Kane}, {Kopparapu}, \&
  {Domagal-Goldman}}]{kane14}
{Kane}, S.~R., {Kopparapu}, R.~K., \& {Domagal-Goldman}, S.~D. 2014, \apjl,
  794, L5, \dodoi{10.1088/2041-8205/794/1/L5}

\bibitem[{Kempton {et~al.}(2018)Kempton, Bean, Louie, Deming, Koll, Mansfield,
  Christiansen, L{\'{o}}pez-Morales, Swain, Zellem, Ballard, Barclay, Barstow,
  Batalha, Beatty, Berta-Thompson, Birkby, Buchhave, Charbonneau, Cowan,
  Crossfield, de~Val-Borro, Doyon, Dragomir, Gaidos, Heng, Hu, Kane, Kreidberg,
  Mallonn, Morley, Narita, Nascimbeni, Pall{\'{e}}, Quintana, Rauscher, Seager,
  Shkolnik, Sing, Sozzetti, Stassun, Valenti, \& von Essen}]{Kempton_2018}
Kempton, E. M.-R., Bean, J.~L., Louie, D.~R., {et~al.} 2018, Publications of
  the Astronomical Society of the Pacific, 130, 114401,
  \dodoi{10.1088/1538-3873/aadf6f}

\bibitem[{{Kite} \& {Barnett}(2020)}]{kite20}
{Kite}, E.~S., \& {Barnett}, M.~N. 2020, Proceedings of the National Academy of
  Science, 117, 18264.
\newblock \doarXiv{2006.02589}

\bibitem[{{Kite} \& {Schaefer}(2021)}]{kite21}
{Kite}, E.~S., \& {Schaefer}, L. 2021, \apjl, 909, L22,
  \dodoi{10.3847/2041-8213/abe7dc}

\bibitem[{{Kl{\"o}ckner} {et~al.}(2012){Kl{\"o}ckner}, {Pinto}, {Lee},
  {Catanzaro}, {Ivanov}, \& {Fasih}}]{pycuda}
{Kl{\"o}ckner}, A., {Pinto}, N., {Lee}, Y., {et~al.} 2012, Parallel Computing,
  38, 157, \dodoi{10.1016/j.parco.2011.09.001}

\bibitem[{{Koll}(2022)}]{koll22}
{Koll}, D. D.~B. 2022, \apj, 924, 134, \dodoi{10.3847/1538-4357/ac3b48}

\bibitem[{{Koll} {et~al.}(2019){Koll}, {Malik}, {Mansfield}, {Kempton}, {Kite},
  {Abbot}, \& {Bean}}]{koll19a}
{Koll}, D. D.~B., {Malik}, M., {Mansfield}, M., {et~al.} 2019, \apj, 886, 140,
  \dodoi{10.3847/1538-4357/ab4c91}

\bibitem[{Kreidberg {et~al.}(2019)Kreidberg, Koll, Morley, Hu, Schaefer,
  Deming, Stevenson, Dittmann, Vanderburg, Berardo, Guo, Stassun, Crossfield,
  Charbonneau, Latham, Loeb, Ricker, Seager, \& Vanderspek}]{Kreidberg_2019}
Kreidberg, L., Koll, D. D.~B., Morley, C., {et~al.} 2019, Nature, 573, 87,
  \dodoi{10.1038/s41586-019-1497-4}

\bibitem[{{Lammer} {et~al.}(2019){Lammer}, {Spro{\ss}}, {Grenfell}, {Scherf},
  {Fossati}, {Lendl}, \& {Cubillos}}]{lammer19}
{Lammer}, H., {Spro{\ss}}, L., {Grenfell}, J.~L., {et~al.} 2019, Astrobiology,
  19, 927, \dodoi{10.1089/ast.2018.1914}

\bibitem[{Li {et~al.}(2015)Li, Gordon, Rothman, Tan, Hu, Kassi, Campargue, \&
  Medvedev}]{li15}
Li, G., Gordon, I.~E., Rothman, L.~S., {et~al.} 2015, The Astrophysical Journal
  Supplement Series, 216, 15.
\newblock \url{http://stacks.iop.org/0067-0049/216/i=1/a=15}

\bibitem[{{Liggins} {et~al.}(2021){Liggins}, {Jordan}, {Rimmer}, \&
  {Shorttle}}]{liggins21}
{Liggins}, P., {Jordan}, S., {Rimmer}, P.~B., \& {Shorttle}, O. 2021, arXiv
  e-prints, arXiv:2111.05161.
\newblock \doarXiv{2111.05161}

\bibitem[{{Luger} \& {Barnes}(2015)}]{luger15}
{Luger}, R., \& {Barnes}, R. 2015, Astrobiology, 15, 119,
  \dodoi{10.1089/ast.2014.1231}

\bibitem[{{Lupu} {et~al.}(2014){Lupu}, {Zahnle}, {Marley}, {Schaefer},
  {Fegley}, {Morley}, {Cahoy}, {Freedman}, \& {Fortney}}]{lupu14}
{Lupu}, R.~E., {Zahnle}, K., {Marley}, M.~S., {et~al.} 2014, \apj, 784, 27,
  \dodoi{10.1088/0004-637X/784/1/27}

\bibitem[{{Madhusudhan} \& {Seager}(2009)}]{madhusudhan09}
{Madhusudhan}, N., \& {Seager}, S. 2009, \apj, 707, 24,
  \dodoi{10.1088/0004-637X/707/1/24}

\bibitem[{Malik {et~al.}(2019{\natexlab{a}})Malik, Kempton, Koll, Mansfield,
  Bean, \& Kite}]{Malik_2019b}
Malik, M., Kempton, E. M.-R., Koll, D. D.~B., {et~al.} 2019{\natexlab{a}}, The
  Astrophysical Journal, 886, 142, \dodoi{10.3847/1538-4357/ab4a05}

\bibitem[{Malik {et~al.}(2019{\natexlab{b}})Malik, Kitzmann, Mendon{\c{c}}a,
  Grimm, Marleau, Linder, Tsai, \& Heng}]{Malik_2019a}
Malik, M., Kitzmann, D., Mendon{\c{c}}a, J.~M., {et~al.} 2019{\natexlab{b}},
  The Astronomical Journal, 157, 170, \dodoi{10.3847/1538-3881/ab1084}

\bibitem[{Malik {et~al.}(2017)Malik, Grosheintz, Mendon{\c{c}}a, Grimm, Lavie,
  Kitzmann, Tsai, Burrows, Kreidberg, Bedell, Bean, Stevenson, \&
  Heng}]{Malik_2017}
Malik, M., Grosheintz, L., Mendon{\c{c}}a, J.~M., {et~al.} 2017, The
  Astronomical Journal, 153, 56, \dodoi{10.3847/1538-3881/153/2/56}

\bibitem[{Mansfield {et~al.}(2019)Mansfield, Kite, Hu, Koll, Malik, Bean, \&
  Kempton}]{Mansfield_2019}
Mansfield, M., Kite, E.~S., Hu, R., {et~al.} 2019, The Astrophysical Journal,
  886, 141, \dodoi{10.3847/1538-4357/ab4c90}

\bibitem[{{Matsuo} {et~al.}(2019){Matsuo}, {Greene}, {Johnson}, {Mcmurray},
  {Roellig}, \& {Ennico}}]{matsuo19}
{Matsuo}, T., {Greene}, T.~P., {Johnson}, R.~R., {et~al.} 2019, \pasp, 131,
  124502, \dodoi{10.1088/1538-3873/ab42f1}

\bibitem[{McCord \& Westphal(1971)}]{McCord_1971}
McCord, T.~B., \& Westphal, J.~A. 1971, The Astrophysical Journal, 168, 141,
  \dodoi{10.1086/151069}

\bibitem[{{Mikhail} \& {Sverjensky}(2014)}]{mikhail14}
{Mikhail}, S., \& {Sverjensky}, D.~A. 2014, Nature Geoscience, 7, 816,
  \dodoi{10.1038/ngeo2271}

\bibitem[{{Morley} {et~al.}(2017){Morley}, {Kreidberg}, {Rustamkulov},
  {Robinson}, \& {Fortney}}]{morley17}
{Morley}, C.~V., {Kreidberg}, L., {Rustamkulov}, Z., {Robinson}, T., \&
  {Fortney}, J.~J. 2017, \apj, 850, 121, \dodoi{10.3847/1538-4357/aa927b}

\bibitem[{Nickolls {et~al.}(2008)Nickolls, Buck, Garland, \& Skadron}]{cuda}
Nickolls, J., Buck, I., Garland, M., \& Skadron, K. 2008, Queue, 6, 40,
  \dodoi{10.1145/1365490.1365500}

\bibitem[{Nittler {et~al.}(2011)Nittler, Starr, Weider, McCoy, Boynton, Ebel,
  Ernst, Evans, Goldsten, Hamara, Lawrence, McNutt, Schlemm, Solomon, \&
  Sprague}]{Nittler_2011}
Nittler, L.~R., Starr, R.~D., Weider, S.~Z., {et~al.} 2011, Science, 333, 1847,
  \dodoi{10.1126/science.1211567}

\bibitem[{{Oliphant}(2007)}]{scipy}
{Oliphant}, T.~E. 2007, {Computing in Science \& Engineering}, 9, 10 ,
  \dodoi{10.1109/MCSE.2007.58}

\bibitem[{{Owen} \& {Wu}(2013)}]{owen13}
{Owen}, J.~E., \& {Wu}, Y. 2013, \apj, 775, 105,
  \dodoi{10.1088/0004-637X/775/2/105}

\bibitem[{Owen \& Wu(2017)}]{Owen_2017}
Owen, J.~E., \& Wu, Y. 2017, The Astrophysical Journal, 847, 29,
  \dodoi{10.3847/1538-4357/aa890a}

\bibitem[{Pieters(1978)}]{Pieters_1978}
Pieters, C.~M. 1978, 9th Lunar and Planetary Science Conference, 2825–2849

\bibitem[{Pieters(1986)}]{Pieters_1986}
---. 1986, Reviews of Geophysics, 24, 557, \dodoi{10.1029/rg024i003p00557}

\bibitem[{{Polyansky} {et~al.}(2018){Polyansky}, {Kyuberis}, {Zobov},
  {Tennyson}, {Yurchenko}, \& {Lodi}}]{polyansky18}
{Polyansky}, O.~L., {Kyuberis}, A.~A., {Zobov}, N.~F., {et~al.} 2018, \mnras,
  480, 2597, \dodoi{10.1093/mnras/sty1877}

\bibitem[{Price-Whelan {et~al.}(2018)Price-Whelan, Sip{\H{o}}cz, Günther, Lim,
  Crawford, Conseil, Shupe, Craig, Dencheva, Ginsburg, VanderPlas, Bradley,
  P{\'{e}}rez-Su{\'{a}}rez, de~Val-Borro, Aldcroft, Cruz, Robitaille, Tollerud,
  Ardelean, Babej, Bach, Bachetti, Bakanov, Bamford, Barentsen, Barmby,
  Baumbach, Berry, Biscani, Boquien, Bostroem, Bouma, Brammer, Bray,
  Breytenbach, Buddelmeijer, Burke, Calderone, Rodr{\'{\i}}guez, Cara, Cardoso,
  Cheedella, Copin, Corrales, Crichton, D'Avella, Deil, Depagne, Dietrich,
  Donath, Droettboom, Earl, Erben, Fabbro, Ferreira, Finethy, Fox, Garrison,
  Gibbons, Goldstein, Gommers, Greco, Greenfield, Groener, Grollier, Hagen,
  Hirst, Homeier, Horton, Hosseinzadeh, Hu, Hunkeler, Ivezi{\'{c}}, Jain,
  Jenness, Kanarek, Kendrew, Kern, Kerzendorf, Khvalko, King, Kirkby, Kulkarni,
  Kumar, Lee, Lenz, Littlefair, Ma, Macleod, Mastropietro, McCully, Montagnac,
  Morris, Mueller, Mumford, Muna, Murphy, Nelson, Nguyen, Ninan, Nöthe, Ogaz,
  Oh, Parejko, Parley, Pascual, Patil, Patil, Plunkett, Prochaska, Rastogi,
  Janga, Sabater, Sakurikar, Seifert, Sherbert, Sherwood-Taylor, Shih, Sick,
  Silbiger, Singanamalla, Singer, Sladen, Sooley, Sornarajah, Streicher,
  Teuben, Thomas, Tremblay, Turner, Terr{\'{o}}n, van Kerkwijk, de~la Vega,
  Watkins, Weaver, Whitmore, Woillez, Zabalza, , \& and}]{astropy}
Price-Whelan, A.~M., Sip{\H{o}}cz, B.~M., Günther, H.~M., {et~al.} 2018, The
  Astronomical Journal, 156, 123, \dodoi{10.3847/1538-3881/aabc4f}

\bibitem[{{Richard} {et~al.}(2012){Richard}, {Gordon}, {Rothman}, {Abel},
  {Frommhold}, {Gustafsson}, {Hartmann}, {Hermans}, {Lafferty}, {Orton},
  {Smith}, \& {Tran}}]{richard12}
{Richard}, C., {Gordon}, I.~E., {Rothman}, L.~S., {et~al.} 2012, \jqsrt, 113,
  1276, \dodoi{10.1016/j.jqsrt.2011.11.004}

\bibitem[{{Rothman} {et~al.}(2010){Rothman}, {Gordon}, {Barber}, {Dothe},
  {Gamache}, {Goldman}, {Perevalov}, {Tashkun}, \& {Tennyson}}]{rothman10}
{Rothman}, L.~S., {Gordon}, I.~E., {Barber}, R.~J., {et~al.} 2010, \jqsrt, 111,
  2139, \dodoi{10.1016/j.jqsrt.2010.05.001}

\bibitem[{{Schaefer} \& {Fegley}(2011)}]{schaefer11}
{Schaefer}, L., \& {Fegley}, Bruce, J. 2011, \apj, 729, 6,
  \dodoi{10.1088/0004-637X/729/1/6}

\bibitem[{{Schaefer} {et~al.}(2012){Schaefer}, {Lodders}, \&
  {Fegley}}]{schaefer12}
{Schaefer}, L., {Lodders}, K., \& {Fegley}, B. 2012, \apj, 755, 41,
  \dodoi{10.1088/0004-637X/755/1/41}

\bibitem[{{Schlawin} {et~al.}(2020){Schlawin}, {Leisenring}, {Misselt},
  {Greene}, {McElwain}, {Beatty}, \& {Rieke}}]{schlawin20}
{Schlawin}, E., {Leisenring}, J., {Misselt}, K., {et~al.} 2020, \aj, 160, 231,
  \dodoi{10.3847/1538-3881/abb811}

\bibitem[{{Schlawin} {et~al.}(2021){Schlawin}, {Leisenring}, {McElwain},
  {Misselt}, {Don}, {Greene}, {Beatty}, {Nikolov}, {Kelly}, \&
  {Rieke}}]{schlawin21}
{Schlawin}, E., {Leisenring}, J., {McElwain}, M.~W., {et~al.} 2021, \aj, 161,
  115, \dodoi{10.3847/1538-3881/abd8d4}

\bibitem[{{Schlichting} {et~al.}(2015){Schlichting}, {Sari}, \&
  {Yalinewich}}]{schlichting15}
{Schlichting}, H.~E., {Sari}, R., \& {Yalinewich}, A. 2015, \icarus, 247, 81,
  \dodoi{10.1016/j.icarus.2014.09.053}

\bibitem[{{Showman} {et~al.}(2013){Showman}, {Wordsworth}, {Merlis}, \&
  {Kaspi}}]{showman13}
{Showman}, A.~P., {Wordsworth}, R.~D., {Merlis}, T.~M., \& {Kaspi}, Y. 2013,
  {Atmospheric Circulation of Terrestrial Exoplanets}, ed. S.~J. {Mackwell},
  A.~A. {Simon-Miller}, J.~W. {Harder}, \& M.~A. {Bullock} (University of
  Arizona Press), 277--327, \dodoi{10.2458/azu\_uapress\_9780816530595-ch012}

\bibitem[{{Sneep} \& {Ubachs}(2005)}]{sneep05}
{Sneep}, M., \& {Ubachs}, W. 2005, \jqsrt, 92, 293

\bibitem[{Thalman {et~al.}(2014)Thalman, J.~Zarzana, Tolbert, \&
  Volkamer}]{thalman14}
Thalman, R., J.~Zarzana, K., Tolbert, M., \& Volkamer, R. 2014, Journal of
  Quantitative Spectroscopy and Radiative Transfer, 147, 171–177,
  \dodoi{10.1016/j.jqsrt.2014.05.030}

\bibitem[{Underwood {et~al.}(2016)Underwood, Tennyson, Yurchenko, Huang,
  Schwenke, Lee, Clausen, \& Fateev}]{underwood16}
Underwood, D.~S., Tennyson, J., Yurchenko, S.~N., {et~al.} 2016, Monthly
  Notices of the Royal Astronomical Society, 459, 3890,
  \dodoi{10.1093/mnras/stw849}

\bibitem[{{van der Walt} {et~al.}(2011){van der Walt}, {Colbert}, \&
  {Varoquaux}}]{numpy}
{van der Walt}, S., {Colbert}, S.~C., \& {Varoquaux}, G. 2011, Computing in
  Science and Engineering, 13, 22, \dodoi{10.1109/MCSE.2011.37}

\bibitem[{Vanderspek {et~al.}(2019)Vanderspek, Huang, Vanderburg, Ricker,
  Latham, Seager, Winn, Jenkins, Burt, Dittmann, Newton, Quinn, Shporer,
  Charbonneau, Irwin, Ment, Winters, Collins, Evans, Gan, Hart, Jensen,
  Kielkopf, Mao, Waalkes, Bouchy, Marmier, Nielsen, Ottoni, Pepe,
  S{\'{e}}gransan, Udry, Henry, Paredes, James, Hinojosa, Silverstein, Palle,
  Berta-Thompson, Crossfield, Davies, Dragomir, Fausnaugh, Glidden, Pepper,
  Morgan, Rose, Twicken, Villase{\~{n}}or, Yu, Bakos, Bean, Buchhave,
  Christensen-Dalsgaard, Christiansen, Ciardi, Clampin, Lee, Deming, Doty,
  Jernigan, Kaltenegger, Lissauer, McCullough, Narita, Paegert, Pal, Rinehart,
  Sasselov, Sato, Sozzetti, Stassun, \& Torres}]{Vanderspek_2019}
Vanderspek, R., Huang, C.~X., Vanderburg, A., {et~al.} 2019, The Astrophysical
  Journal, 871, L24, \dodoi{10.3847/2041-8213/aafb7a}

\bibitem[{Wagner \& Kretzschmar(2008)}]{wagner08}
Wagner, W., \& Kretzschmar, H.-J. 2008, International Steam Tables - Properties
  of Water and Steam Based on the Industrial Formulation IAPWS-IF97 (Springer,
  Berlin, Heidelberg), \dodoi{https://doi.org/10.1007/978-3-540-74234-0}

\bibitem[{Wordsworth \& Pierrehumbert(2014)}]{Wordsworth_2014}
Wordsworth, R., \& Pierrehumbert, R. 2014, The Astrophysical Journal, 785, L20,
  \dodoi{10.1088/2041-8205/785/2/l20}

\bibitem[{Wordsworth {et~al.}(2018)Wordsworth, Schaefer, \&
  Fischer}]{Wordsworth_2018}
Wordsworth, R.~D., Schaefer, L.~K., \& Fischer, R.~A. 2018, The Astronomical
  Journal, 155, 195, \dodoi{10.3847/1538-3881/aab608}

\bibitem[{{Yurchenko} {et~al.}(2017){Yurchenko}, {Amundsen}, {Tennyson}, \&
  {Waldmann}}]{yurchenko17}
{Yurchenko}, S.~N., {Amundsen}, D.~S., {Tennyson}, J., \& {Waldmann}, I.~P.
  2017, \aap, 605, A95, \dodoi{10.1051/0004-6361/201731026}

\bibitem[{{Zahnle} \& {Catling}(2017)}]{zahnle17}
{Zahnle}, K.~J., \& {Catling}, D.~C. 2017, \apj, 843, 122,
  \dodoi{10.3847/1538-4357/aa7846}

\bibitem[{{Zahnle} {et~al.}(2020){Zahnle}, {Lupu}, {Catling}, \&
  {Wogan}}]{zahnle20}
{Zahnle}, K.~J., {Lupu}, R., {Catling}, D.~C., \& {Wogan}, N. 2020, \psj, 1,
  11, \dodoi{10.3847/PSJ/ab7e2c}

\bibitem[{{Zhang} {et~al.}(2019){Zhang}, {Chachan}, {Kempton}, \&
  {Knutson}}]{zhang19}
{Zhang}, M., {Chachan}, Y., {Kempton}, E. M.~R., \& {Knutson}, H.~A. 2019,
  \pasp, 131, 034501, \dodoi{10.1088/1538-3873/aaf5ad}

\bibitem[{{Zhang} {et~al.}(2020){Zhang}, {Chachan}, {Kempton}, {Knutson}, \&
  {Chang}}]{zhang20}
{Zhang}, M., {Chachan}, Y., {Kempton}, E. M.~R., {Knutson}, H.~A., \& {Chang},
  W.~H. 2020, \apj, 899, 27, \dodoi{10.3847/1538-4357/aba1e6}

\bibitem[{{Zhang} {et~al.}(2018){Zhang}, {Zhou}, {Rackham}, \&
  {Apai}}]{zhang18}
{Zhang}, Z., {Zhou}, Y., {Rackham}, B.~V., \& {Apai}, D. 2018, \aj, 156, 178,
  \dodoi{10.3847/1538-3881/aade4f}

\end{thebibliography}
\bibliographystyle{aasjournal}




\appendix{}

\section{Updated numerical iteration in \helios{} with added surface}
\label{app:surface}

For this work we have updated the numerical forward stepping algorithm of \helios{} that iterates towards radiative-convective equilibrium and, in contrast to the previous implementation where the surface temperature was calculated directly from the energy balance across the surface boundary, see Eq.~(4) in \citet{Malik_2019b}, we have also included the surface temperature in the numerical iteration. 

The original expression for the temperature iteration, see Eq.~(24) in \citet{Malik_2017}, required the knowledge of local atmospheric properties, such as the density and heat capacity. Furthermore, the radiative timescale was used, see Eq.~(27) in \citet{Malik_2017}, to make the forward stepping independent of local atmospheric inertia, allowing for a faster and more stable convergence towards equilibrium. However, since the steady-state radiative equilibrium merely requires a vanishing net flux divergence across each atmospheric layer, we have found the inclusion and calculation of such a large number of atmospheric properties not necessary. That is why we have simplified the original formalism to the following expressions. Note that we denote the wavelength-integrated ``bolometric'' flux with $\mathcal F$ ($[\mathcal F] =$ erg s$^{-1}$ cm$^{-2}$) and the spectral flux with $F$ ($[F] =$ erg s$^{-1}$ cm$^{-3}$). The change in temperature of atmospheric layer $i$ ,$\Delta T_i$, between successive iteration steps is calculated as
\begin{equation}
    \Delta T_i = - f_{{\rm pre}, i} \frac{P_i}{\Delta P_i} \Delta \mathcal F_{{\rm net}, i},
\end{equation}
where $P_i$ is the pressure in the center of layer $i$, $\Delta P_i$ is the pressure difference across layer $i$, $\Delta \mathcal F_{{\rm net}, i}$ is the net flux difference across layer $i$ and $f_{{\rm pre}, i}$ is a dimensional prefactor given by
\begin{equation}
\label{eq:prefactor}
f_{{\rm pre}, i} = \frac{{\rm K}\,{\rm s}\,{\rm cm}^2\,{\rm erg}^{-1}}{\left[|(\Delta \mathcal F_{{\rm net}, i})|/({\rm erg}\,{\rm s}^{-1}\,{\rm cm}^{-2})\right]^{0.9}}.
\end{equation}
The goal of this prefactor is first to make the iteration more stable by dampening the impact of $\Delta \mathcal F_{{\rm net}, i}$, thus making the temperature iteration smoother between neighboring layers. Second, $f_{{\rm pre}, i}$ is used to adapt the temperature step during the iteration. If the temperature is close to equilibrium, $f_{{\rm pre}, i}$ becomes increasingly smaller. This prevents numerical oscillations around the equilibrium temperature without ever reaching it (details on the exact evolution of $f_{{\rm pre}, i}$ are given in \citealt{Malik_2017}). Analogous to the atmospheric layer temperatures, the surface temperature starts from the planetary dayside-averaged temperature at the beginning of the run and then advances with each iteration step as
\begin{equation}
\Delta T_{\rm surf} = - f_{{\rm pre, surf}} \frac{P_0}{\Delta P_0} \Delta \mathcal F_{{\rm net, surf}},
\end{equation}
with 
\begin{equation}
\label{eq:surf_flux}
\Delta \mathcal F_{{\rm net, surf}} = \mathcal F_{{\rm net}, 0} - \mathcal F_{\rm intern},
\end{equation}
where $\mathcal F_{\rm intern}$ is the internal heat flux. The index ``0'' denotes the first atmospheric layer or interface from the bottom. The prefactor $f_{{\rm pre, surf}}$ is calculated with Eq.~(\ref{eq:prefactor}), but using $\Delta \mathcal F_{{\rm net, surf}}$ in the denominator. The net flux at the surface boundary is given by $\mathcal F_{{\rm net}, 0} = \mathcal F_{\uparrow, 0} - \mathcal F_{\downarrow, 0}$. One caveat of including the surface in the numerical iteration of the atmosphere is that the surface and the first atmospheric layer sometimes become ``stuck'', oscillating back and forth between iterations, prone to happen when the near-surface atmosphere is optically thick. We employ a numerical trick to solve that issue. As long as the first atmospheric layer is not in radiative equilibrium, we use $\mathcal F_{{\rm net}, 1}$ in place of $\mathcal F_{{\rm net}, 0}$ in Eq.~(\ref{eq:surf_flux}), which essentially includes the first atmospheric layer in the energy balance of the surface. Once the first atmospheric layer has converged, we switch back to the correct energy balance expression for the surface, as given by Eq.~(\ref{eq:surf_flux}).

Finally, the wavelength-varying surface albedo is included in the upward flux from the surface. Namely, the upward pointing spectral flux at the surface is calculated as
\begin{equation}
F_{\uparrow, 0, \lambda} = A_{\rm surf, \lambda} F_{\downarrow, 0, \lambda} + (1 - A_{\rm surf, \lambda}) \pi B_{\lambda}(T_{\rm surf}), 
\end{equation}
where $A_{\rm surf, \lambda}$ is the wavelength-dependent surface albedo, $B_{\lambda}$ is the Planck function and $T_{\rm surf}$ is the surface temperature.

\section{Role of the emissivity on the inferred albedo}
\label{app:emissivity}

The effect of assuming different emissivity values on the inferred albedo is shown in Fig.~\ref{fig:inf_albedo_emissivity}. When fitting a blackbody curve to a spectrum, the total area under the curve integrated from zero to infinity, analogous to the total energy emitted, is not conserved. This discrepancy is exacerbated if the assumed emissivity of the blackbody model significantly differs from the one of the realistic surface spectrum. Moreover, the farther the wavelength region for the fitting is separated from the peak of the planetary thermal emission, the larger is the discrepancy in the areas below the two curves. In the MIRI case, if the blackbody fit assumes an emissivity higher than the one used in the physical model the total energy of the blackbody emission is smaller than in the physical model, resulting in a too low blackbody temperature and consequently a too high inferred albedo. This is best visible for the three grey albedo models with $\alpha=0$, $\alpha=0.1$ and $\alpha=0.3$. When the fit for the grey albedo models is conducted with the correct emissivity the inferred albedo matches the real bond albedo.

When using NIRSpec, this emissivity effect is somewhat smaller and also non-monotonic. This is because in this case the instrument bandpass overlaps with the peak planetary emission making the areas between the blackbody curve and the planetary spectrum more similar. Furthermore, there appears to be a dependence on individual albedo variation. The surfaces showing an inverse trend have an overall higher albedo and a more strongly varying albedo within the NIRSpec bandpass.

\section{Additional Figures}

\renewcommand{\thefigure}{C\arabic{figure}}
\setcounter{figure}{0}

\begin{figure*}[b!]
\plottwo{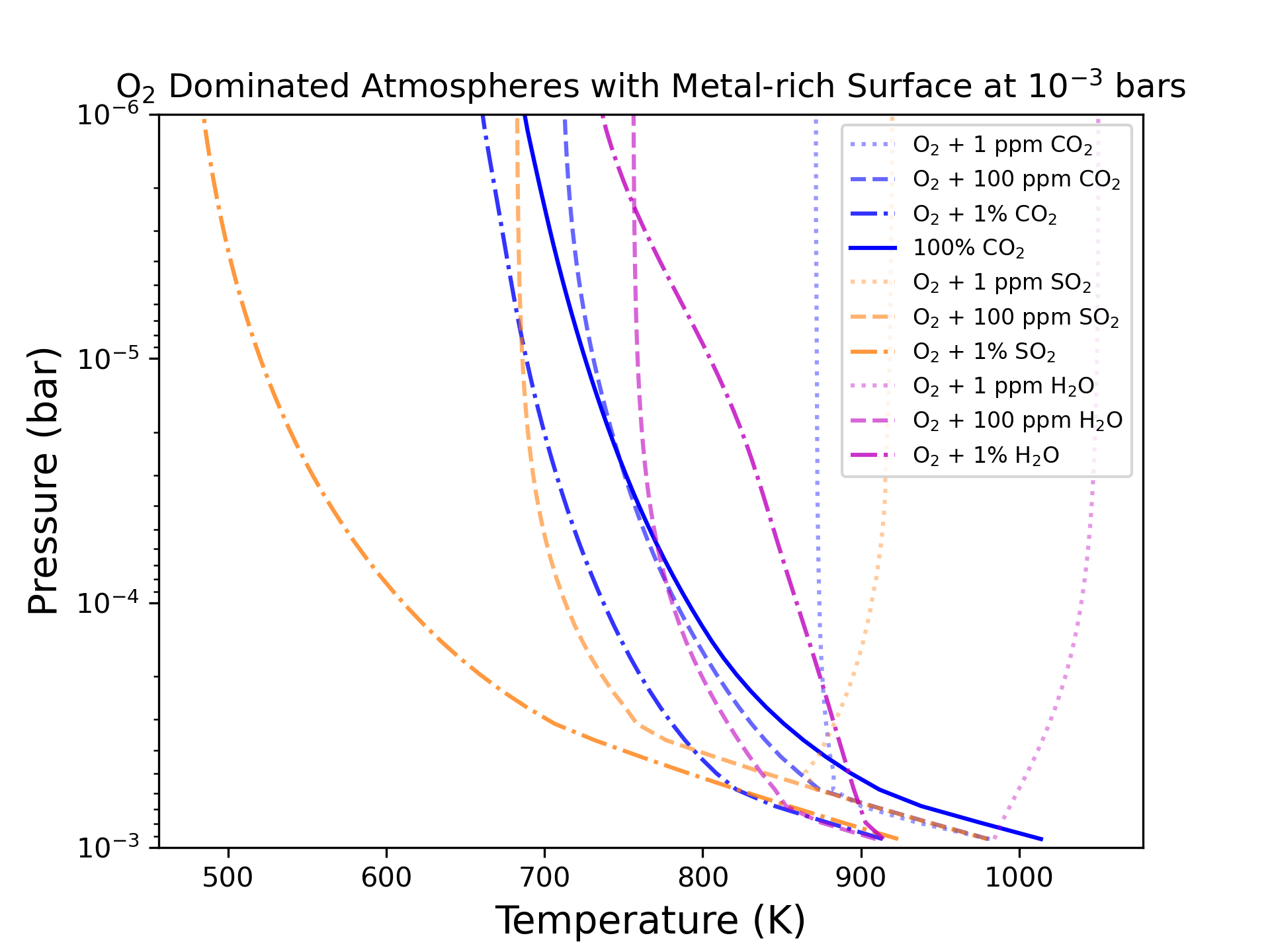}{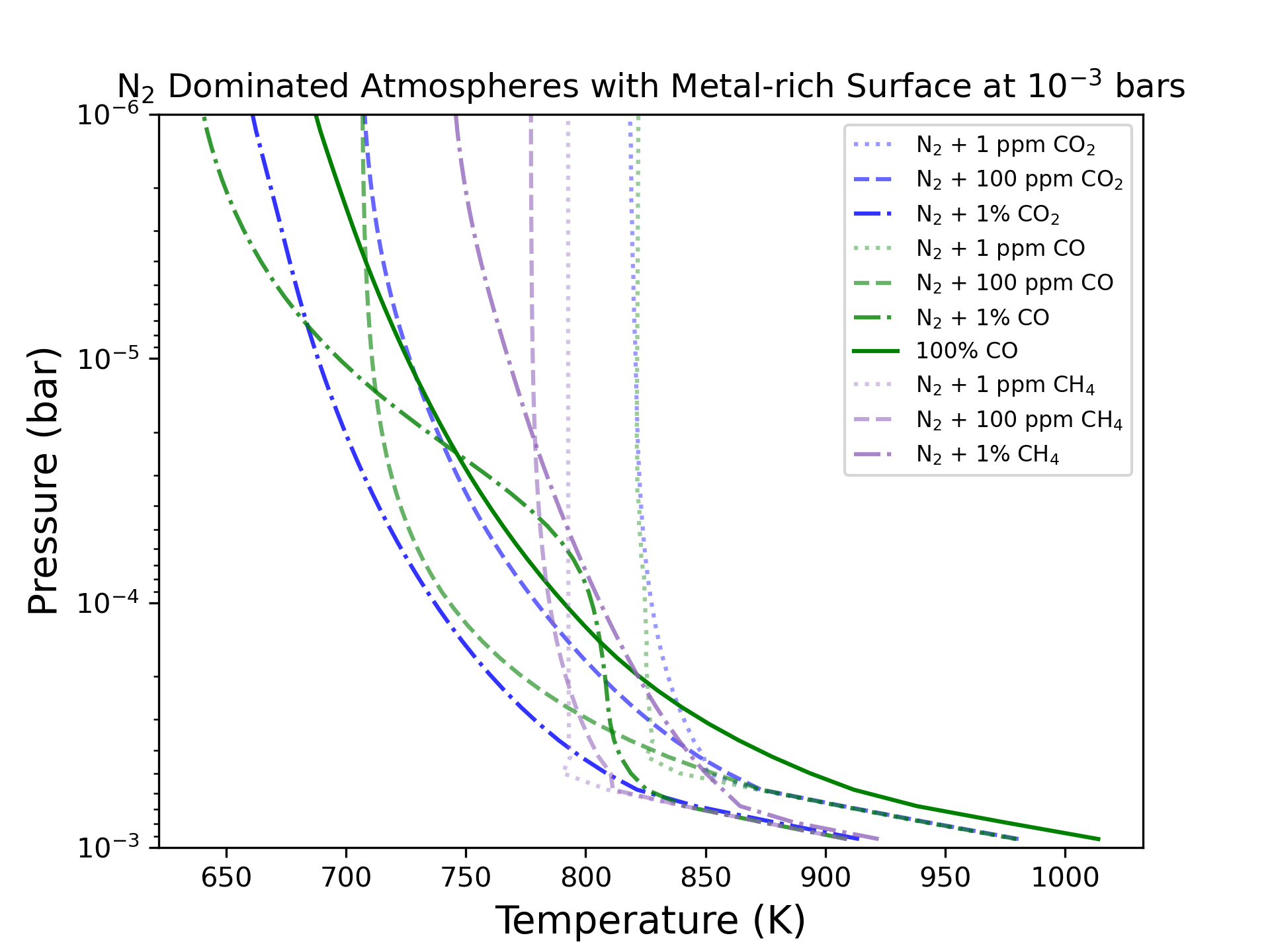}
\plottwo{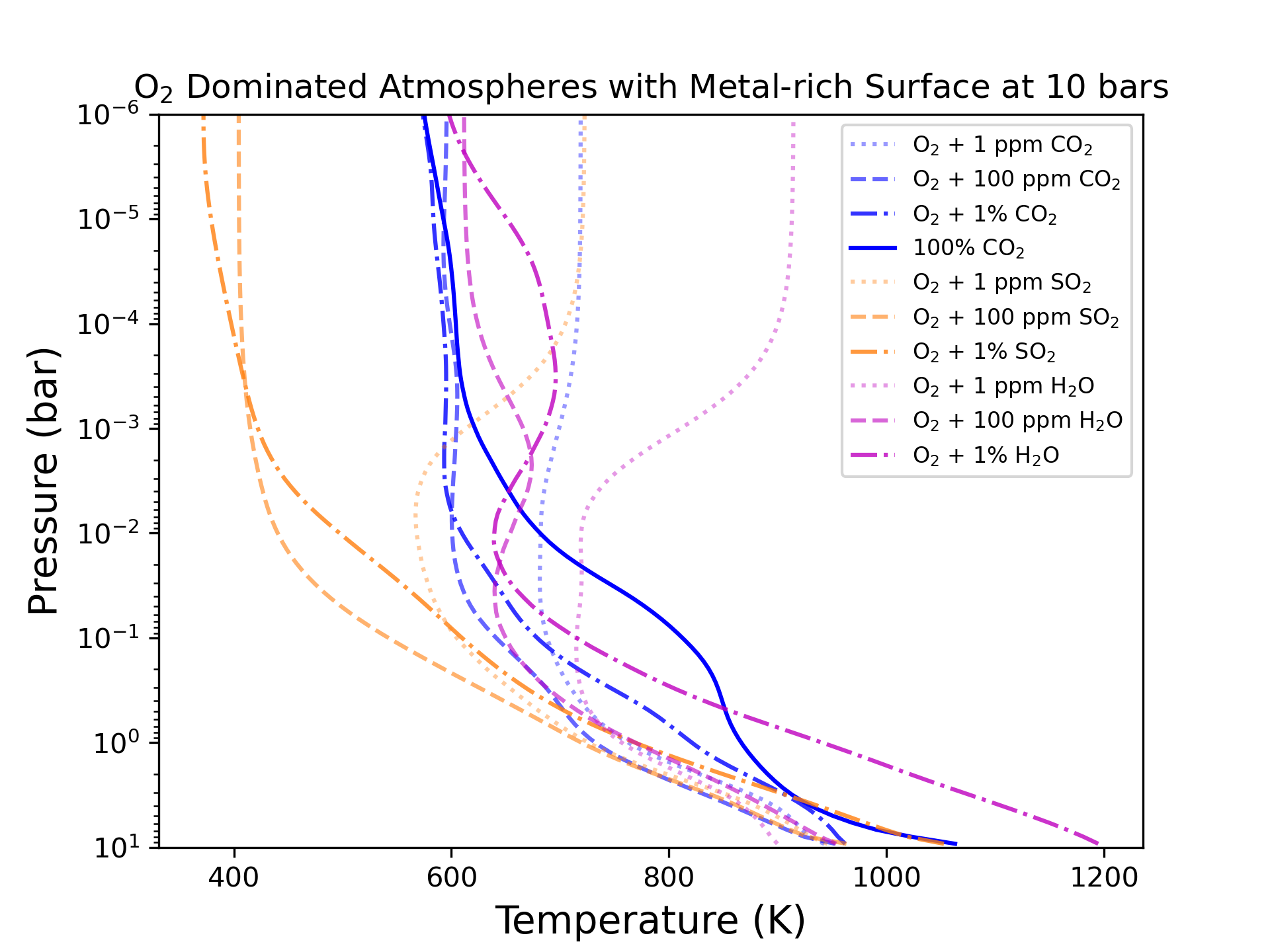}{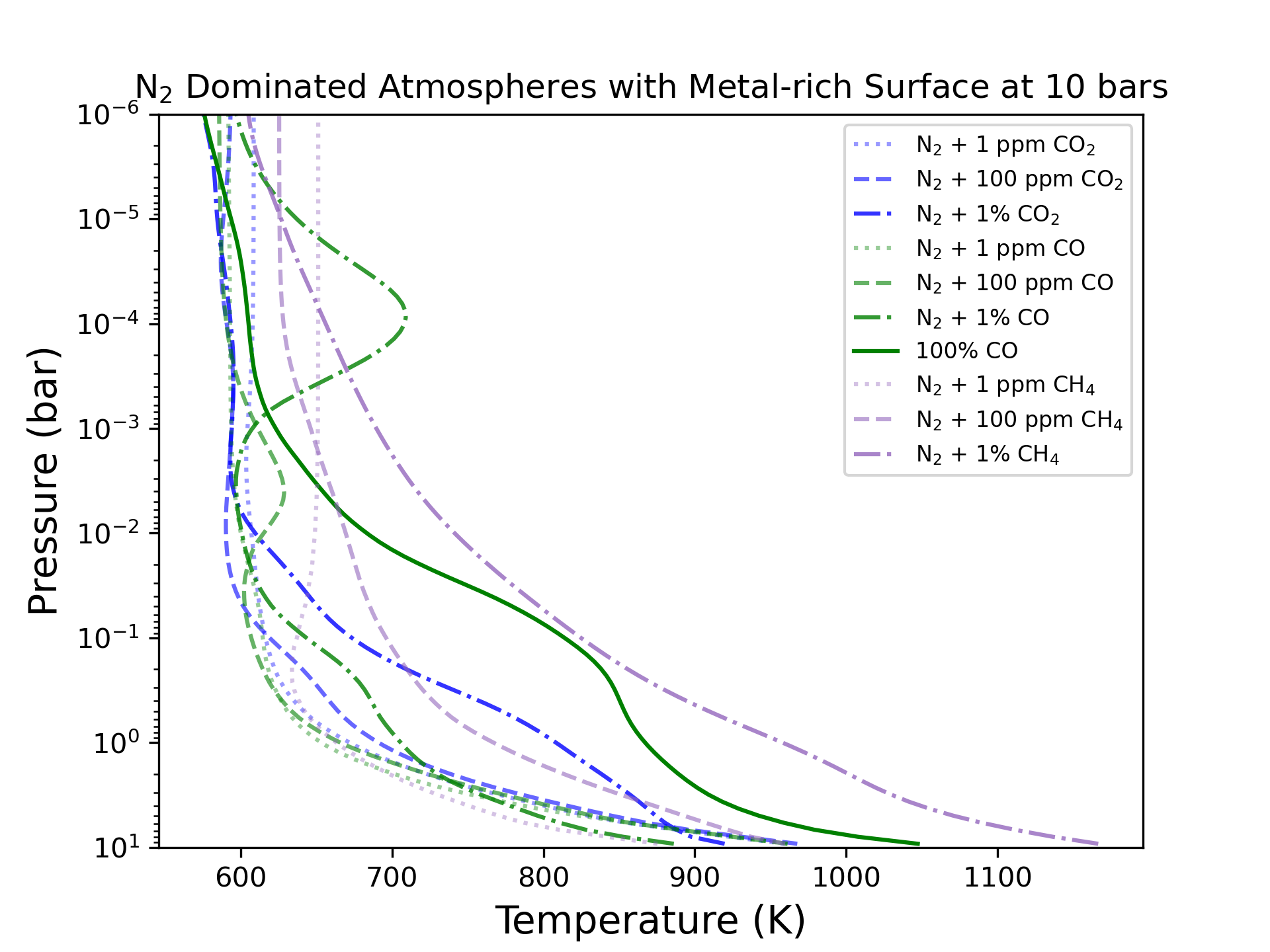}
\centering
\caption{Temperature-pressure (T-P) profiles for oxidizing/O$_2$-dominated atmosphere models (left) and reducing/N$_2$-dominated atmosphere models (right) assuming a surface pressure of 10$^{-3}$ bars (top) and 10 bars (bottom).}
\label{fig:tp}
\end{figure*}

\begin{figure*}
\centering\includegraphics[width=17cm]{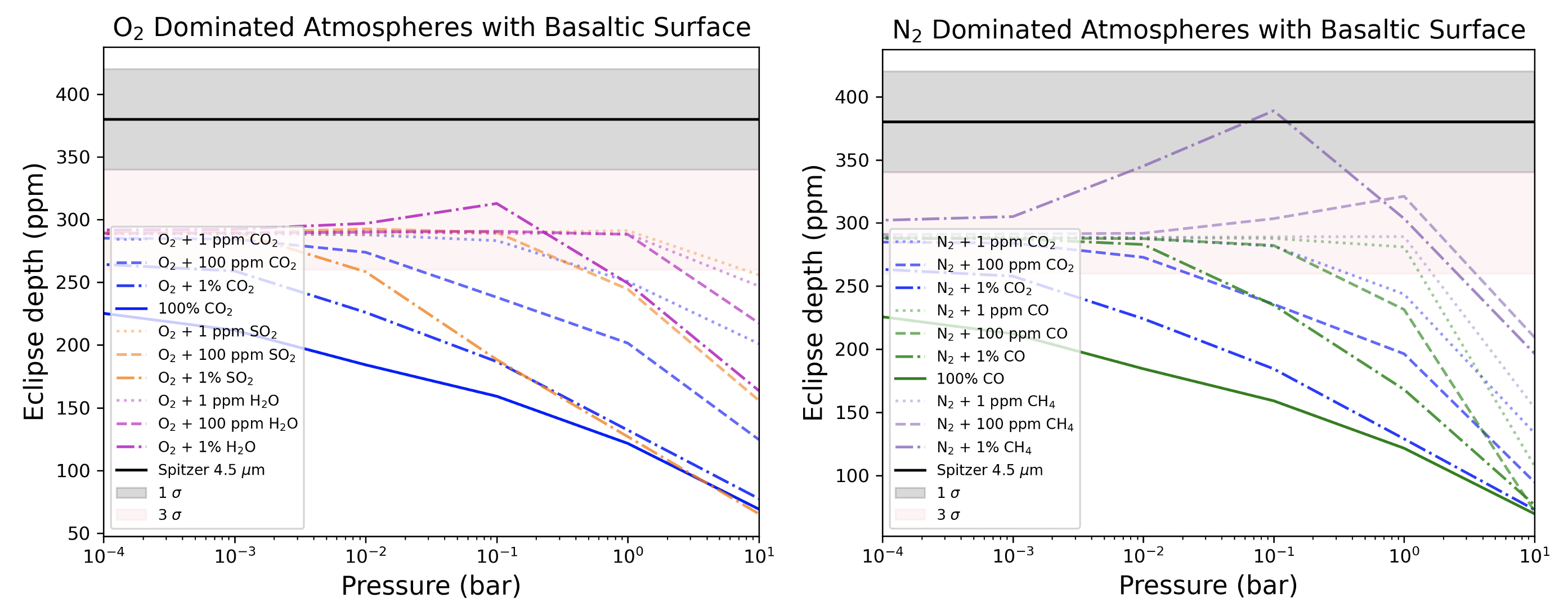}
\caption{Predicted Spitzer 4.5 $\mu$m eclipse depth for the explored atmospheric models in combination with a basaltic surface as a function of surface pressure compared to the Spitzer measurement (black horizontal line). Oxidizing/O$_2$-dominated atmospheres are on the left and reducing/N$_2$-dominated atmospheres on the right. The grey shaded area corresponds to 1 $\sigma$ uncertainty and the pink shaded area to 3 $\sigma$ uncertainty in the negative direction from the observed value.}
\label{fig:kreidberg_plot_basaltic}
\end{figure*}

\begin{figure*}[b]
\centering\includegraphics[width=12.6cm]{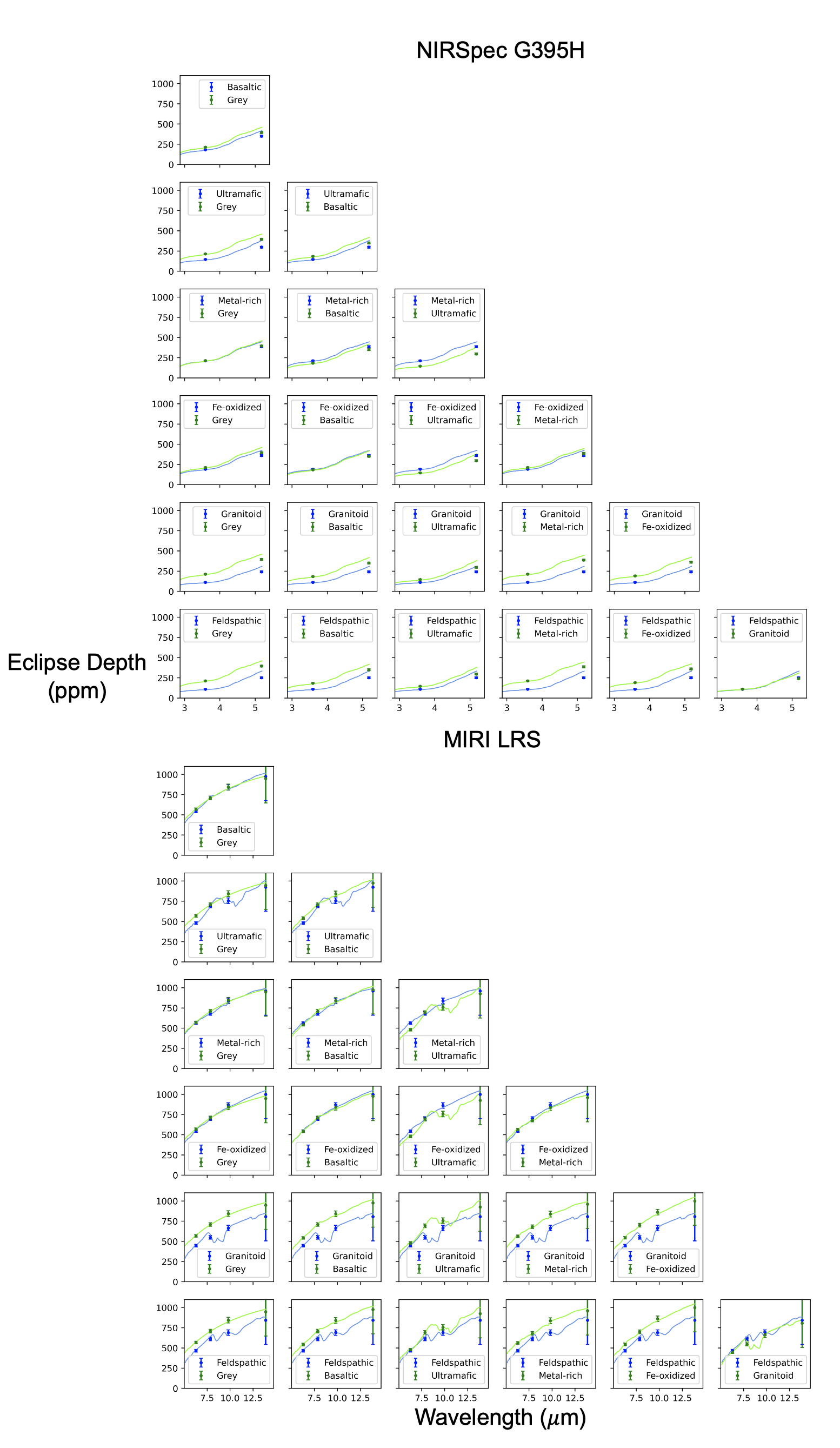}
\vspace{-0.3cm}
\caption{Secondary eclipse spectra of surface-only model pairs over the full range of the NIRSpec G395H (top) and MIRI LRS (bottom) bandpasses with overlaid mock data points at $R$ = 3. The error bars correspond to 5 secondary eclipse observations. The model spectra are downsampled to $R$ = 100 for clarity.}
\label{fig:distinguish_surf}
\end{figure*}

\begin{figure*}[b]
\centering\includegraphics[width=16cm]{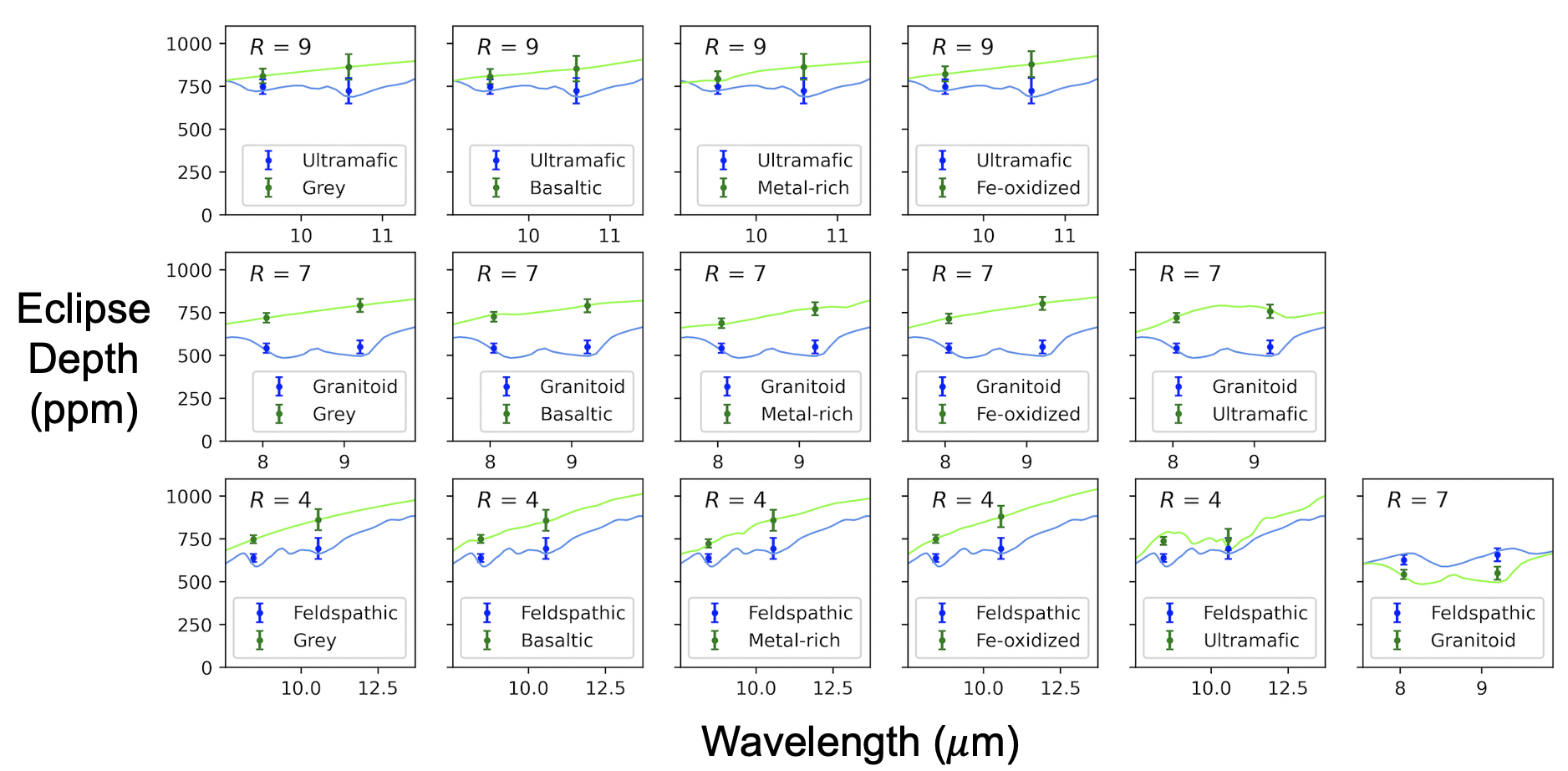}
\caption{Secondary eclipse spectra of surface-only model pairs over the range of selected surface features within the MIRI LRS bandpass with overlaid mock data points. Each case uses a resolution (shown in top left corner) to optimally match the spectral features. The error bars correspond to 5 secondary eclipse observations. The model spectra are downsampled to $R$ = 100 for clarity.}
\label{fig:distinguish_surf_isolated}
\end{figure*}

\begin{figure*}
\centering\includegraphics[width=17cm]{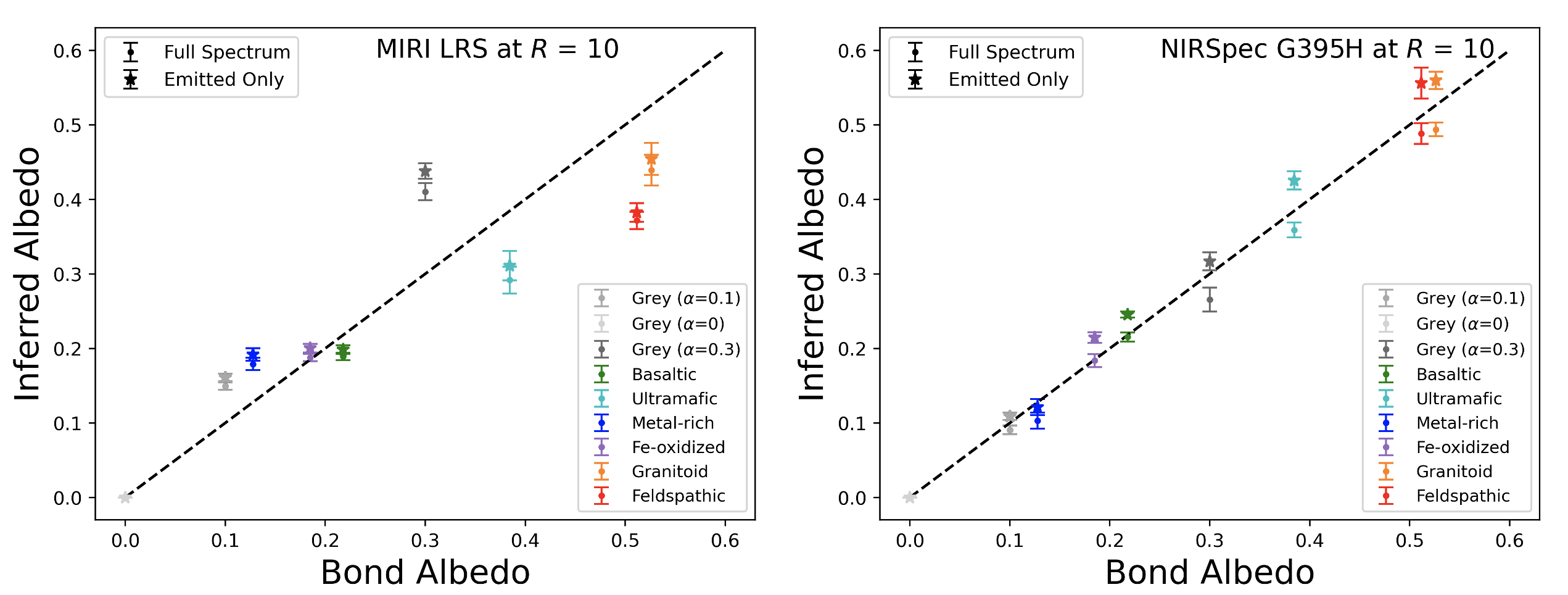}
\caption{{\bf Left:} Bond Albedo vs. Inferred Albedo for simulated MIRI LRS data at $R$ = 10. The black dotted line represents the location where the inferred and Bond albedos are equal. The inferred albedo is determined using once the full spectrum (i.e., both planetary emission and reflected light contributions) and once the emitted radiation only. The nominal case uses the full spectrum. {\bf Right:} Analogous to the left panel but using simulated NIRSpec G395H data at $R$ = 10.}
\label{fig:inf_albedo_emitted_only}
\end{figure*}

\begin{figure*}
\centering\includegraphics[width=17cm]{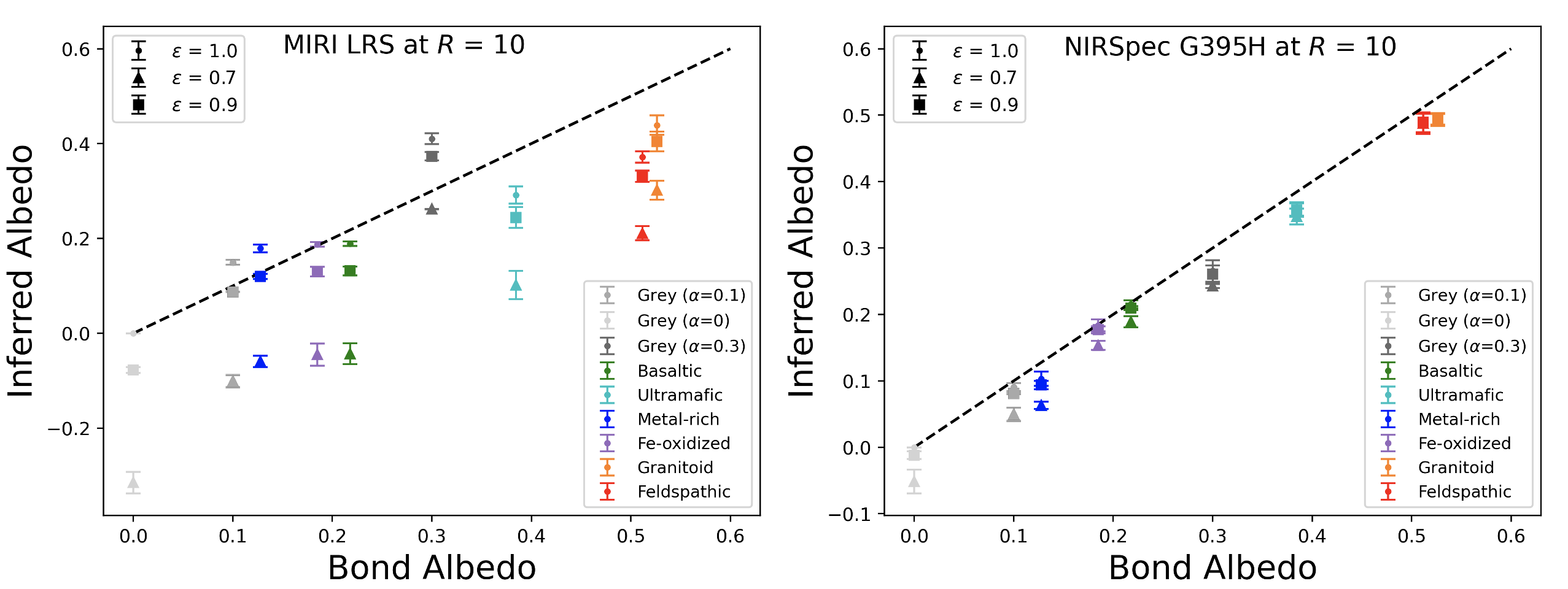}
\caption{{\bf Left:} Bond Albedo vs. Inferred Albedo for simulated MIRI LRS data at $R$ = 10. The black dotted line represents the location where the inferred and Bond albedos are equal. Three values of emissivity, 1.0, 0.9, and 0.7, are assumed for the planetary emission when inferring the temperature and the albedo. The nominal value used is 1.0. {\bf Right:} Analogous to the left panel but using simulated NIRSpec G395H data at $R$ = 10.}
\label{fig:inf_albedo_emissivity}
\end{figure*}

\begin{figure*}
\centering\includegraphics[width=7cm]{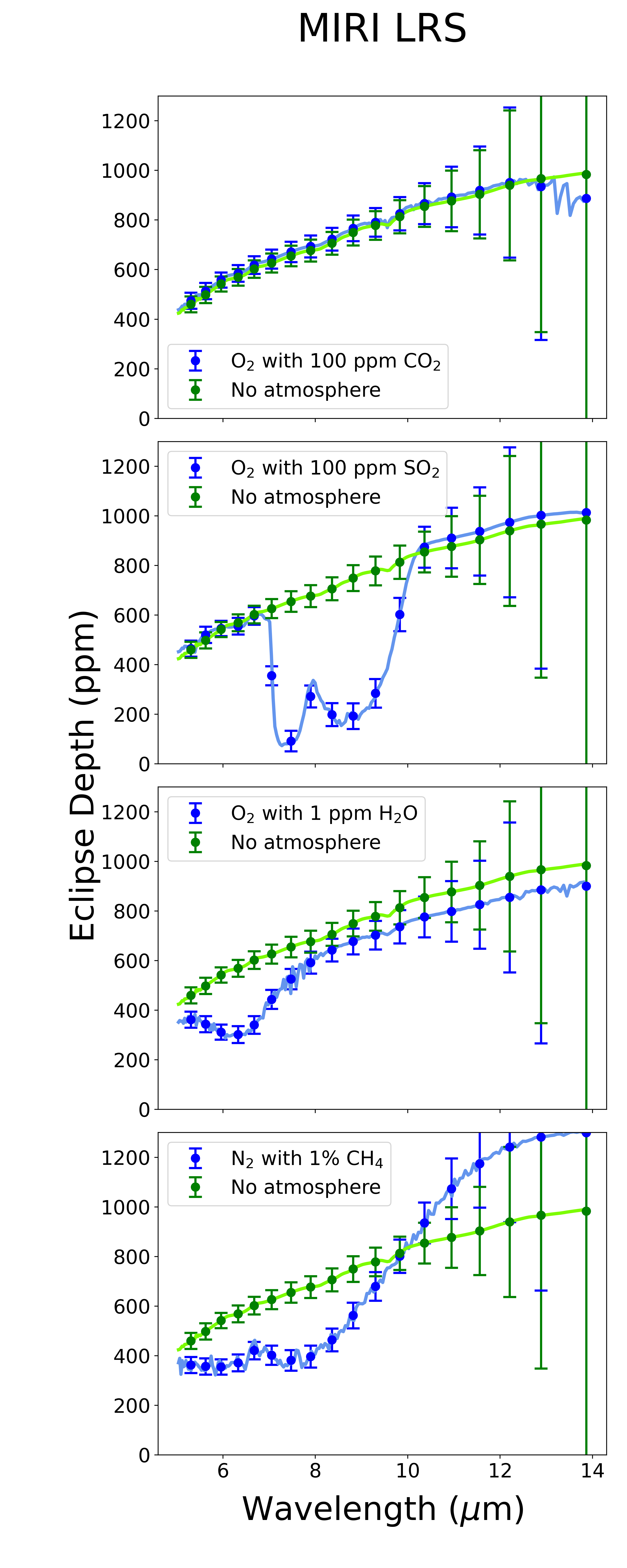}
\centering\includegraphics[width=7cm]{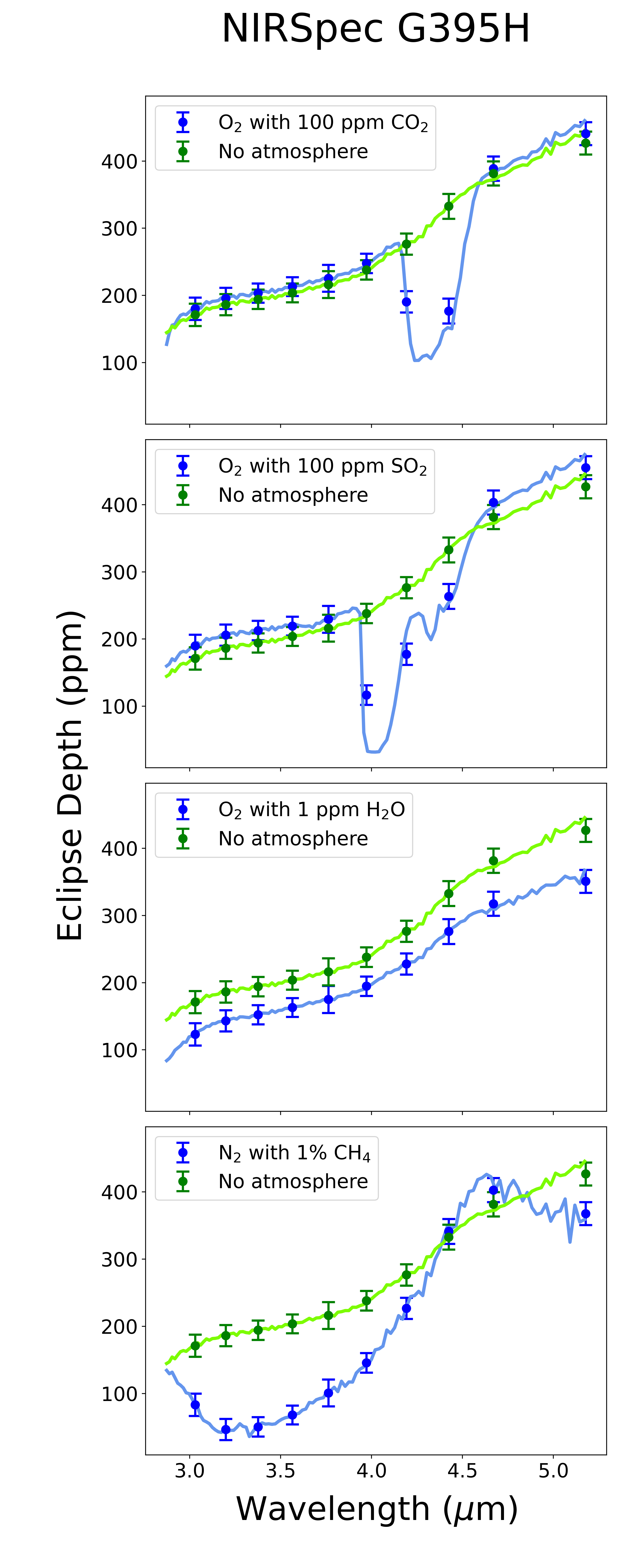}
\caption{Secondary eclipse spectra of atmospheric models vs. surface-only models, overlaid with MIRI LRS (left) and NIRSpec G395H (right) simulated data points at $R$ = 10 over the whole instrument bandpass. The metal-rich surface is used for this test. Error bars show uncertainties for 5 eclipse observations. Model spectra are downsampled from the native resolution to $R$ = 100 for clarity.}
\label{fig:atmo_v_noatmo}
\end{figure*}

\begin{figure*}
\centering
\plotone{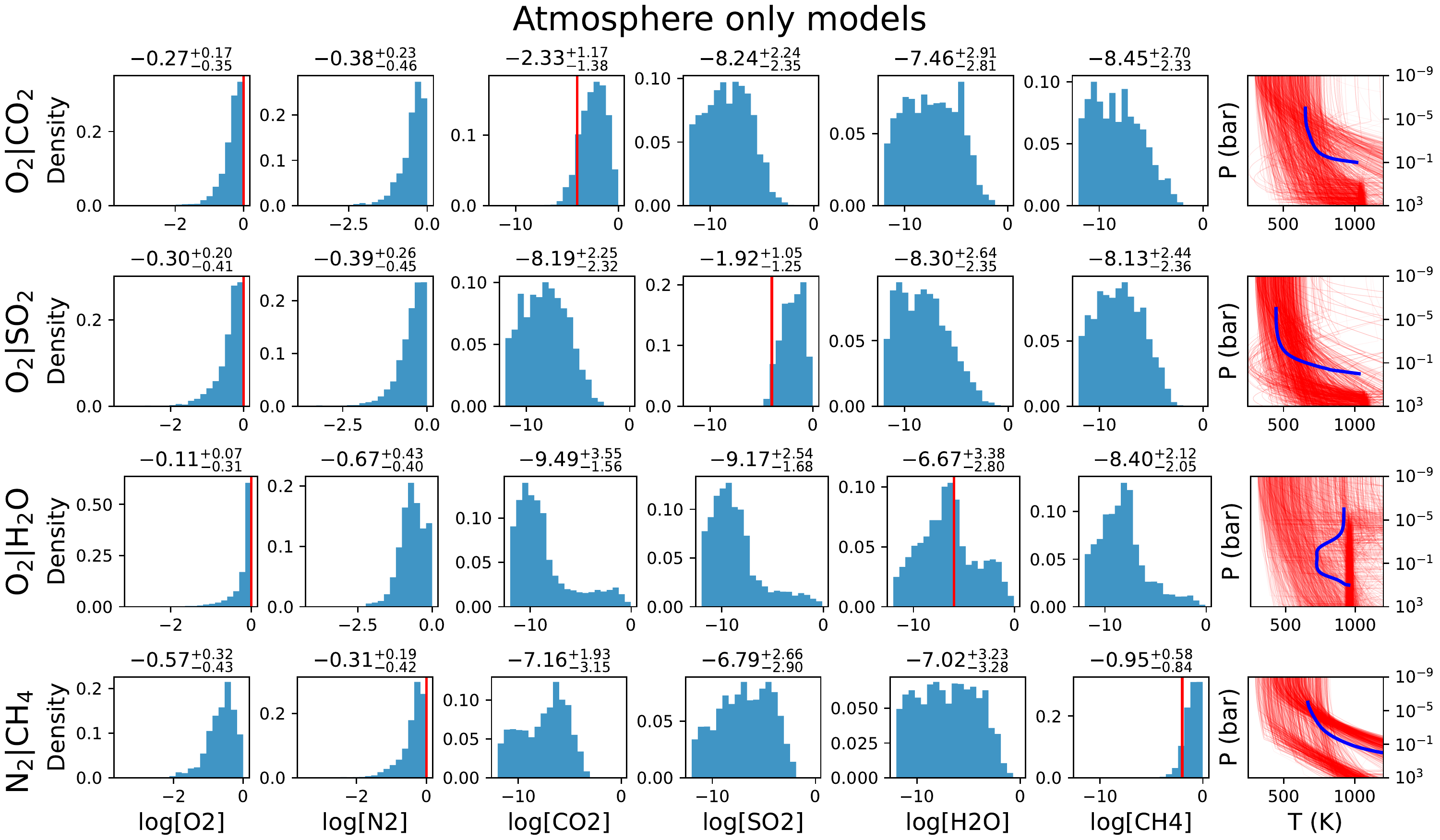}
\caption{Posterior distributions of retrieved volume mixing ratios using atmospheric forward models without a non-grey surface. The retrieved mean and 1-sigma width is given on top of each panel and the input mixing ratio (the ``true'' value) is shown as red, vertical line. The corresponding T-P profiles are shown on the right with the input profile (from the forward model) in blue and the profiles realizations from the retrieval modeling as red lines.}
\label{fig:retrieval_grid_atmo_only}
\end{figure*}

\begin{figure*}
\centering
\plotone{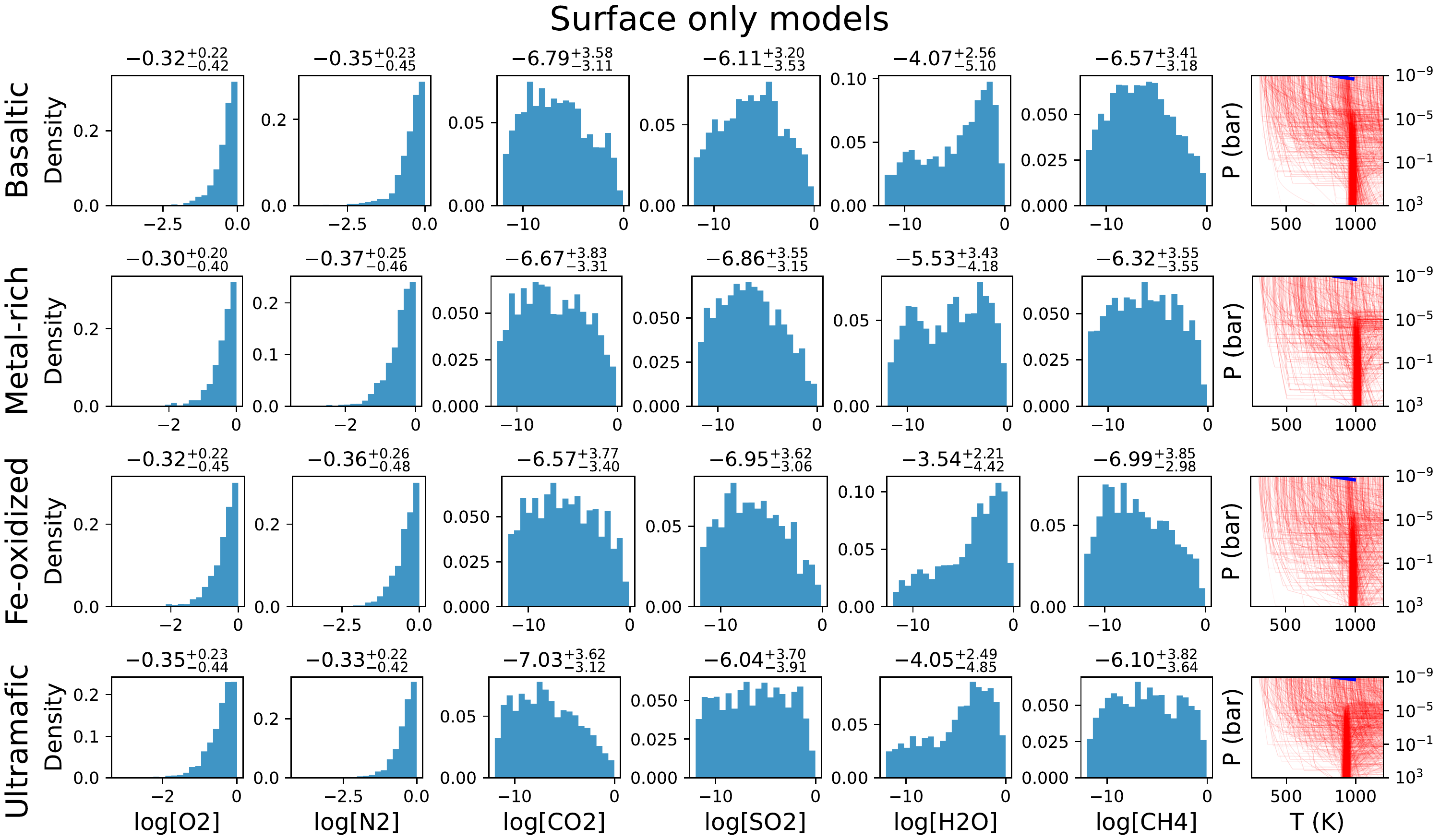}
\caption{Same as Fig.~\ref{fig:retrieval_grid_atmo_only} but showing here the retrieval results using the surface-only (no atmosphere) forward models with different surface crusts.}
\label{fig:retrieval_grid_surf_only}
\end{figure*}

\end{document}